\input epsf.sty
\overfullrule=0pt

\def\ctr#1{\noindent{\hfill#1\hfill}}

\def\simless{\mathbin{\lower 3pt\hbox
   {$\rlap{\raise 5pt\hbox{$\char'074$}}\mathchar"7218$}}}   
\def\simgreat{\mathbin{\lower 3pt\hbox
   {$\rlap{\raise 5pt\hbox{$\char'076$}}\mathchar"7218$}}}   

\font\bigbf=cmbx10 scaled\magstep1

%

\ctr{\bigbf LIGHT PROPAGATION IN INHOMOGENEOUS UNIVERSES}

\medskip

\ctr{\bigbf II. COSMOLOGICAL PARAMETER SURVEY}

\bigskip\bigskip

\ctr{PREMANA PREMADI$^{1,2}$, HUGO MARTEL$^3$,
RICHARD MATZNER$^{4,5}$, \& TOSHIFUMI FUTAMASE$^2$}

\bigskip\bigskip

\noindent $^1$ Department of Astronomy \& Bosscha Observatory,
               Bandung Institute of Technology, Indonesia

\noindent $^2$ Astronomical Institute, Tohoku University, Sendai, Japan

\noindent $^3$ Department of Astronomy, University of Texas, Austin, TX 78712

\noindent $^4$ Center for Relativity, University of Texas, Austin, TX 78712

\noindent $^5$ Department of Physics, University of Texas, Austin, TX 78712

\bigskip\bigskip\bigskip

\ctr{\bf ABSTRACT}

\bigskip

Using a multiple-lens plane algorithm, 
we study light propagation in inhomogeneous universes, for 43 different
{\it COBE}-normalized Cold Dark Matter models, with various values
of the density parameter $\Omega_0$, cosmological constant~$\lambda_0$, 
Hubble constant $H_0$, and rms density fluctuation $\sigma_8$. 
This is the largest cosmological
parameter survey ever done in this field. 
We performed a total of 3,798 experiments, each experiment consisting of
propagating a square  beam of angular size $21.9''\times21.9''$ composed of 
116,281 light rays from the observer up to redshift $z=3$. These experiments
provide statistics of the magnification, shear, and multiple imaging of
distant sources. The results of these
experiments can be compared with observations, and eventually
help constraining the possible values of the cosmological parameters. 
Additionally, they provide insight into the gravitational lensing process and
its complex relationship with the various cosmological parameters.

Our main results are the following:
(1) The magnification distribution depends mostly
upon $\lambda_0$ and $\sigma_8$. As $\sigma_8$ increases, the low-tail
of the magnification distribution shifts toward lower magnifications, 
while the high-tail is
hardly affected. The magnification
distribution also becomes wider as 
$\lambda_0$ increases. This 
effect is particularly large for models with $\lambda_0=0.8$.
(2) The magnification probability $P_m$ is almost independent of 
$\sigma_8$, for
any combination of $\Omega_0$, $\lambda_0$, $H_0$, indicating that
$P_m$ does not depend strongly on the amount
of large-scale structure.
(3) The shear distribution, like the magnification distribution,
depends mostly upon $\lambda_0$ and $\sigma_8$.
The shear distribution
becomes wider with increasing $\sigma_8$ and increasing $\lambda_0$. 
The similarities between the properties of the magnification
and shear distributions suggests that both phenomena
are caused by weak lensing. 
(4) About 0.3\% of sources have multiple images. 
The double-image probability $P_2$ increases strongly with $\lambda_0$
and is independent of $\Omega_0$, $H_0$,
and $\sigma_8$. 
(5) The distribution of image separations depends
strongly upon $\lambda_0$, and is independent
of $\sigma_8$.  
Summarizing these results, we find that (1) 
The properties of gravitational lensing, both weak and strong, depend
much more strongly upon $\lambda_0$ than any
other cosmological parameter, (2) magnification and shear are examples
of weak lensing caused primarily by the distribution of background
matter, with negligible contribution form galaxies, while
multiple images and rings are examples of strong lensing,
caused by direct interaction with galaxies, with negligible
contribution from the background matter. 
Observations of weak lensing can be used to determine
the cosmological constant and the density structure of the
universe, while observations of strong lensing
can be used to determine the cosmological constant and the internal
structure of galaxies and clusters. Gravitational lensing depends much
more weakly upon $\Omega_0$ and $H_0$ than $\sigma_8$ and
$\lambda_0$, making a determination of these parameters from
observations more difficult.

\vfill

\ctr{Accepted for publication in {\sl The Astrophysical Journal Supplement}
(June 2001)}

\eject

%

\ctr{\bf 1.\quad INTRODUCTION}

\medskip

\ctr{\bf 1.1.\quad Importance of Gravitational Lensing in Cosmology}

\medskip

The evolution of a homogeneous, isotropic, expanding universe comprised
of nonrelativistic matter
can be described in terms of three
parameters: the Hubble constant $H_0$, the density parameter $\Omega_0$,
and the cosmological constant $\lambda_0$.\footnote{$^6$}{Additional parameters 
must be included if the universe contains additional components such as radiation or quintessence.} Any combination
of these three parameters corresponds to one particular cosmological model.
The large-scale structure of the universe, galaxies, clusters, 
superclusters, and voids, represents the deviations from this
overall homogeneity and isotropy. The most conservative assumption
is that these structures originate from 
primordial fluctuations that grow with time as a result of 
gravitational instability. 
The fluctuations originate from a Gaussian random process, and
are characterized entirely by a density fluctuation
power spectrum.  
Determining the value of the cosmological
parameters, and the correct model of large-scale structure formation,
is the most important challenge of observational and physical cosmology.

Observing the present, nearby universe can only provide partial information
about the value of the cosmological parameters, and the power spectrum
of density fluctuations (see, e.g. Martel 1995).
To unambiguously determine the correct cosmological model, it is crucial 
to observe the universe
at high redshift (i.e. at large distances), to study its {\it past} 
structure and evolution, as well as its {\it global} geometry. Of all
known redshift-dependent observational tests, gravitational lensing of distant
sources is certainly the most powerful and most promising one. 

Because sources must be located at cosmological distance for lensing
effects to be important,\footnote{$^7$}{We are not considering here the
possibility of microlensing by nearby
massive objects (like MACHO's).}
most early work on gravitational lensing
has focused on QSO's, being the farthest objects in the universe that
were sufficiently luminous to be observable. However
recent developments in instrumentation enable us to consider other
sources at cosmological distances
besides QSO's. The {\sl Hubble Space Telescope} has deepened our
field of view tremendously. The {\sl Hubble Deep Field} (Williams et al. 1996)
and {\sl Southern Hubble Deep Field} (Williams et al. 1998)
are the deepest sets of exposures even taken. Supernovae at
redshifts $z=0.95$ and $z=1.32$
have been observed in these fields (Gilliland, Nugent, \& Phillips 1999).
These are the farthest supernovae ever seen, and
as new instruments become available, it will be possible to observe
supernovae at even larger redshifts, possibly up to $z=20$
(Marri \& Ferrara 1998).

High-$z$ Type Ia supernovae have the potential to be much more useful
than QSO's for gravitational lens studies. Unlike QSO's, Type Ia SNe
are {\it nearly standard candles} (if we exclude the low-luminosity
ones, which have limited cosmological use anyway). Although there are some
variations in luminosity among Type Ia SNe, these variations are
well-understood, and luminosities can be corrected (see the review
by Filippenko \& Riess 1998, and references therein). Having standard
candles at cosmological distances has major implications for gravitational
lensing studies. When a distant source is lensed by the intervening
large-scale structure of the universe, it is usually magnified or
demagnified, without multiple imaging. Only a small fraction of magnified
sources are lensed sufficiently strongly to have multiple images, as we
will show in this paper. Without
multiple imaging (and without the resolution necessary to resolve
the actual shape of the image), a magnified source may be mistaken for
a source with a different luminosity, {\it unless the source is a standard
candle}. If it is a standard candle, we can estimate its brightness from
its known luminosity and estimated distance (based on redshift measurements).
If the estimated brightness does not match the observed one, then the
source is magnified, the magnification factor $\mu$ is given by the ratio
of brightnesses, and the combination of the factor $\mu$ and the redshift
$z$ of the source yields information on the nature of the structures
responsible for the lensing.

In recent years, gravitational lenses have been used to estimate or put
limits on the values of the cosmological parameters. These studies have 
focused on the cosmological constant $\lambda_0$ 
(Fukugita, Futamase, \& Kasai 1990; Turner 1990; 
Kochanek 1992, 1996a, 1996b; 
Krauss \& White 1992; Maoz \& Rix 1993; Bloomfield Torres \& Waga 1996;
Im, Griffiths, \& Ratnatunga 1997; Chiba \& Yoshii 1997, 1999), the
density parameter $\Omega_0$ (Yoshida \& Omote 1992; Mart\'\i nez-Gonz\'alez,
Sanz, \& Cay\'on 1997), 
the $(\Omega_0,\lambda_0$) plane (Asada 1997; Park \& Gott 1997;
Falco, Kochanek, \& Mu\~noz 1998), the deceleration parameter $q_0$ 
(Wambsganss et al. 1997), the difference $\Omega_0-\lambda_0$
(Cooray, Quashnock, \& Miller 1999), or the Hubble constant $H_0$ 
(Watanabe, Sasaki, \& Tomita 1992; Falco et al. 1997;
Keeton \& Kochanek 1997; Kundi\'c et al. 1997). 
All these studies were focussed on one or two particular cosmological
parameters.
In any such study, some assumption must be made about the value of the 
cosmological parameters that are {\it not} being determined.
For instance, studies focusing on the cosmological constant all assume a 
flat universe ($\Omega_0+\lambda_0=1$), while the studies focusing on
the density parameter all assume a vanishing cosmological constant
($\lambda_0=0$). These assumptions are motivated more
by theory than observations. The flatness of the universe
is a requirement of the standard inflationary scenario
(Guth 1981; Linde 1982; Albrecht \& Steinhardt 1982), while a vanishing
cosmological constant is the only known way to solve the
cosmological constant problem (see, e.g. Weinberg 1989).
As we shall see, several of these assumptions must be reconsidered
in the light of recent observations.

\bigskip\smallskip

\ctr{\bf 1.2.\quad The Need for a Full Cosmological Parameter Survey}

\medskip

During the 1980's, there was a strong theoretical prejudice
in favor of the Einstein-de~Sitter model ($\Omega_0=1$, $\lambda_0=0$).
This model was particularly appealing because it
satisfied the flatness requirement of
the standard inflationary scenario, did not require a cosmological 
constant, and had fewer free parameters than any
other model. Numerical simulations of large-scale structure formation 
(Davis et al. 1985; White et al. 1987) 
showed that an Einstein-de~Sitter model in which 
the bulk of the matter is in form of Cold Dark Matter (CDM) 
could satisfy all the 
observational constrains known at the time. The theoretical prejudice
for the Einstein-de~Sitter model, supported by the numerical simulations
of large-scale structure formation,
lead to the quasi-universal acceptance of the {\it Standard Cold Dark Matter
Model} as the model of choice during the late 1980's. The only real
difficulty that this model faced was the so-called age problem. 
In the standard CDM model (or any Einstein-de~Sitter model), $H_0t_0=2/3$.
Independent measurements of the Hubble constant $H_0$ and the age of
the universe $t_0$ showed that this constraint could be satisfied only
marginally, as one needed to select the smallest possible values for
both $H_0$ and $t_0$. In particular, the value of $H_0$ had to
be very near $50\rm\,km\,s^{-1}Mpc^{-1}$. Still, this difficulty was not 
regarded as serious enough to discard the standard CDM model.

Since then, there has been a tremendous increase in the number and quality
of cosmological observations. These observations fall into three
categories: low redshift observations (age of the universe, Hubble constant,
deuterium abundance, baryon fraction in clusters, abundance of
clusters, shape of the cluster mass function, large-scale velocity field),
intermediate-redshift observations (Type Ia supernovae, gravitational lensing,
cluster evolution, Ly$\alpha$ clouds), and high-redshift observations
(cosmic microwave background).
For details, we refer the reader to the recent reviews by
Steigman, Hata, \& Felten (1999), 
Martel \& Matzner (2000), Ross \& Harun-or-Rashid (2000), and Wang et al. 
(2000). These new observations impose numerous 
constrains on the cosmological parameters and the possible models of
large-scale structure formation, and invariably argue against the
standard CDM model. While the existence of cold dark matter is not ruled out,
and indeed remains the best approach to explain galaxy and large-scale 
structure formation, the values of the cosmological parameters in
the standard CDM model are essentially ruled out, as nearly all observations
now support a universe with $\Omega_0<1$.

This forces us to consider alternatives to the standard CDM model.
A low-density CDM model with $\Omega_0<1$, known as ``Open CDM'' or OCDM 
is in much better agreement with most observations, but
does not satisfy the flatness requirement of inflation.
However, the possibility of ``open inflation,'' or
inflation without the flatness requirement, 
has been suggested by several authors (Ratra \& Peebles 1994;
Bucher, Goldhaber, \& Turok 1995; Yamamoto, Sasaki, \& Tanaka
1995; Linde 1995; Linde \& Mezhlumian 1995), and is supported
by some recent observations (Filippenko \& Riess 1998).
But if we stick to the flatness requirement of the standard inflationary 
scenario, then we must include in the model an additional smooth
component to account for the difference between $\Omega_0$ and 1.
Numerous candidates
have been proposed (see, e.g. Fry 1985; Charlton \& Turner 1987), but the
most promising ones at present are the cosmological constant and the 
quintessence (see Turner 1999 for a recent review). The corresponding models,
$\Lambda$CDM and QCDM, have $\Omega_0+\Omega_{\rm S}=1$,
where $\Omega_{\rm S}$ is the effective density parameter associated with
the smooth component.

In the standard CDM model, the primordial power spectrum
of density fluctuations is assumed to have the
Harrison-Zel'dovich form $P(k)\propto k$ at large scales. 
This power spectrum can be modified by introducing a ``tilt.'' 
In this Tilted CDM, or TCDM model, the primordial power spectrum $P(k)$ 
varies as $P(k)\propto k^n$ at large scales, where the primordial exponent
$n$ can differ from unity. With an exponent $n$ of order $0.8-0.9$, it is
possible to reconcile the constrains imposed by the CMB anisotropies and
the present abundance of clusters with a density parameter $\Omega_0<1$,
without the addition of a cosmological constant or quintessence.

To discriminate between these various models, we must determine the
values of the cosmological parameters. The standard CDM model had no
free parameter: the density parameter was unity, and the Hubble constant was
severely constrained by the age problem. These alternative CDM models have 
many free parameters: the density parameter $\Omega_0$, the cosmological
constant $\lambda_0$ (or the effective density parameter
associated with quintessence), the Hubble constant $H_0$, 
and the tilt $n$ of the primordial 
power spectrum. Gravitational lensing of distant
sources provides a powerful tool for studying the structure and evolution
of the large-scale structure in the universe, and eventually constrain the 
value of these cosmological parameters. However, all the previous
studies listed in \S1.1 focused on only one or two parameters, while 
assuming some particular values for the other parameters. 
If the correct cosmological model has several free parameters, it is clearly 
impossible to determine the value of these parameters separately. This
would be like looking for a fugitive in a city, but limiting the search to
one street and one avenue only. A full survey of the cosmological
parameter space is required in order to determine or limit the values of 
all cosmological parameters {\it simultaneously}. In this paper, we present 
the first study of light propagation in inhomogeneous universe that surveys
the full 4-parameter phase-space formed by $\Omega_0$, $\lambda_0$,
$H_0$, and $n$.

\bigskip\smallskip

\ctr{\bf 1.3.\quad Objectives}

\medskip

There are two distinct objectives to this study. The first one is to 
determine the properties of lensed sources, such as their
magnification distributions, shear distributions, image separations, and so on;
properties which depend upon the value of the cosmological parameters.
By directly comparing the predictions of the simulations
with current and future observations of gravitational lens systems, 
we hope to eventually constrain the values of the 
cosmological parameters, and the possible scenarios of large-scale
structures and galaxy formation. 

The second objective is more theoretical in nature. 
By studying the properties of gravitational lenses, and how these properties
depends upon the values of the cosmological parameters,
individually or in combination, we can gain insight into the phenomenon of 
light propagation in inhomogeneous universe and its relationship to the 
underlying cosmological model. 
This objective is more ambitious than the first one: instead of 
merely determining which particular
model reproduces observations better, we wish
to understand the {\it reason} which favors this particular model.
Achieving this second objective requires that we
extend the parameter survey to regions of 
the parameter space that are not particularly favored by observations.
This forced us to consider a record number of cosmological models
in this study, and even with 43 models, our covering of the 4-dimensional
parameter space is patchy at best. 

The various effects of gravitational lensing can be divided into two broad
categories: weak lensing and strong lensing.\footnote
{$^8$}{Microlensing would be
a third category, which we are not considering in this paper.} 
Weak lensing is caused by the
smooth distribution of matter, with moderate density contrast, located
between the source and the observer. The magnification or demagnification
of sources, or the shear --- circular sources having elliptical images
--- are examples of weak lensing. Weak lensing provides an unbiased
information about the matter distribution in the universe, as well
as the underlying geometry, that can be used to constrain cosmological
models (Bacon, Refregier, \& Ellis 2000; van Waerbeke et al. 2000;
Munshi \& Coles 2000; Jain, Seljak, \& White 2000). 
Strong lensing involves direct interaction
between the beam and large mass concentrations such as galaxies and cluster 
of galaxies. Strong lensing causes spectacular events such as multiple
images, giant
arcs, and Einstein rings, providing mostly information
about the density structure of the lens itself
(Schneider, Ehlers, \& Falco 1992, hereafter SEF, chapter 8), though 
some properties of strong lensing can be used to determine the cosmological
parameters as well. For instance, the time delay between multiple images
of a single source can be used to determine the Hubble constant.
Most previous studies have focused either on weak lensing or strong lensing.
One important goal of our work is to study the properties of weak and
strong lensing simultaneously. To achieve this, we must use an algorithm 
that can resolve cosmic structures over a very large dynamical range in
length, from the size of cluster and superclusters down to the central
cores of galaxies. This lead to the development of a new version of the
multiple lens-plane algorithm, based on earlier work by Jaroszy\'nski
(see \S2 below). 

Because of the nature of these objectives, this project involved an
amount of effort that is quite substantial compared with similar
studies that have been performed and published by various authors in
recent years. We are studying a record number of cosmological models, 43.
We have performed a total of 3,798 ray-tracing experiments.\footnote{$^9$}{An
additional 101 experiments were performed afterward, to answer some specific
questions raised by an anonymous referee, concerning the angular size of
the sources.} In order
to study the properties of weak and strong lensing simultaneously, 
we used beams composed of a very large number of light rays, 116,281.
Overall, we have simulated the propagation in inhomogeneous
universes of 441,635,238 light rays. We have generated the images
of 3,137,675 extended sources located at redshift $z=3$. For each and
every one of these images, we have computed the magnification and
aspect ratio, and whenever we encountered special kinds of images, such as
multiple images and rings, we studied their properties, such as 
image separations, brightness ratios, and hole diameters. The
calculations started on a Cray J90 supercomputer, and later moved to a
more powerful Cray SV1 supercomputer. The $N$-body simulations used to 
generate
the large-scale structure for the various cosmological models 
(3 simulations for each of the 43 models, for a total
of 129 $\rm P^3M$ simulations with one quarter million
particles each) took about 2000 CPU hours,
while the ray-tracing experiments took about 600 CPU hours. 

The remainder of this paper is organized as follows: in \S2, we describe
the numerical algorithm used for the simulations. In \S3, we describe the
cosmological models included in the study, the ray-tracing experiments,
and the technique used for analyzing the results. 
In \S4, we review the various elements that affect gravitational lensing.
Results are presented in \S5, including the magnification distributions
(\S5.1), the magnification probability (\S5.2), the shear distributions
(\S5.4), the multiplicity of images (\S5.5), the distribution of
image separations (\S5.6), and the properties of Einstein Rings (\S5.7).
In \S6, we discuss the various approximations implied by the algorithm
and their possible effects on the results.
Summary and conclusions are presented in \S7.

\bigskip\smallskip

\ctr{\bf 2.\quad THE MULTIPLE LENS-PLANE ALGORITHM} 

\medskip

Our numerical algorithm was described in detail by
Premadi, Martel, \& Matzner (1998, hereafter Paper I), and convergence
tests of the algorithm were presented by Martel, Premadi, \& Matzner (2000).
In this section, we give a brief summary of the method, 
and describe some minor 
refinements that have been introduced into the algorithm since 
the publication of Paper~I.

A light ray traveling from a distant source to the observer is affected
continuously by the distribution of matter it encounters along its trajectory.
The multiple lens-plane algorithm (SEF, and references therein) 
consists of approximating this continuous effect
by a finite number of instantaneous deflections, caused by the matter 
distribution encountered at various locations along the trajectory of the ray.
To implement this method, we
divide the space between the source and the observer into 
redshift intervals, and project the matter inside each interval onto 
a plane normal to the line of sight, called a lens plane. 
Every lens plane deflects the light rays that go through it, and
the deflection angles can be computed using geometrical optics.
We can then follow the evolution of a light beam
propagating through the universe, by
adding successively the contributions of each lens plane to the deflection 
and deformation of the beam. During the past 
decade, this method has been one of the main tools for studying the properties
of gravitational lenses located at cosmological distances 
(Blandford \& Nayaran 1986; Blandford \& Kochanek
1987; Schneider \& Weiss 1988a, b; 
Jaroszy\'nski et al. 1990; Jaroszy\'nski 1991, 1992;
Babul \& Lee 1991; Bartelmann \& Schneider 1991; 
Wambsganss, Cen, \& Ostriker 1996; 
Bartelmann et al. 1998; Couchman, Barber, \& Thomas 1999;
van Waerbeke, Bernardeau, \& Mellier 1999;
Hamana, Martel, \& Futamase 2000; Jain et al. 2000;
see also Kochanek \& Apostolakis 1988; Paczy\'nski \& Wambsganss 1989;
for an interesting alternative, see Fluke, Webster, \& Mortlock 1999).

What usually distinguishes a particular version of the multiple lens-plane
algorithm from other versions is the method used for
representing the surface density on the lens planes.
The issue of length resolution is critical.
The magnification of the images of distant sources depends essentially on
the amount of matter located near the beam, along the line of sight. 
However, the deformation (or shear) of the images results primarily
from the tidal influence of distant matter. Therefore, to accurately 
simulate the effect of both magnification and shear, we must reproduce
the surface density of the lens planes over the largest possible dynamical
range in length. Our algorithm achieves a very high length resolution
by combining numerical simulations of large-scale structure formation with
a Monte-Carlo method for locating galaxies inside these
structures. This approach was pioneered by Jaroszy\'nski (1991, 1992).

\bigskip\smallskip

\ctr{\bf 2.1.\quad Large-Scale Structure Formation}

\medskip

We use a
Particle-Particle/Particle-Mesh ($\rm P^3M$) code 
(Hockney \& Eastwood 1981) to
simulate the formation and evolution of large-scale structure in 
the universe. The algorithm produces snapshots of the large-scale
structure at
various redshifts. If we interpret these redshifts in terms of distances
from the observer, we can treat each snapshot as representing a different
region of the universe, and by combining them, we can build a chain of
cubic boxes representing the large-scale structure over distances of
Gigaparsecs, from the observer to distant sources. We can then project
the matter distribution inside each box onto one lens plane.
Since different boxes represent different regions of the universe,
the large-scale structure inside neighboring boxes should be uncorrelated.
This is clearly a problem if the boxes originate from
one single simulation, since they would then
represent the same large-scale structure at various evolutionary stages.
To solve this problem, we perform three independent
calculations for each cosmological model, by using three different sets of
initial conditions. We then combine the results of these simulations,
such that the first simulation provides boxes 1, 4, 7, $\ldots$ along the
line of sight, the second simulation provides boxes 2, 5, 8, $\ldots$, and
the third simulation provides boxes 3, 6, 9, $\ldots$, thus ensuring that
two consecutive boxes never come
from the same calculation. To reduce correlations even more, we make
use of the periodic boundary conditions of the simulations, by giving to the
matter distribution in each box a random shift. 

For all $\rm P^3M$ simulations, we use $64^3$ particles and a $128^3$ grid,
inside a cubic box of comoving size $L_{\rm box}=128\,\rm Mpc$.
The comoving softening length of the algorithm is $300\,\rm kpc$.
The total mass of the system is
$M_{\rm sys}=3H_0^2\Omega_0L_{\rm box}^3/8\pi G=5.821\times10^{17}
(\Omega_0h^2)M_\odot$. The mass per particle is
$m=M_{\rm sys}/64^3=2.220\times10^{12}(\Omega_0h^2)M_\odot$.
We performed a total of 129 simulations
(3 per model for 43 models).\footnote{$^{10}$}{These simulations now constitute
the core of the {\it Texas $P^3M$ Database} (Martel \& Matzner 2000).}

\bigskip\smallskip

\ctr{\bf 2.2.\quad The Galaxies Distributions}

\medskip

The large-scale structure simulations described in \S2.1 can
be used to compute the effect of distant matter on
the propagation of the beam. However, at distances less than a
few megaparsecs, we cannot ignore the fact that matter has collapsed
to form galactic-size objects which are much smaller than the
resolution of the P$^3$M algorithm.\footnote{$^{11}$}
{The universe also contains 
virialized objects at larger scales, but these are properly simulated
by the $\rm P^3M$ algorithm.}
We cannot extend 
the resolution of the P$^3$M code down to galactic scales, because
simulating the galaxy formation process would require additional physics
besides gravity, such as hydrodynamical and radiative processes.
Instead, we complement the P$^3$M algorithm with an empirical
Monte Carlo method for locating galaxies inside the computational volume,
based on the underlying distribution of dark matter
(Jaroszy\'nski 1991, 1992; Paper~I; Martel, Premadi, \& Matzner 1998). 

First, we need to determine the number of galaxies present in the
computational volume. 
We assume that the present galaxy luminosities follow
a Schechter luminosity function,
$$n(L)dL={n_*\over L_*}\left({L\over L_*}\right)^{\alpha}e^{-L/L_*}dL\,,
\eqno{(1)}$$

\noindent
where $n(L)$ is the number density of galaxies per unit luminosity. 
We use the values
$\alpha=-1.10$, $n_*=0.0156\,h^{3}{\rm Mpc}^{-3}$, and $L_*=1.3\times
10^{10}h^{-2}L_{\odot}$, where $h$ is the
Hubble constant in units of $\rm 100\,km\,s^{-1}Mpc^{-1}$ 
(Efstathiou, Ellis, \& Peterson 1988).
There is a fourth parameter, 
the luminosity $L_{\min}$ of the faintest galaxies,
which must be introduced to prevent the total number of galaxies
from diverging. The value of this parameter is not well-known.  
We assume a value of $L_{\rm min}=0.01L_*$

Equation~(1) allows us to directly compute the present number density
$n_0$, and luminosity density $j_{0}$, 
$$\eqalignno{
n_0&=n_*\int_{x_{\min}}^{\infty} x^{\alpha} e^{-x}dx=n_*\Gamma(\alpha+1,x_{\min})
\,,&(2)\cr
j_0&=n_*L_*\int_{x_{\min}}^{\infty}x^{\alpha+1}e^{-x}dx
     =n_*L_*\Gamma(\alpha+2,x_{\min})\,,&(3)\cr}$$

\noindent
where $x\equiv L/L_*$, $x_{\min}\equiv L_{\min}/L_*$, and $\Gamma$ is
the incomplete Gamma function. For $x_{\rm min}=0.01$, equations~(2) and
(3) give $n_0=0.0808\,h^3\rm Mpc^{-3}$ and 
$j_0=2.13\times10^8hL_\odot\,\rm Mpc^{-3}$.
The total number of galaxies
in the computational volume is given by 
$$N_{\rm gal}=n_0L_{\rm box}^3=n_*L_{\rm box}^3\Gamma(\alpha+1,x_{\min})\,.
\eqno{(4)}$$

\noindent
For $L_{\rm box}=128\,\rm Mpc$, equation~(4) gives
$N_{\rm gal}=28200$,
46500, 71500, and 104000, respectively, for the
values $h=0.55$, 0.65, 0.75, and 0.85 considered in this paper
(see \S3.1 below).

We use a Monte-Carlo rejection method for determining
the location of the galaxies in the computational volume 
at present ($z=0$). We divide the computational volume in cubic cells
of size $(1\,\rm Mpc)^3$,
and locate a certain number of galaxies in each cell. That number is
chosen from a Gaussian distribution with a standard deviation
proportional to the total matter density in that cell, 
determined from the distribution of particles in the $\rm P^3M$ 
simulation.\footnote{$^{12}$}{The use of a Gaussian distribution is 
an improvement 
over the original algorithm presented in Paper~I, because it allows
for the presence of galaxies in low-density regions.}
The proportionality constant 
is chosen in order to reproduce the number of galaxies $N_{\rm gal}$ given
by equation~(4). The actual location of each galaxy is chosen to be
the center of the cell, plus a random offset of order of the cell size,
also chosen from a Gaussian distribution. With our particular choice of 
cell size, this method naturally creates compact groups of galaxies,
with separations comparable to the ones found in the
Local Group, for instance.
Once the position of each galaxy in the computational volume at 
present is known, we reconstruct the trajectories of galaxies, and 
determine their locations at any redshift, by following the
trajectory of the $\rm P^3M$ particle nearest to each galaxy
(a similar method was used in Jaroszy\'nski 1991, 1992). 

Tests have shown that this method produces realistic galaxy distributions
(Paper I). In particular, the observed 2-point correlation function 
of galaxies is well reproduced down to separations of order of
the cell size, $1\,\rm Mpc$.

Each galaxy is modeled by a truncated, non-singular isothermal sphere,
whose parameters depend upon the galaxy luminosity and morphological type.
We adopt the galaxy models described by Jaroszy\'nski (1991, 1992). 
The projected surface density of each galaxy is given by
$$\sigma(r) = \cases{\displaystyle
              {v^2\over4G(r^2+r_c^2)^{1/2}}\,, & $r<r_{\max}\,;$ \cr
              \noalign{\vskip5pt}
              0\,,                             & $r>r_{\max}\,;$ \cr}
\eqno{(5)}$$

\noindent where $r$ is the projected distance from the center.
The parameters $r_c$, $r_{\max}$, and $v$ are the core radius, maximum radius,
and rotation velocity, respectively, and are given by
$$\eqalignno{
r_c&=r_{c0}\left({L\over L_*}\right)\,,&(6)\cr
r_{\max}&=r_{\max0}\left({L\over L_*}\right)^{1/2}\,,&(7)\cr
v&=v_0\left({L\over L_*}\right)^\gamma\,,&(8)\cr}$$

\noindent where the parameters $r_{c0}$, $r_{\max0}$, $v_0$, and $\gamma$ are
given in Table 1 (Chiba \& Futamase [1999], used a similar approach, with
different values for the parameters).
We use a Monte-Carlo method to generate for each galaxy a luminosity
$L\geq L_{\min}$, with a probability $P(L)$ proportional to
$n(L)$. We determine the morphological type
of each galaxy by using the observed {\it morphology-density
relation} (Dressler 1980;
Postman \& Geller 1984). Regions of the sky with high concentration of
galaxies contain on average more early-type galaxies (ellipticals and S0's)
and fewer late-type galaxies (spirals) than
regions with lower concentration of galaxies. By combining
this relation with a Monte-Carlo method, we can ascribe a morphological 
type to each galaxy. This can be quite important. Several authors
(e.g. Krauss \& White 1992; Kochanek 1996a) have found that the
lensing effect of galaxies is much more important for ellipticals
than for spirals, mostly because ellipticals have a smaller core radius.
Consequently, it is not sufficient to have a realistic distribution of
galaxies: the distribution must be realistic {\it within each morphological
type}, otherwise, some effects, such as strong double lensing by pairs
of elliptical galaxies, would be underestimated.

\bigskip

\ctr{TABLE 1: Galaxy Parameters}

\bigskip

\ctr{\vbox{
\halign{\strut#\hfil&\quad\hfil#\hfil&\quad\hfil#\hfil&\quad\hfil#\hfil&
        \quad\hfil#\hfil\cr
\noalign{\hrule\smallskip\hrule\smallskip}
Type & $r_0$ ($h^{-1}\rm kpc$) & $r_{\max0}$ ($h^{-1}\rm kpc$) & 
$v_0$ ($\rm km\,s^{-1}$) & $\gamma$ \cr
\noalign{\smallskip\hrule\smallskip}
Elliptical & 0.1 & 30 & 390 & 0.250 \cr
S0         & 0.1 & 30 & 357 & 0.250 \cr
Spiral     & 1.0 & 30 & 190 & 0.381 \cr
\noalign{\smallskip\hrule}
}}}

\bigskip

By combining the distribution of background matter
simulated by the P$^3$M algorithm with the distribution and surface
densities of galaxies, we are effectively describing the surface
density of the lens planes over 8 orders of magnitude in length, from the
size of the largest superclusters and voids, $\sim100\,\rm Mpc$, down
to the core radii of the smallest galaxies, $\sim1\,\rm pc$.
The combination of fully nonlinear large-scale structure formation,
galaxy distributions that reproduce the observed 
2-point correlation function, 
morphological type distributions that reproduce the observed morphology-density
relation, and galaxy surface density profiles, gives to the matter distribution
in our algorithm a level of realism that was not present in any
of the previous studies.

Figure~1 illustrates the dynamical range that can be achieved with this method.
The top left panel shows the large scale structure of the universe as
simulated by the P$^3$M code, for a box of 128 Megaparsecs in size. The dots
are the particles used by the P$^3$M algorithm, and are therefore 
mass-tracers. This panel shows
a complex network of clusters, filaments, and voids. We enlarge
a region of size 10 Mpc
and display it on the top right panel. At that scale, we see several irregular
clusters. The dots still represent the particles used in the P$^3$M code.
We enlarge the central cluster, and display it on the middle right panel.
In addition to the P$^3$M particles, we also plot the galaxies (large dots),
whose locations were determined by the Monte-Carlo part of the algorithm. 
We isolate a group of 5 galaxies in this cluster, and plot them at a smaller
scale on the middle left panel. This group, which is about 3 times smaller
than our own ``local group,'' is composed of 3 spiral galaxies, one S0
galaxy, and one elliptical galaxy. We enlarge the central galaxy, and display
it on the bottom left panel. The large circle represents the edge of
the dark-matter halo (radius $r=r_{\rm max}$). The central dot represents
the galactic core. We enlarge this core, which has a radius
$r=r_c$, and display it in the bottom right panel. There are 7 orders of
magnitudes in length between the diameter of this core and the size of
the largest structures shown on the top left panel, and some galaxies
in the simulation have a core which is an order of magnitude smaller than
the one represented here. 

\bigskip\smallskip

\ctr{\bf 3.\quad THE COSMOLOGICAL PARAMETER SURVEY}

\medskip

\ctr{\bf 3.1.\quad The Cosmological Models}

\medskip

All models considered in this paper are Tilted Cold Dark Matter models (TCDM),
normalized according to the results of the {\it COBE} DMR experiment.
The density fluctuation power spectrum for these models is described in
great detail in Bunn \& White (1997). 
The power spectrum at the initial redshift $z_i$ is given by
$$P(k,z_i)=2\pi^2\biggl({c\over H_0}\biggr)^{3+n}\delta_H^2
{\cal L}^{-2}(z_i,0)k^nT_{\rm CDM}^2(k)\,.\eqno{(9)}$$

\noindent where $c$ is the speed of light, $H_0$ is the Hubble constant,
and ${\cal L}(z_i,0)$ is the linear growth factor between
the initial redshift $z_i$ and the present, and $T_{\rm CDM}$ is the transfer function, given by
$$T_{\rm CDM}(q)={\ln(1+2.34q)\over2.34q}\big[1+3.89q+(16.1q)^2+(5.46q)^3
+(6.71q)^4\big]^{-1/4}\eqno{(10)}$$

\epsfysize=19cm
\hskip2cm\epsfbox{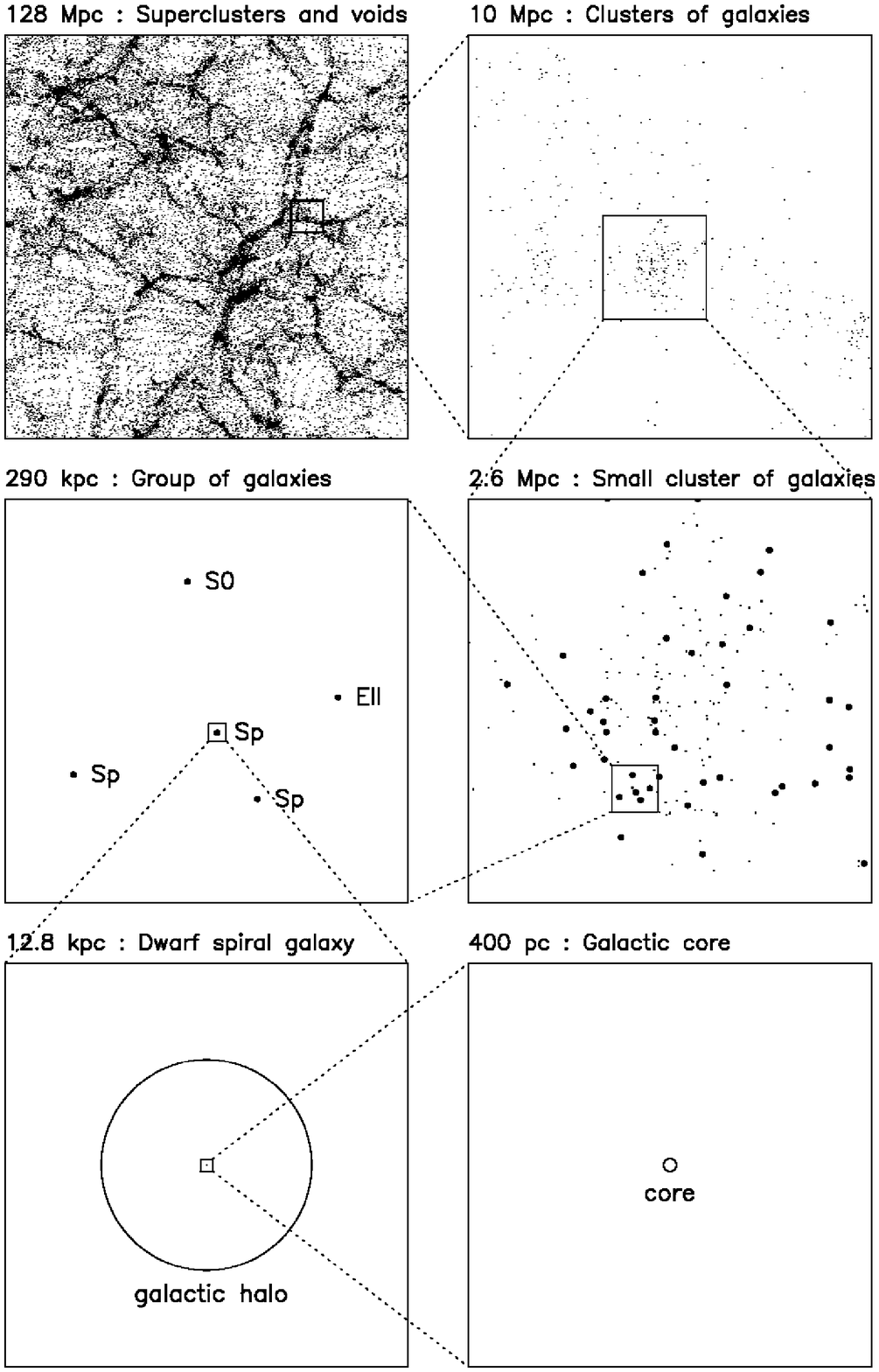}

\medskip 

{\narrower\noindent
Figure 1: Series of zooms illustrating the dynamical range of the algorithm.
Top left panel: Distribution of background matter projected on a lens plane
at $z=0$; Top right panel: Enlargement of a dense region, showing
several clusters of galaxies; Middle right panel: A particular cluster 
of galaxies. Small dots represent P$^3$M, dark matter particles. Large
dots represent actual galaxies; Middle left panel: A small group of
galaxies inside the cluster, composed of 3 spirals, one S0, and one elliptical.
Bottom left panel: A particular spiral galaxy, with a halo radius $r_{\max}$
given by equation~(7); Bottom right panel: The core of the spiral galaxy,
with a radius $r_c$ given by equation~(6).\par}

\noindent (Bardeen et al. 1986), with $q$ defined by
$$\eqalignno{
q&=\biggl({k\over{\rm Mpc}^{-1}}\biggr)\alpha^{-1/2}(\Omega_0h^2)^{-1}
\Theta_{2.7}^2\,,&(11)\cr
\alpha&=a_1^{-\Omega_{\rm B0}/\Omega_0}a_2^{-(\Omega_{\rm B0}/\Omega_0)^3}\,,
&(12)\cr
a_1&=(46.9\Omega_0h^2)^{0.670}\big[1+(32.1\Omega_0h^2)^{-0.532}\big]\,,&(13)\cr
a_2&=(12.0\Omega_0h^2)^{0.424}\big[1+(45.0\Omega_0h^2)^{-0.582}\big]\,,&(14)
\cr}$$

\noindent (Hu \& Sugiyama 1996, eqs.~[D-28] and [E-12]), where 
$\Omega_{\rm B0}$ is the density parameter of
the baryonic matter, and $\Theta_{2.7}$
is the temperature of the cosmic microwave background in units of 2.7K.
The density fluctuation $\delta_H$ at horizon scale is given by
$$\delta_H=\cases{
1.95\times10^{-5}\Omega_0^{-0.35-0.19\ln\Omega_0-0.17\tilde n}
e^{-\tilde n-0.14\tilde n^2}
\,,& $\lambda_0=0$;\cr
\noalign{\vskip5pt}
1.94\times10^{-5}\Omega_0^{-0.785-0.05\ln\Omega_0}
e^{-0.95\tilde n-0.169\tilde n^2}\,,
& $\lambda_0=1-\Omega_0$;\cr}\eqno{(15)}$$

\noindent where $\tilde n\equiv n-1$.

This power spectrum is characterized by 6 independent parameters:
$\Omega_0$, $\Omega_{\rm B0}$, $\lambda_0$, $H_0$, $T_{\rm CMB}$, and $n$.
The normalization of the power spectrum is often described in terms of the
rms density fluctuation $\sigma_8$ at a scale of $8h^{-1}\rm Mpc$, defined by
$$\sigma_8^2={1\over2\pi^2}\int_0^\infty P(k)W^2(k\ell)k^2dk\,,\eqno{(16)}$$

\noindent where $\ell=8h^{-1}\rm Mpc$, and $W$ is the window function, given by
$$W(x)={3\over x^3}(\sin x -x\cos x)\,.\eqno{(17)}$$

The value of $\sigma_8$ is a function of the 6 aforementioned parameters.
We invert this relation, treating $\sigma_8$ as an independent parameter, and
the tilt $n$ as a dependent one. We also set $T_{\rm CMB}=2.7\,\rm K$ and 
$\Omega_{\rm B0}=0.015h^{-2}$ for all models, thus reducing the dimensionality
of the parameter-space from 6 to 4.\footnote{$^{13}$}{Recent results support
$\Omega_{\rm B0}=0.0193h^{-2}$
(Burles \& Tytler 1998), a value slightly larger than the one we assumed. 
The difference is too small to affect the
power spectrum in any significant way, for any of
the models considered.} The independent parameters in 
this parameter
space are therefore $\Omega_0$, $\lambda_0$, $H_0$, and $\sigma_8$.

We survey this parameter space by considering
43 different cosmological models. This constitutes the largest parameter
survey ever done in this field. The values of the parameters are listed in
the first 4 columns of Table 2, while the values of
the dependent parameter $n$ are listed in the fifth column. The values
of $\sigma_8$ were chosen by imposing that $n$ remains in the range 
$[0.7,1.3]$. 

\bigskip\smallskip

\ctr{\bf 3.2.\quad The Ray-Tracing Experiments}

\medskip

For each model, we performed numerous ray-tracing experiments. 
The number of experiments for each model is listed in the
sixth column of Table~2. In each  
experiment, we compute the propagation of a beam consisting
of $341^2=116,281$ light rays forming a square lattice on
the image plane. The size of the beam is $21.9''\times21.9''$, and
the separation between rays is $21.9''/341=0.064''$. This is significantly
smaller than the typical size of an emitting region. A source of angular
diameter $1''$ will contain 190 rays (more if the source is magnified),
enough to resolve details such as multiple images. We locate the source plane 
at a redshift $z_{\rm S}=3$, which is a reasonable choice. The effect of 
lensing would be more important for sources at larger redshifts, but these
sources would be more difficult to observe. By choosing $z_S=3$, we
hope to obtain results that can be compared with current observations.
For a list of the source redshifts used in similar calculations, see Table~3
in Martel et al. (2000). The redshift intervals between the lens
planes were chosen as in Paper I, \S3.2.2.

\vfill\eject

\def \Aa {0.2 & 0.0 & 65 & 0.5 & 1.3188 & 104}
\def \Ab {0.2 & 0.0 & 75 & 0.5 & 1.2190 &  70}
\def \Ac {0.2 & 0.0 & 65 & 0.3 & 1.0966 &  92}
\def \Ad {0.2 & 0.0 & 75 & 0.3 & 0.9993 &  47}
\def \Ba {0.5 & 0.0 & 65 & 1.0 & 1.0439 &  77}
\def \Bb {0.5 & 0.0 & 75 & 1.0 & 0.9656 &  50}
\def \Bc {0.5 & 0.0 & 65 & 0.8 & 0.9457 &  48}
\def \Bd {0.5 & 0.0 & 75 & 0.8 & 0.8686 &  51}
\def \Ca {0.7 & 0.0 & 65 & 1.1 & 0.9346 &  60}
\def \Cb {0.7 & 0.0 & 75 & 1.1 & 0.8648 &  50}
\def \Cc {0.7 & 0.0 & 65 & 0.9 & 0.8461 &  96}
\def \Cd {0.7 & 0.0 & 75 & 0.9 & 0.7773 &  50}
\def \Ga {1.0 & 0.0 & 55 & 1.0 & 0.8465 & 145}
\def \Gb {0.2 & 0.0 & 55 & 0.3 & 1.2187 &  99}
\def \Gc {0.2 & 0.8 & 55 & 0.8 & 1.2057 &  99}
\def \Ha {1.0 & 0.0 & 85 & 1.0 & 0.6605 &  51}
\def \Hb {0.2 & 0.0 & 85 & 0.3 & 0.9191 & 101}
\def \Hc {0.2 & 0.8 & 85 & 0.8 & 0.8749 &  45}
\def \Ka {0.2 & 0.8 & 65 & 0.8 & 1.0702 & 108}
\def \Kb {0.2 & 0.8 & 75 & 0.8 & 0.9629 &  51}
\def \Kc {0.2 & 0.8 & 65 & 0.6 & 0.9326 &  70}
\def \Kd {0.2 & 0.8 & 75 & 0.6 & 0.8273 &  51}
\def \La {0.5 & 0.5 & 65 & 1.0 & 0.8807 &  62}
\def \Lb {0.5 & 0.5 & 75 & 1.0 & 0.8024 &  51}
\def \Lc {0.5 & 0.5 & 65 & 0.8 & 0.7808 & 105}
\def \Ld {0.5 & 0.5 & 75 & 0.8 & 0.7049 &  51}
\def \Ma {0.7 & 0.3 & 65 & 1.1 & 0.8601 &  77}
\def \Mb {0.7 & 0.3 & 75 & 1.1 & 0.7912 &  46}
\def \Mc {0.7 & 0.3 & 65 & 0.9 & 0.7720 &  64}
\def \Md {0.7 & 0.3 & 75 & 0.9 & 0.7042 &  51}
\def \Sa {1.0 & 0.0 & 65 & 0.9 & 0.7234 & 184}
\def \Sb {1.0 & 0.0 & 65 & 1.1 & 0.8120 & 175}
\def \Sc {1.0 & 0.0 & 65 & 1.3 & 0.8861 & 176}
\def \Ta {0.2 & 0.0 & 75 & 0.4 & 1.1228 & 199}
\def \Tb {0.2 & 0.0 & 75 & 0.6 & 1.2979 &  51}
\def \Tc {0.2 & 0.0 & 75 & 0.7 & 1.3648 & 199}
\def \Ua {0.2 & 0.8 & 65 & 0.7 & 1.0062 &  51}
\def \Ub {0.2 & 0.8 & 65 & 0.9 & 1.1269 &  50}
\def \Uc {0.2 & 0.8 & 65 & 1.0 & 1.1568 &  67}
\def \Xa {1.0 & 0.0 & 65 & 1.0 & 0.7698 & 156}
\def \Xb {1.0 & 0.0 & 75 & 1.0 & 0.7094 & 142}
\def \Xc {1.0 & 0.0 & 65 & 1.2 & 0.8506 & 168}
\def \Xd {1.0 & 0.0 & 75 & 1.2 & 0.7893 &  58}

\ctr{TABLE 2: Parameters of the Models}

\bigskip

\ctr{\vbox{
\halign{\strut\hfil#\hfil&\quad\hfil#\hfil&\quad\hfil#\hfil&
        \quad\hfil#\hfil&\quad\hfil#\hfil&\quad\hfil#\cr
\noalign{\hrule\smallskip\hrule\smallskip}
$\Omega_0$ & $\lambda_0$ & $H_0$ & $\sigma_8$ & $n$ & $N_{\rm exp}$ \cr
\Gb \cr
\Ac \cr \Aa \cr 
\Ad \cr \Ta \cr \Ab \cr \Tb \cr \Tc \cr 
\Hb \cr
\noalign{\smallskip\hrule\smallskip}
\Gc \cr 
\Kc \cr \Ua \cr \Ka \cr \Ub \cr \Uc \cr 
\Kd \cr \Kb \cr
\Hc \cr 
\noalign{\smallskip\hrule\smallskip}
\Bc \cr \Ba \cr \Bd \cr \Bb \cr
\noalign{\smallskip\hrule\smallskip}
\Lc \cr \La \cr \Ld \cr \Lb \cr
\noalign{\smallskip\hrule\smallskip}
\Cc \cr \Ca \cr \Cd \cr \Cb \cr
\noalign{\hrule}
\Mc \cr \Ma \cr \Md \cr \Mb \cr
\noalign{\smallskip\hrule\smallskip}
\Ga \cr 
\Sa \cr \Xa \cr \Sb \cr \Xc \cr \Sc \cr 
\Xb \cr \Xd \cr
\Ha \cr 
\noalign{\smallskip\hrule}
}}}

\vfill\eject

\ctr{\bf 3.3.\quad Analyzing the Distribution of Light Rays}

\medskip

To analyze the results of the experiments, we lay down on the source plane
a square grid composed of $31\times31=961$ square cells. The location
and size of the grid are adjusted such that, in the absence of lensing,
the edges of the grid would correspond to the edges of the beam. Figure~2
shows the beam configuration and the grid on the source plane, for a 
typical experiment. In the absence of lensing, each cell would contain 121
rays. A source located in a cell containing more than 121 rays would be 
magnified, whereas a source located in a cell containing less than 121 rays
would be demagnified. By counting the number of rays in each cell, we can then
compute a magnification map. The magnification $\mu_{e,i}$ in cell $i$
for experiment $e$ is then given by
$$\mu_{e,i}={N_{e,i}\over\langle N\rangle}\,,\eqno{(18)}$$

\epsfysize=14cm
\hskip0cm\epsfbox{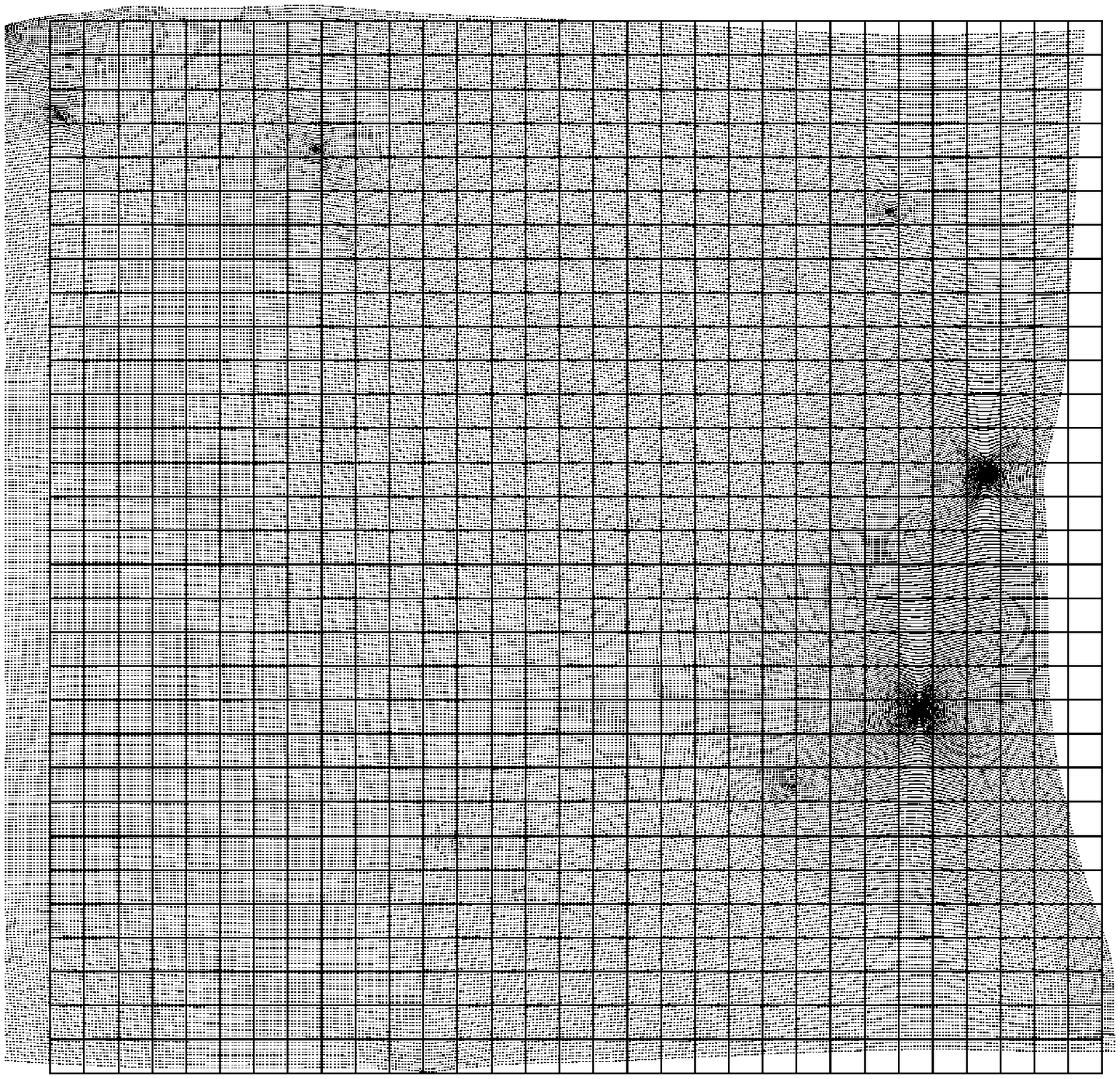}

\medskip 

{\narrower\noindent 
Figure 2: Configuration of the beam on the source plane, for a 
typical experiment. Each dot represents one light ray. The square
grid that we lay down on the source plane is also indicated. 
In the absence of lensing, the beam would coincide with the grid, and each
cell would contain $11\times11$ light rays.\par}

\bigskip

\noindent where $N_{e,i}$ is the number of rays in cell $i$ for
experiment $e$, and $\langle N\rangle$ is the average number of rays
per cell. 
Looking at Figure~2, we see many cells located along the edges
of the grid that are empty or partially empty. This is of course a
numerical artifact caused by the finite size of the beam. These cells should
be excluded from the analysis, otherwise they would artificially bias
the magnification distributions toward low values. Our original approach was
to exclude all cells located along the edges of the grid, using only
the inner $29\times29$ cells instead of the full $31\times31$. However, this
proved insufficient in cases of large beam deformation, such as the
case shown in Figure~2. For this reason, we exclude all cells that are
along the edges of the grid, plus any cell that is adjacent to an empty cell,
since that cell might be half-empty, or even completely empty. In Figure~3,
we show an enlargement of the right edge of Figure~2. The cells located on the 
left of the thick line are included in the analysis, but the ones on the
right are excluded: 6 of them are along the edge of the grid, and 3 others
are adjacent to empty cells.

\bigskip

\epsfysize=8.8cm
\hskip3cm\epsfbox{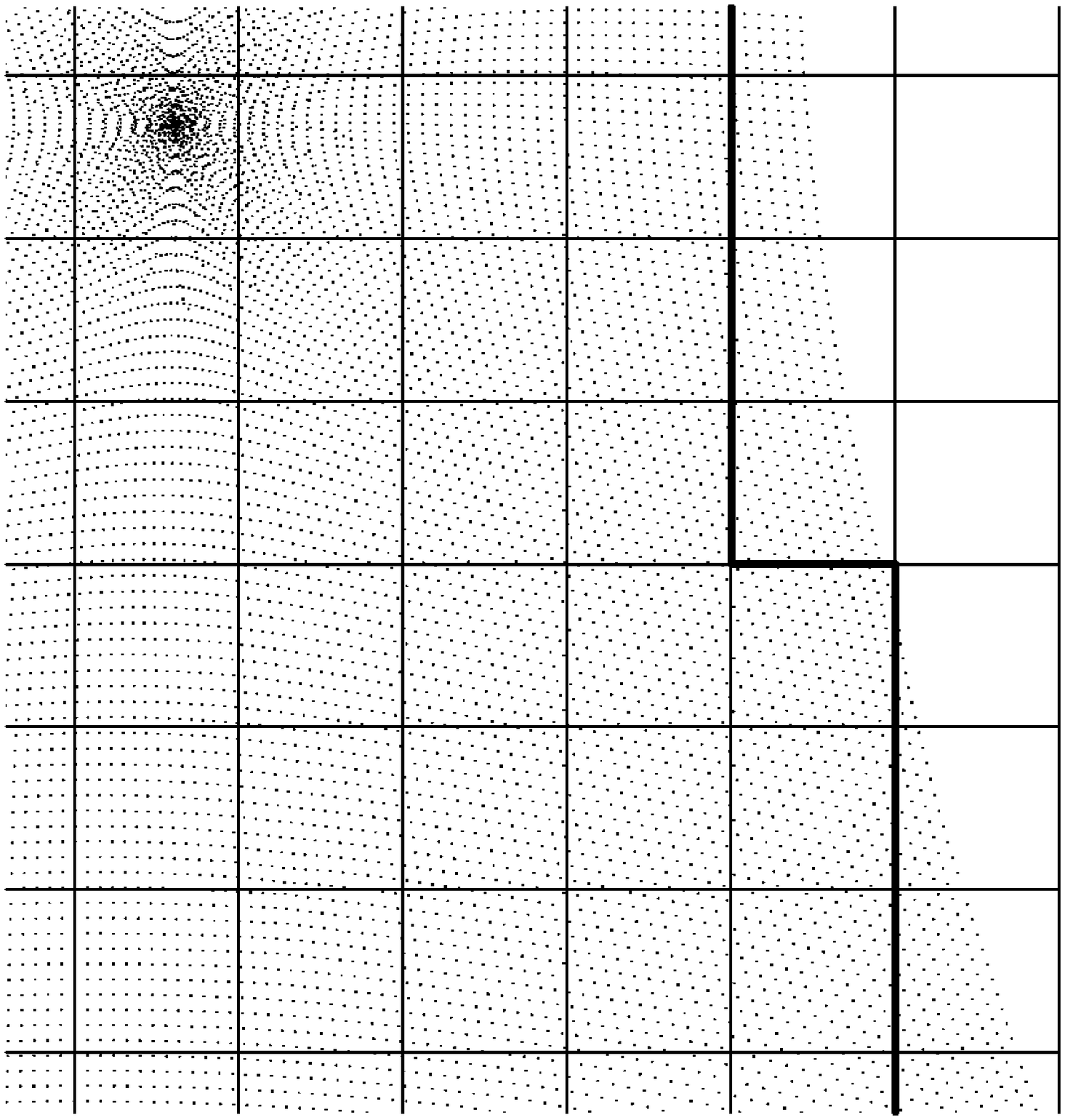}

\medskip 

{\narrower\noindent 
Figure 3: Enlargement of the right edge of Figure~2. The thick line 
separates cells that are included in the analysis (on the left) and
cells that are excluded (on the right), either because they are along
the edge of the grid or because they are adjacent to an empty cell.\par}

\bigskip

The number of rays exceeds the number of cells by a factor of 121. There
are two reasons for using such a large number of rays. First, if the number
of rays were comparable to the number of cells, the magnification map would
be very sensitive to the actual location of the grid. With an average of
121 rays per cell, the sensitivity to grid location is significantly reduced.
Second, with a large number of rays per cell, we can study the properties
of the images. Each cell on the source plane constitutes a potential
location for a source. Assuming that 
a particular cell contains a source, we can identify the light rays 
contained in that cell and trace back these rays on the image plane.
This enables us to study the properties of images, such as multiplicity,
angular separations, brightness ratios, and shapes (simple images,
arcs, rings, $\ldots$). An obvious problem is that the cells are squares,
while sources are expected to be circular. We solve this problem by superposing
over each square cell a circular cell whose diameter is equal to the 
diagonal of the square cell, as shown in Figure~4. These circular cells
have an angular diameter of $1''$, and in the absence of lensing, each 
one would contain $\langle N\rangle=121(\pi/2)=190$ rays.

\bigskip\smallskip

\ctr{\bf 4.\quad THE ELEMENTS OF GRAVITATIONAL LENSING}

\medskip

Before we discuss the results of the experiments, we first review
the various elements that enter into gravitational lensing, and the 
relationship between these elements and the cosmological parameters.
This will facilitate the interpretation of the results presented in the
following section. 

\epsfysize=6cm
\hskip3.1cm\epsfbox{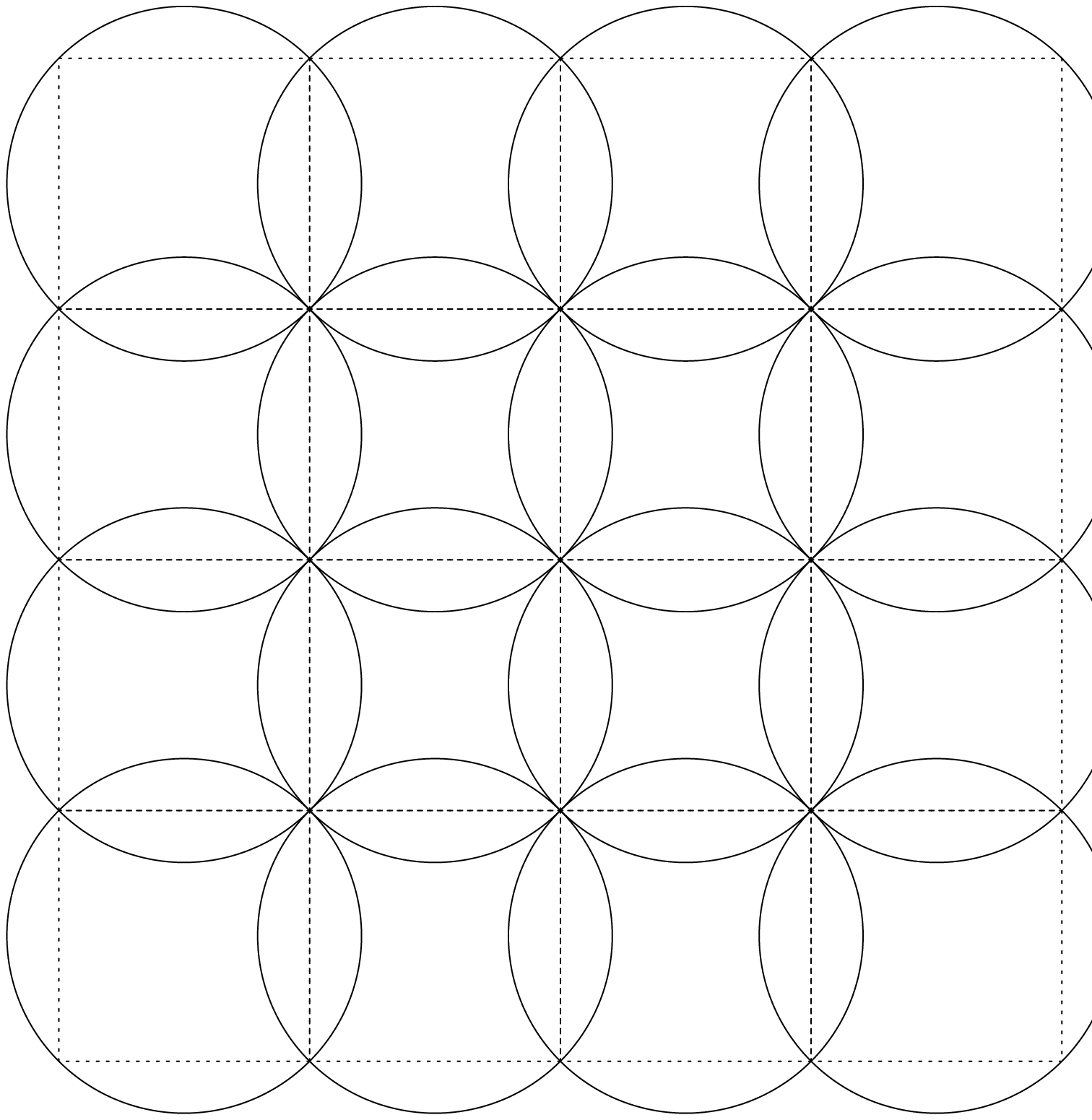}

{\narrower\noindent 
Figure 4: Dashed lines: portion of the square grid laid down on the
source plane. Solid circles: Circular cells superposed on the square cells.
\par}

\bigskip\smallskip

\ctr{\bf 4.1.\quad The Cosmological Distances}

\medskip

The angular displacement of rays caused by lensing depends critically
on the angular diameter distances between the source and the
observer, $D_{\rm S}$, the source and the lens, $D_{\rm LS}$,
and the lens and the observer, $D_{\rm L}$. In our simulations, there is
a continuous distribution of matter between the source and the observer 
(approximated by a finite number of lens planes). In this context, the term
``lens'' refers to the fraction of that matter distribution that is
primarily responsible for lensing. The amount of lensing depends on the
product $D_{\rm L}D_{\rm LS}$, which is maximum for
$D_{\rm L}\simeq D_{\rm LS}$. Hence, lensing is caused primarily by the matter
located about half-way between the source and the observer. As we
discussed in Paper~I, this distance effect is competing with two
others effects: First, the large-scale structure, which is responsible 
for lensing, grows with time, favoring lenses that are closer to the observer.
Second, for a fixed number of galaxies, the number
density of galaxies decreases with time as the universe expands, and 
therefore the beam is more likely to hit a galaxy at locations that are 
closer to the source. We found in Paper I that the distance 
effect dominates over the other effects, and therefore the matter located
half-way between the source and the observer is responsible for most
of the lensing. The only exception is the Einstein-de~Sitter model,
in which the large-scale structure keeps growing all the way to the
present. The matter responsible for most of the lensing
tends to be located somewhat closer to the observer than the half-way point.
This effect is negligible for other models because the
large-scale structure does not grow all the way to the present,
but instead freezes-out at some redshift of order $z_{\rm fr}\sim1/\Omega_0$.

If the distance effect dominates, then the source redshift determines the
``lens'' redshift, and therefore only one distance, say $D_{\rm S}$, is
independent. This distance is given by
$$D_{\rm S}={c\over H_0}f(\Omega_0,\lambda_0,0,z_s)\,,\eqno{(19)}$$

\noindent where the function $f$ is given by:
$$f(\Omega_0,0,z_i,z_j)={2\Big[
(2-\Omega_0+\Omega_0z_j)(1+\Omega_0z_i)^{1/2}
-(2-\Omega_0+\Omega_0z_i)(1+\Omega_0z_j)^{1/2}\Big]
\over\Omega_0^2(1+z_i)(1+z_j)^2}\,,\eqno{(20)}$$

$$f(\Omega_0,1-\Omega_0,z_i,z_j)={1\over1+z_j}
\int_{z_i}^{z_j}dz\Big[\Omega_0(1+z)^3
+(1-\Omega_0)\Big]^{-1/2}\eqno{(21)}$$

\noindent
(Fukugita et al. 1992). Figure~5 shows the angular diameter distances 
for the cosmological models considered in this paper. The
distances depend on $\Omega_0$, $\lambda_0$, and $H_0$, but not $\sigma_8$.
The dependence upon $\lambda_0$ can be quite strong, which is why
gravitational lenses can impose very stringent limits on the value of the
cosmological constant (e.g. Fukugita et al. 1990; Kochanek 1996a).

\epsfysize=7.5cm
\hskip3cm\epsfbox{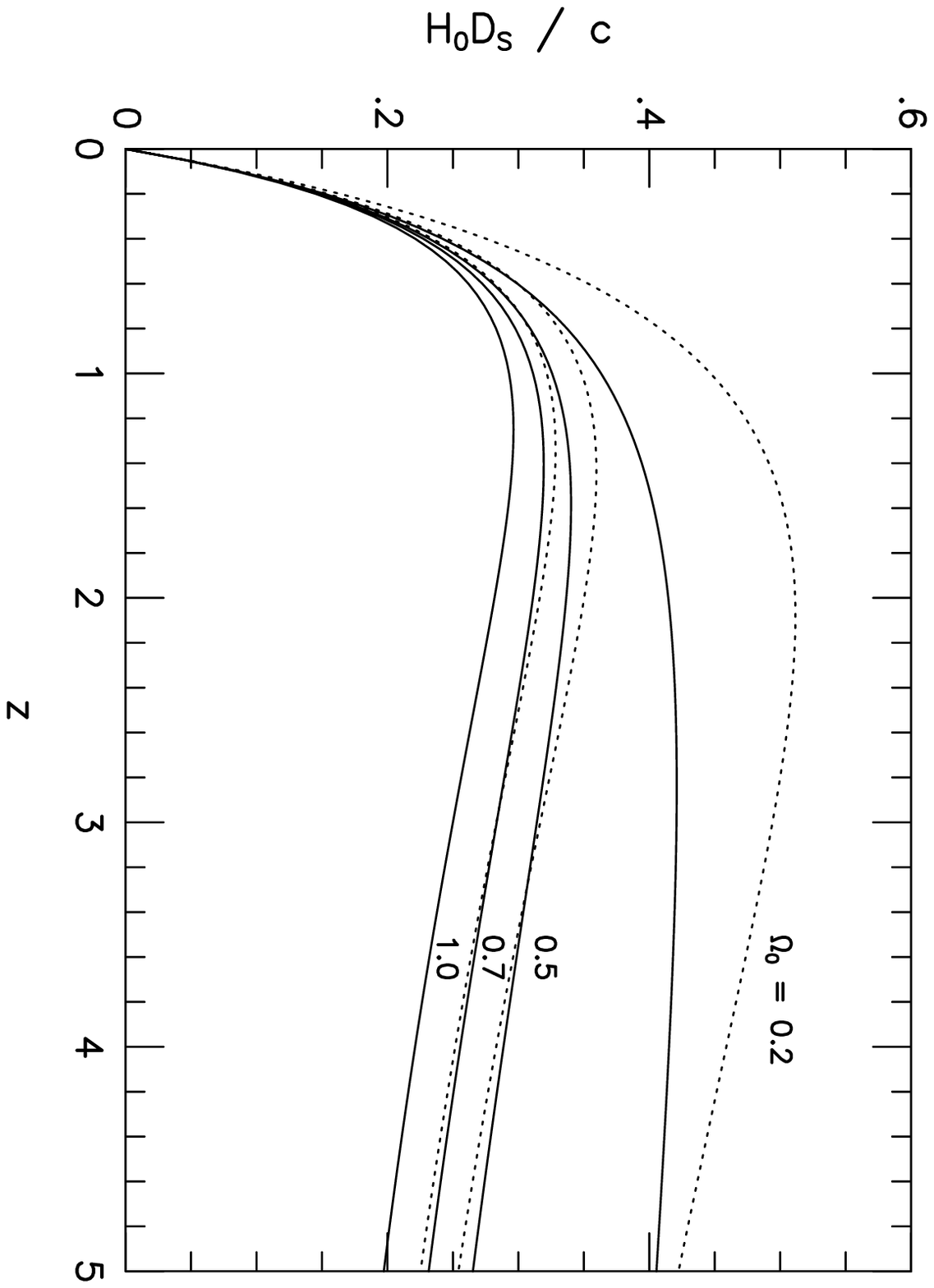}

{\narrower\noindent 
Figure 5: Angular diameter distance $D_s$ in units of the Hubble radius
$c/H_0$, versus redshift $z$, 
for the various models considered in this paper. 
Solid curves:
matter-dominated models ($\lambda_0=0$); dotted curves:
flat, cosmological constant models
($\Omega_0+\lambda_0=1$). 
Curves are labeled with the values of $\Omega_0$.\par}

\bigskip\smallskip

\ctr{\bf 4.2.\quad The Mean Background Density}

\medskip

The importance of gravitational lensing depends on the mean density
of matter between the source and the observer, which is proportional
to the present mean density of the universe,
$$\bar\rho_0={3H_0^2\Omega_0\over8\pi G}\,.\eqno{(22)}$$

\noindent The mean density depends on $H_0$ and $\Omega_0$, but
not $\lambda_0$ and $\sigma_8$.

\bigskip\smallskip

\ctr{\bf 4.3.\quad The Large-Scale Structure}

\medskip

If the universe was perfectly homogeneous, there would be no gravitational
lensing. It is the presence of density inhomogeneities, resulting from
the growth of the large-scale structure, that is responsible for lensing. 
The amount of structure in the present universe is characterized by
the parameter $\sigma_8$. However, what is relevant is not the
present large-scale structure, but the large-scale structure at
redshift $z_{\rm L}$ corresponding to an angular diameter distance
$D_{\rm L}\simeq D_{\rm S}/2$ where most of the matter responsible
for lensing is located. Using linear perturbation theory, we can
estimate the rms density fluctuation $\sigma_{8,\rm L}$ at that redshift,
$$\sigma_{8,\rm L}={\sigma_8\over{\cal L}(z_{\rm L},0)}\,,\eqno{(23)}$$

\noindent where ${\cal L}(z_{\rm L},0)$ is the linear growth factor between
redshifts $z_{\rm L}$ and 0, which depends on $\Omega_0$ and $\lambda_0$.
Thus, $\sigma_{8,\rm L}$ depends on $\sigma_8$, $\Omega_0$, 
and $\lambda_0$, but not $H_0$. 
For the Einstein-de~Sitter model, ${\cal L}(z,0)=1+z$,
hence $\sigma_8$ decreases monotonically with increasing $z_L$.
For $\Omega_0<1$ models, the perturbation ``freezes-out'' at
redshift $z_{\rm fr}\sim1/\Omega_0$, and grows very
slowly between $z_{\rm fr}$ and the present. Hence, at fixed $\sigma_8$, 
$\sigma_{8,\rm L}$ increases with decreasing $\Omega_0$,
and if $z_{\rm fr}\gg z_{\rm L}$, $\sigma_{8,\rm L}$
should be comparable to $\sigma_8$. 

\bigskip\smallskip

\ctr{\bf 5.\quad RESULTS}

\medskip

\ctr{\bf 5.1.\quad The Magnification Distributions}

\medskip

We compute the magnification distributions by combining all experiments
within each cosmological model, and binning the cells on the magnification
maps according to the value $\mu_{e,i}$ of the magnification in each cell.
We use magnification bins of equal width $\Delta\mu=0.02$. The probability
$P(\mu)\Delta\mu$ that a source has a magnification between $\mu$ and 
$\mu+\Delta\mu$ is given by
$$P(\mu)\Delta\mu={\sum_e n_e(\mu,\mu+\Delta\mu)\over\sum_e n_e(0,\infty)}\,,
\eqno{(24)}$$

\noindent where $n_e(\mu_1,\mu_2)$ is the number of cells in experiment $e$
with magnification between $\mu_1$ and $\mu_2$, and the sums are over
the experiments. Notice that $n_e(0,\infty)$ is the number of cells 
for experiment $e$ that are included in the analysis, which can vary among
experiments (see discussion of Fig.~3). Equation~(24) implies 
$$\int_0^\infty P(\mu)d\mu=1\,,\eqno{(25)}$$

\noindent where the ``integral'' is actually a sum over bins. By 
substituting equation~(18) into equation~(24), we can easily show that
$$\int_0^\infty P(\mu)\mu\,d\mu=1\,.\eqno{(26)}$$

\noindent Equations (25) and (26) express the conservation of probability 
and the conservation of  flux, respectively.

To estimate the accuracy of the distributions, 
we used a convergence criterion. 
The magnification and shear distributions were always updated each
time new experiments were performed in a given model.
We also computed, in each magnification bin, the uncertainty
on the mean $\Delta P=\sigma/N_{\rm exp}^{1/2}$, where $\sigma$ is the
standard deviation in that bin, and $N_{\rm exp}$ is the number of
experiments included in the average. 
Eventually, we reached a point where adding new experiments
did not change the distribution, in which case convergence has been 
achieved. 
For models with large $\Omega_0$, the convergence is reached much more slowly,
and for a few models  the especially large number of experiments done
was still not sufficient to reach a tight convergence.
We refer to models for which convergence
has or has not been achieved as having ``good statistics'' or
``poor statistics,'' respectively, and we shall be careful when drawing 
conclusions for models which have poor statistics. In Figures 6--9,
we compare the magnification distributions for various models.
Error bars indicate the uncertainty $\Delta P$ on the mean (for clarity, we
display error bars for only 1/4 to 1/3 of the bins).
The important thing to point out is that the
error bars are shorter than the separations between curves,
except in regions where the curves are nearly identical.
The differences between the various curves in Figures 6--9 are therefore real,
and not a consequence of insufficient accuracy (except when stated otherwise).

\bigskip\smallskip

\ctr{\sl 5.1.1.\quad The $\sigma_8$ Dependence}

\medskip

Figure~6 shows the magnification distributions for three different 
models with various combinations of $\Omega_0$, $\lambda_0$,
and $H_0$. The various curves represent various values of $\sigma_8$.
For these models, we actually have results for
5 different values of $\sigma_8$, but
for clarity we only plot the distribution for the smallest, median, 
and largest values of $\sigma_8$. In the absence of lensing, these
distributions would be $\delta$-functions centered at $\mu=1$.
These distributions reveal that most sources are slightly demagnified,
few sources are strongly magnified, and the effect
of lensing increases with increasing $\sigma_8$.  
For the open and $\Lambda$
models, we see a clear trend: as $\sigma_8$ increases, the peak of
the distribution decreases, the low edge of the distribution moves
to even lower values (more demagnification), but the right edge is
hardly affected. We see a similar trend for the Einstein-de~Sitter model,
but more experiments are needed to improve the statistics. 
The explanation resides in the fact that the magnification
is caused primarily by the matter located near the beam, whereas the matter
located far from the beam is primarily responsible for the shear. 
Each lens plane contains a certain distribution of matter,
with overdense and underdense regions. If the beam 
propagates through an underdense 
region, it will diverge, resulting in demagnification, and if it 
propagates through an 
overdense region, it will converge, resulting in magnification. In a
cosmological model like CDM, structure formation proceeds
hierarchically. Small structures form first, then merge to form bigger
structures, which merge into even bigger structures, and so on. 
As time goes on, clusters become more massive and less abundant,
and the voids between them become larger. A larger
$\sigma_8$ implies that this hierarchical merging process is more advanced,
and this affects the matter distribution in
two ways: first, the underdense regions become more underdense and the 
overdense regions become more overdense, and second, the fraction of the 
plane covered by underdense regions (the ``filling factor'')
increases while the fraction covered by overdense 
regions decreases. In the case of demagnification, these two effects act in 
the same direction: As $\sigma_8$ increases, the beam is more likely to 
propagate through 
an underdense region (because of the larger filling factor), and if it does,
it will result in
stronger demagnification, because these regions are more underdense.
In the case of magnification, these effects act in opposite directions:
as $\sigma_8$ increases,
the beam is less likely to propagate through 
an overdense region, but if it does, the
magnification will be stronger. As we see, these two effects are almost
perfectly cancelling each other, and the distributions at values of
$\mu>1$ are essentially independent of $\sigma_8$, at least for the
open and $\Lambda$ models. For the Einstein-de~Sitter model, the statistics
are poor, even though we performed significantly more experiments for
this model than the other ones. 

\medskip

\epsfysize=14.3cm
\hskip2cm\epsfbox{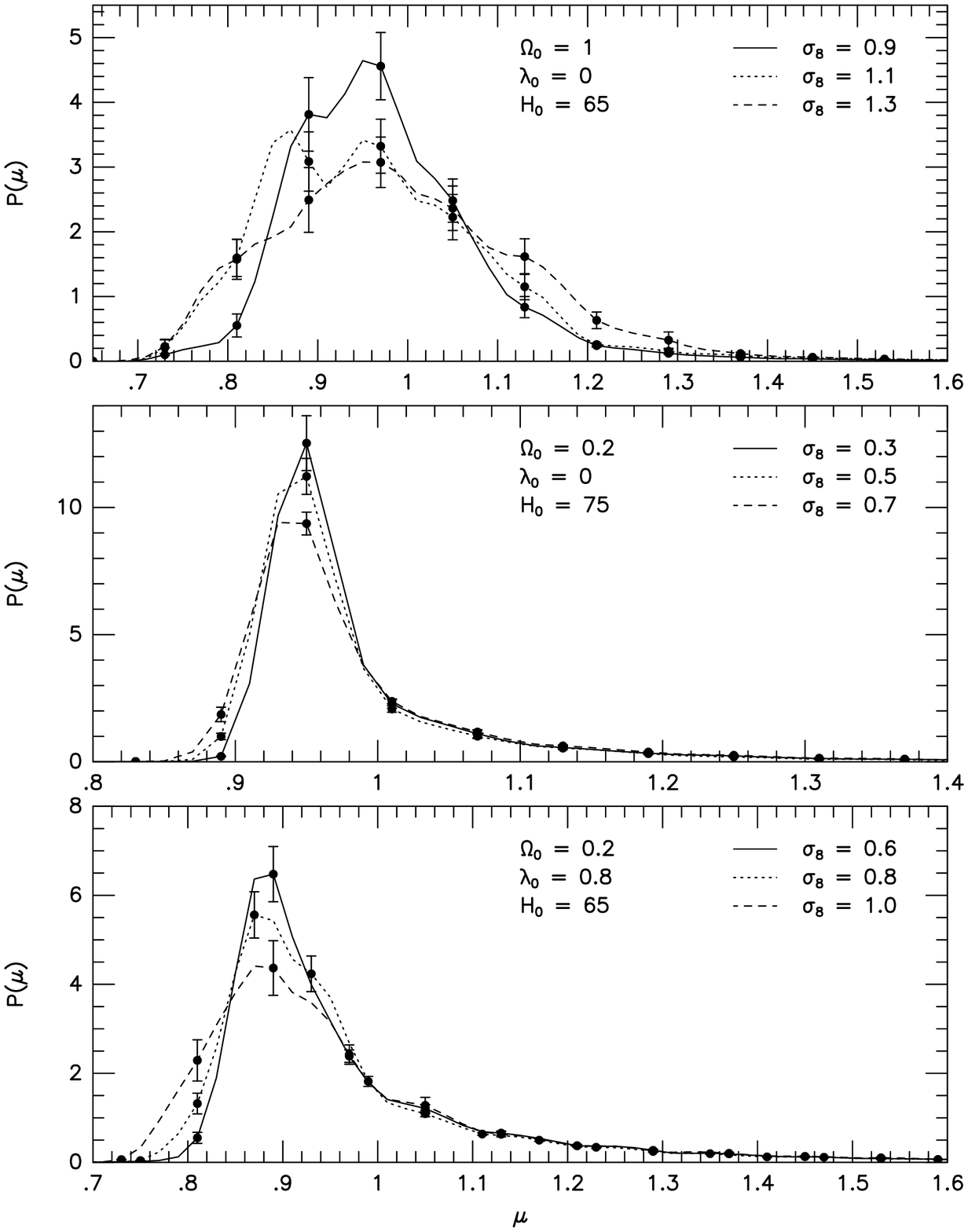}

{\narrower\noindent 
Figure 6: Magnification distributions for various combinations
of $\Omega_0$, $\lambda_0$, and $H_0$, showing the effect of varying 
$\sigma_8$. Error bars indicate the $1-\sigma$ uncertainty $\Delta P$ on 
the mean, for a few representative bins. See text for details.\par}

\bigskip

\ctr{\sl 5.1.2.\quad The $H_0$ Dependence}

\medskip

Figure~7 shows the magnification distributions for three different 
models with various combinations of $\Omega_0$, $\lambda_0$,
and $\sigma_8$. The various curves represent various values of $H_0$.
For the Einstein-de~Sitter model (top panel) and $\Lambda$-model
(bottom panel),
the distributions become narrower as $H_0$ increases
(the distributions of the top panel are quite noisy, but if one
focuses on the ordering of the curves along the
left edge of the distribution, the trend becomes clear). This trend was
expected, since the cosmological distances are shorter in models with large
$H_0$, resulting in a weakening of the effect of gravitational lensing. 
Notice that this effect is partially compensated by the fact that at fixed
$\Omega_0$, the mean background density increases like $H_0^2$. From the
point of view of the algorithm, a model with larger $H_0$ has a {\it smaller
number} of {\it denser} lens planes between the source and the observer.
For the open model (middle panel), these two effect almost perfectly
cancel each other, and the distributions are essentially independent of
$H_0$.

\medskip

\epsfysize=14.3cm
\hskip2cm\epsfbox{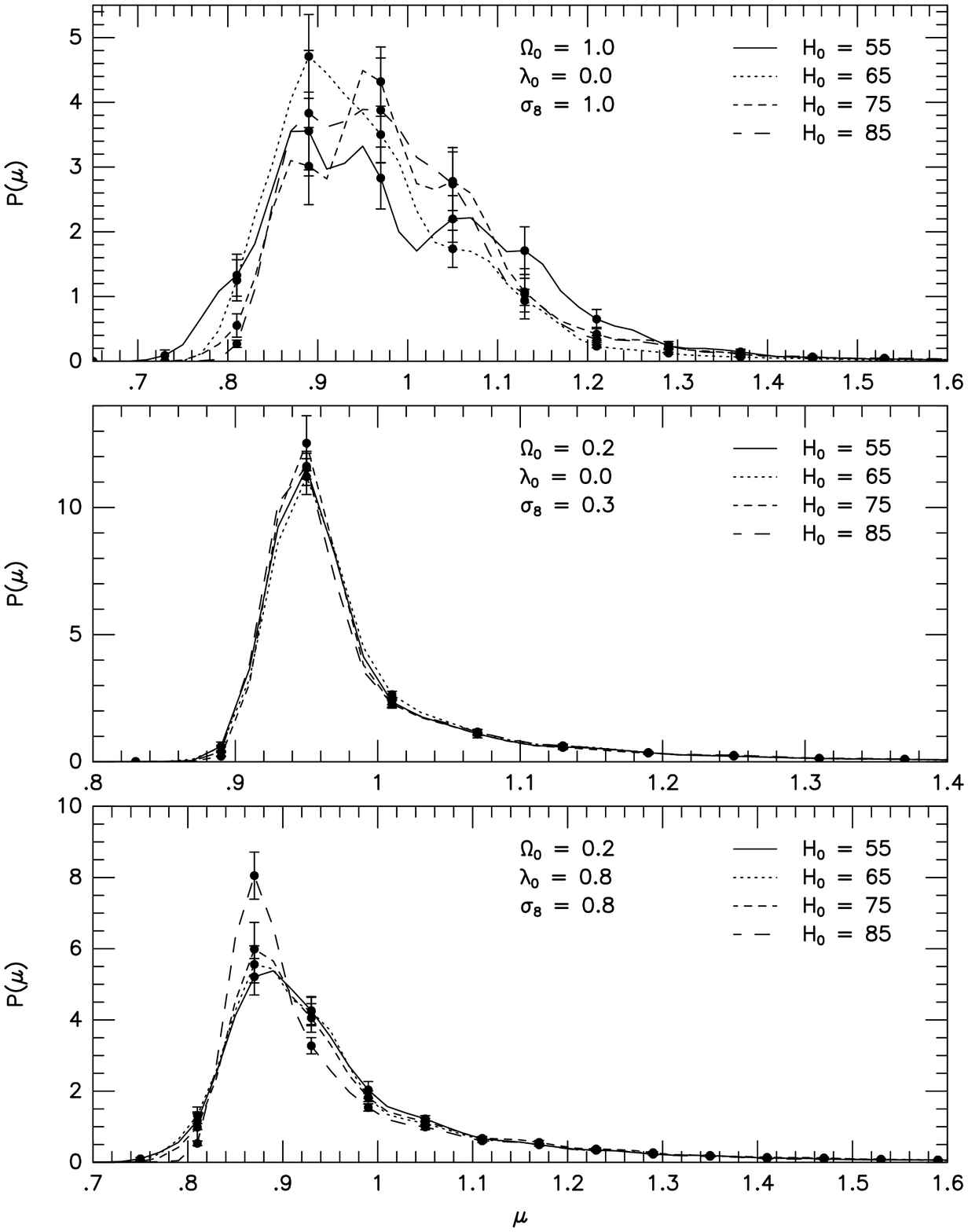}

{\narrower\noindent 
Figure 7: Magnification distributions for various combinations
of $\Omega_0$, $\lambda_0$, and $\sigma_8$, showing the effect of varying
$H_0$. Error bars indicate the $1-\sigma$ uncertainty $\Delta P$ on 
the mean, for a few representative bins. See text for details.\par}

\bigskip

\ctr{\sl 5.1.3.\quad The $\Omega_0$ Dependence}

\medskip

Figure~8 shows the magnification distributions for two different 
models with various combinations of $\lambda_0$, $H_0$
and $\sigma_8$. The top panel shows matter-dominated
models ($\lambda_0=0$), while the bottom panel shows flat
models ($\Omega_0+\lambda_0=1$).
The various curves represent various values of $\Omega_0$. 
Of all the various dependences, the $\Omega_0$ dependence is the most
difficult one to interpret. The reason is that $\Omega_0$ is the only
parameter which all elements of lensing, distance, mean background density,
and structures, depend on. We are therefore dealing with three
concurrent effects. As $\Omega_0$ increases, the mean background density
increases, favoring stronger lensing effects. However, the
cosmological distances decrease, favoring
weaker lensing effects, and at
fixed $\sigma_8$, the large-scale structure at high-redshift is
less developed because freeze-out occurs later, also
favoring weaker lensing effects. For 
matter-dominated models (top panel), the importance of lensing increases with
$\Omega_0$, resulting in a shift of the distribution toward lower values.
The dominant effect in this regime is therefore the mean background density.
The bottom panel shows the magnification distributions for
flat models, with $\lambda_0$ increasing
as $\Omega_0$ decreases. As Figure~5 shows, at redshifts $z\leq3$
the angular diameter distances increase with $\lambda_0$ for all models, 
and this reinforces the dependence upon $\Omega_0$.
The distances get even larger with smaller $\Omega_0$, and this
helps overcoming the mean background density effect.
The bottom panel in Figure~8
shows that, in the limit of large $\lambda_0$, the effect of having large
distances dominates over competing effects. However, for values
$\Omega_0\geq0.5$ (or $\lambda_0\leq0.5$), there is no clear dependence
of the magnification distribution on $\Omega_0$, and we would need much better
statistics to determine whether there is an effect. In any case, the
effect would probably be quite small.
 
\bigskip

\epsfysize=9.7cm
\hskip2cm\epsfbox{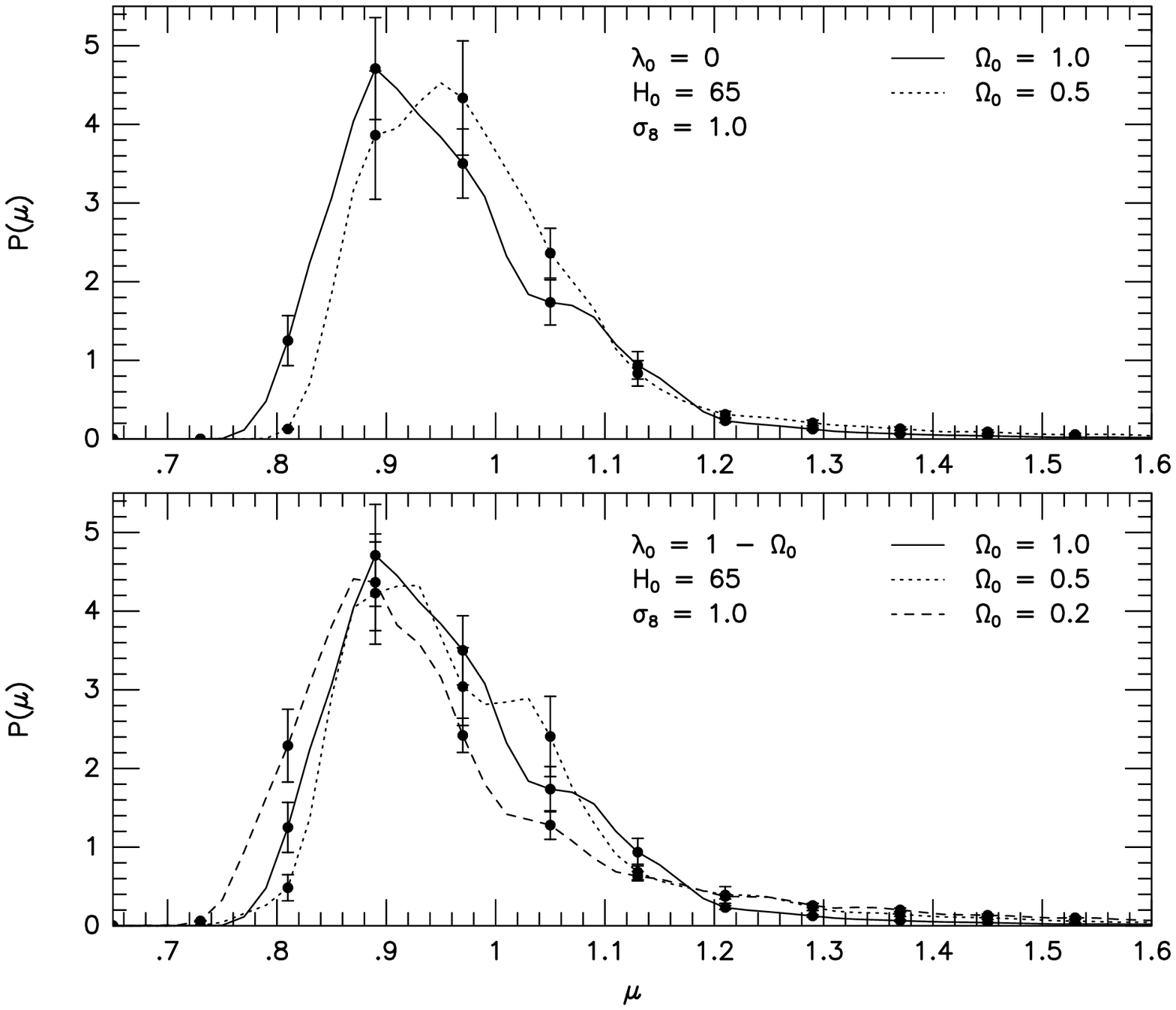}

{\narrower\noindent 
Figure 8: Magnification distributions for various combinations
of $\lambda_0$, $H_0$, and $\sigma_8$, showing the effect of varying
$\Omega_0$. Error bars indicate the $1-\sigma$ uncertainty $\Delta P$ on 
the mean, for a few representative bins. See text for details.\par}

\bigskip

\ctr{\sl 5.1.4.\quad The $\lambda_0$ Dependence}

\medskip

Figure~9 shows the magnification distributions for two different 
models with various combinations of $\Omega_0$, $H_0$,
and $\sigma_8$. The various curves represent various values of $\lambda_0$.
The presence of a cosmological constant increases the effect
of magnification by increasing the cosmological distances.
For the low-density model ($\Omega_0=0.2$, top panel), the
cosmological constant results in a widening of the distribution,
and a shift toward lower magnifications. The effect is much less important
for the $\Omega_0=0.5$ model (bottom panel). This indicates that the effect
of the cosmological constant is small for values $\lambda_0\leq0.5$, and
becomes important only for larger values. As Figure~5 shows, the effect
of $\lambda_0$ on the angular diameter distances is relatively small
for models with $\Omega_0\geq0.5$ (or $\lambda_0\leq0.5$), but
the difference between the $\Omega_0=0.2$, $\lambda_0=0$,
and the $\Omega_0=0.2$, $\lambda_0=0.8$ models is very large.

\bigskip\smallskip

\ctr{\sl 5.1.5.\quad Large-Scale Structure versus Galaxies}

\medskip

In our simulations, the density inhomogeneities responsible for lensing
consists of the large-scale structure of the universe and the
galaxies that are embedded inside that structure. Including large-scale
structure and galaxies in the algorithm enables us to study both the
effect of weak and strong lensing. While most of the weak lensing results 
from the presence of the large-scale structure, the rare, high-magnification
events usually result from the presence of a massive galaxy along
the line of sight. One may wonder if strong lensing is affected by
the cosmological parameters, $\sigma_8$ in particular. This parameter
measures the amplitude of the density fluctuations in the large-scale 
structure, but since the algorithm chooses galaxy locations according to the
background density, galaxies are more clustered in models with a
larger $\sigma_8$. This does not change the probability that a given
source will be lensed by a galaxy, but it could increase the
probability of a source being lensed by two or several galaxies
clustered together, resulting in a very large magnification. Lensing by 
multiple galaxies must be invoked to explain the properties of
some of the images (see Fig.~13 below). However, the high-tail of the
magnification distributions shown in Figure~6 appears to be 
independent of the value of $\sigma_8$. In particular, we do not find the
dramatic increase in the values of $\mu$ with higher $\sigma_8$ that
one would expect if sources are lensed by compact groups of galaxies.

\bigskip

\epsfysize=9.7cm
\hskip2cm\epsfbox{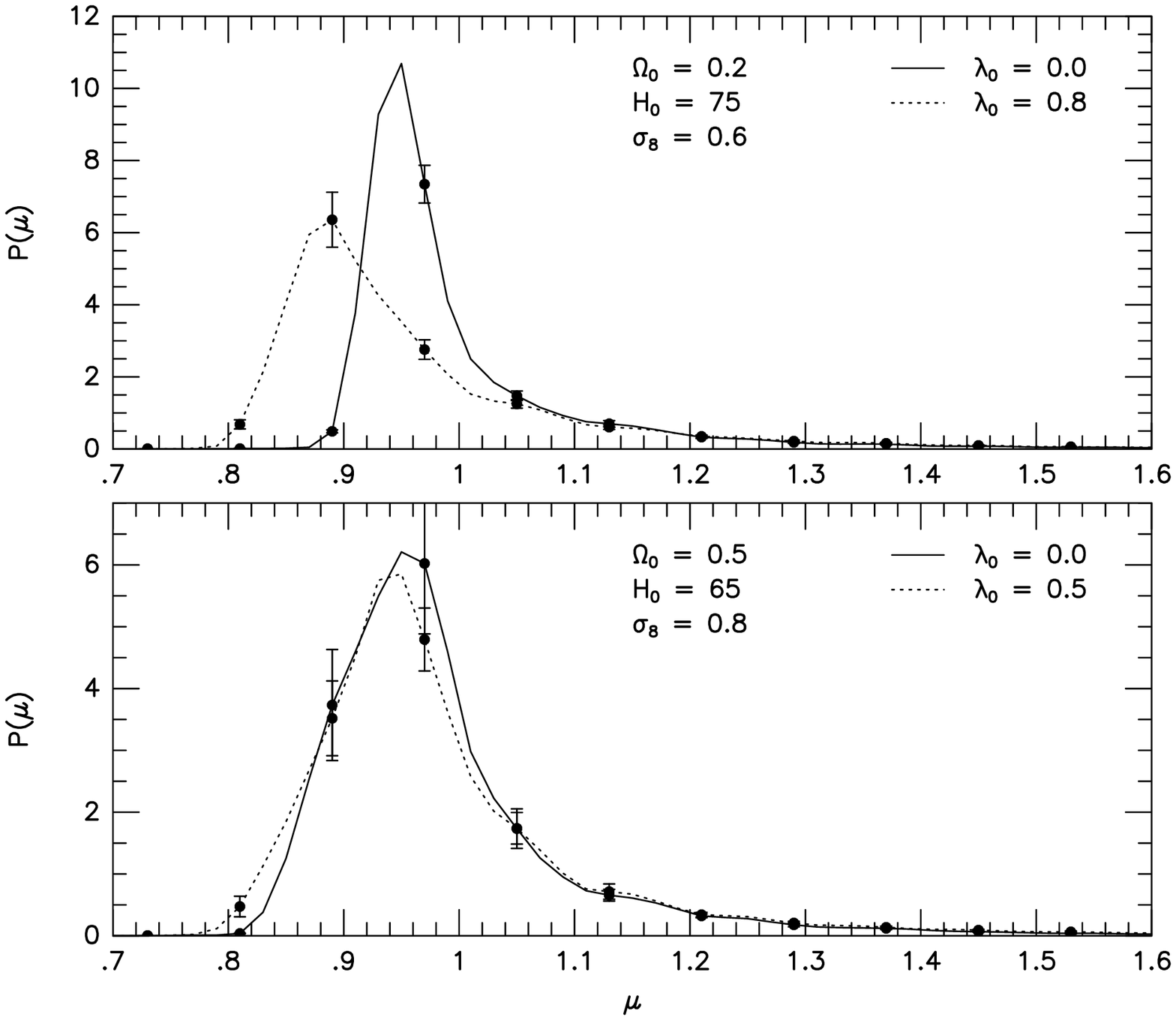}

{\narrower\noindent 
Figure 9: Magnification distributions for various combinations
of $\Omega_0$, $H_0$, and $\sigma_8$, showing the effect of varying
$\lambda_0$. Error bars indicate the $1-\sigma$ uncertainty $\Delta P$ on 
the mean, for a few representative bins. See text for details.\par}

\bigskip

We investigate this question by estimating the likelihood that a source
will be lensed by two galaxies on the same lens plane. Our algorithm models
galaxies as truncated isothermal spheres with a mass-dependent radius 
$r_{\max}$ given by equation~(7).
We computed the geometric cross section $\pi r_{\max}^2$  for all galaxies, 
and added them up, to find out what fraction $f_{\rm gal}$
of the lens plane is covered by galaxies. For lens planes located near $z=0$, 
we obtain $0.00326\leq f_{\rm gal}\leq0.00503$ (the value depends on $H_0$;
see comment following eq.~[4]). 
Hence, only a small fraction of the plane is covered by galaxies. 
However, for sources at $z=3$, the lensing is caused
primarily by planes located near $z=1$. 
Since at $z=1$ the planes are smaller but the galaxies 
have the same size,\footnote{$^{14}$}{Our current algorithm neglects
galaxy evolution, and assumes that the galaxy parameters given
by equations.~(6)--(8) are independent of redshift.}
we gain a factor $(1+z)^2=4$, and the fraction
covered by galaxies is then in the range 
$0.013\leq f_{\rm gal}\leq0.020$, still quite small. 
Galaxy clustering will create regions where the covering is
larger. If we assume that clustering could increase the space density
of galaxies in some regions by a factor of 1000 (quite optimistic at $z=1$),
the surface density would go up by 100, and there could be some galaxies
overlapping.

\bigskip

\epsfysize=18cm
\hskip4.5cm\epsfbox{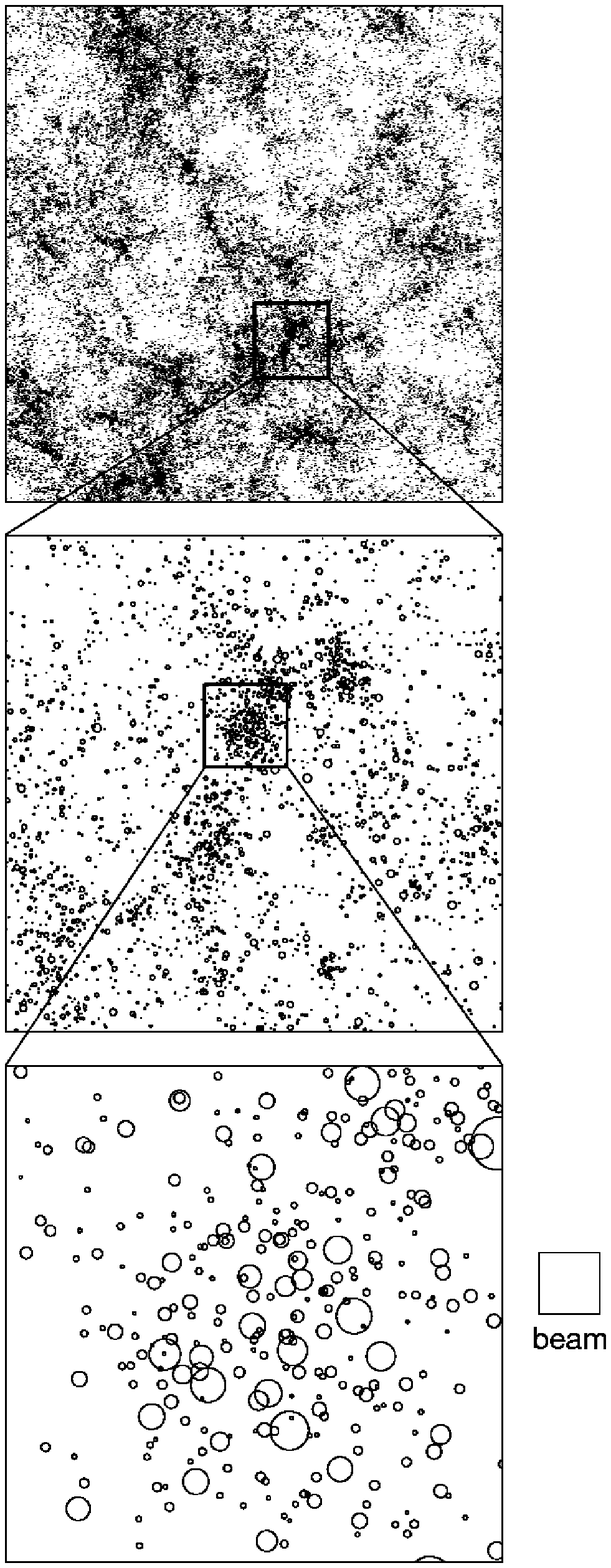}

\medskip

{\narrower\noindent 
Figure 10: Top panel: Galaxy distribution at $z=0.964$ for
$\Omega_0=0.2$, $\lambda_0=0.8$, $H_0=65\,\rm km\,s^{-1}Mpc^{-1}$,
$\sigma_8=0.8$ model. Each galaxy is represented by a dot.
Middle panel: enlargement of a galaxy-rich region. Bottom
panel: enlargement of a rich cluster of galaxies. Galaxies in the
middle and bottom panels are represented by circles of radius
$r_{\rm max}$, as defined by equation~(7).\par}

\bigskip

Figure~10 shows the galaxy distribution on a lens plane located
at $z=0.964$, for the model $\Omega_0=0.2$, $\lambda_0=0.8$, 
$H_0=65\rm\,km\,s^{-1}Mpc^{-1}$,
$\sigma_8=0.8$. The top panel shows the galaxy distribution over the
entire lens plane, with galaxies represented as dots. Because of freeze-out,
the galaxies at that redshift are almost as clustered as at redshift $z=0$.
We zoom-in on one of the densest, most galaxy-rich region, and display that
region in the middle panel of Figure~10, 
where galaxies are now represented as 
circles with radii equal to $r_{\max}$. Because the galaxies follow
a Schechter distribution, most galaxies have small radii and few have
large radii. We zoom-in on the richest galaxy cluster in this region, and
display it in the bottom panel of Figure~10. 
There are some galaxies overlapping, 
but not many. The beam, displayed next to the bottom panel of Figure~10, 
is larger  than the galaxies at that redshift, so it might hit several 
galaxies, but the light coming for one individual cell in the 
beam is unlikely to hit more than one galaxy, in spite of the fact that
galaxies are strongly clustered in this region.
Of course, we have considered only one plane. A 
light ray can hit a galaxy on one
plane, and then another galaxy on another plane. But since the structures
on neighboring planes are uncorrelated, the probability of occurrence of
such double hit is unaffected by clustering. The high tail of
the magnification distributions depends on the distribution of
galaxy properties (masses, truncation radii, core radii), but
not on their actual locations or level of clustering.

\bigskip\smallskip

\ctr{\bf 5.2.\quad The Magnification Probability}

\medskip

The magnification probability is defined as
$$P_m=\int_1^\infty P(\mu)d\mu\,.\eqno{(27)}$$

\noindent
In a large, representative region of the sky, $P_m$ represents
the fraction of sources that are magnified. In Figure~11, we
plot $P_m$ vs. $\sigma_8$, for all models. We do not find any
particular trend. Instead, $P_m$ is essentially independent of
$\sigma_8$. This could have been anticipated from Figure~6,
which showed that for most models, $P(\mu)$ is independent
of $\sigma_8$ in the integration range of equation~(27).
However, this result is in conflict with the argument we
presented in \S5.1.1. We argued that as $\sigma_8$ increases,
the filling factor of overdense regions decreases, so
a particular source is less likely to be magnified, but if it is,
the magnification will be larger. However, $P_m$ measures the
fraction of sources that are magnified, and does not depend
upon how large the magnifications are. Hence, according to this
argument, $P_m$ should simply be a measure of the filling factor of
overdense regions, and therefore should decrease with
increasing $\sigma_8$ at fixed $\Omega_0$, $\lambda_0$, and $H_0$.
We do not see this trend in Figure~11. We even see the opposite
trend ($P_m$ increasing with $\sigma_8$) in
a few cases, such as $\Omega_0=1$, $\lambda_0=0$, 
$H_0=65\rm\,km\,s^{-1}Mpc^{-1}$
(open triangles in bottom panel). This shows that the
interpretation of the magnification distributions in terms of
filling factors of overdense and underdense regions might be
sufficient to explain the properties of $P(\mu)$ for $\mu<1$,
as we did in \S5.1.1, but is too simplistic in the regime $\mu>1$.

The basic flaw in this argument is the assumption that $P_m$
depends only on whether or not sources are magnified
(by having the beam going through a high-density region),
and not by how much. If there was only one lens plane between
the source and the observer, this argument 
would be correct: $P_m$ would be equal to the filling factor of the 
overdense regions on that plane. However, there are many planes, and even
though there is an optimal redshift $z_{\rm L}$ where most
of the matter responsible for lensing is located, there should be several
planes at redshift $z\sim z_{\rm L}$ that contribute to lensing
significantly. Suppose that the light coming from a source
hits a high-density region on plane $i$ and then a low-density region
on plane $j$. Plane $i$ will magnify the source and plane $j$ will demagnify 
it. Whether the demagnification caused by plane $j$ is sufficient to
bring the value of $\mu$ below unity will depend on how large the
magnification by plane $i$ was. For small values of $\sigma_8$, the filling 
factor of overdense regions on plane $i$ is large, and many sources
are magnified, but they are only slightly magnified ($\mu\simgreat1$),
and the demagnification caused by plane $j$ will bring many of these sources
down to $\mu<1$. With larger $\sigma_8$, fewer sources are magnified,
but they are magnified to larger values of $\mu$, and more of them can
``survive'' the demagnification caused by plane $j$. This can also
be regarded as a consequence of the absence of correlation between planes.
Increasing $\sigma_8$ increases the density contrast on each plane, but
by stacking uncorrelated planes along the line of sight, the resulting mean
density contrast is much smoother than the density contrast on any individual
plane, and the effect of increasing $\sigma_8$ is mostly averaged-out.

\bigskip

\epsfysize=16cm
\hskip1.5cm\epsfbox{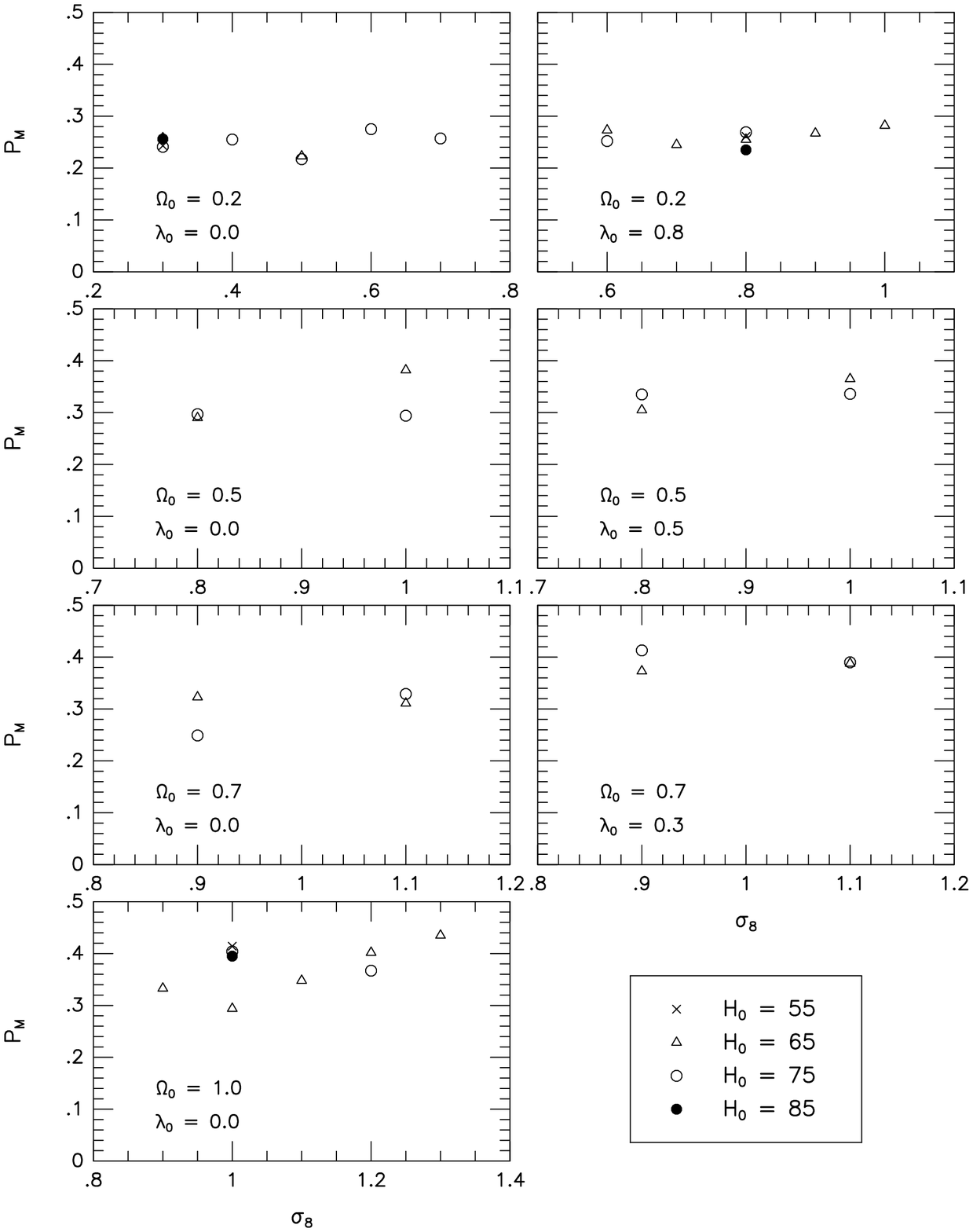}

\medskip

{\narrower\noindent 
Figure 11: Magnification probability $P_m$ vs. $\sigma_8$. The values
of $\Omega_0$ and $\lambda_0$ are indicated in each panel. The various
symbols correspond to various values of $H_0$.\par}

\bigskip

The only significant trend we found is that $P_m$ increases with
$\Omega_0$, as shown in Figure~12. In spite of the large scatter, the
trend is significant. For $\Omega_0=0.2$, $P_m$ is between 0.2 and 0.3,
while for $\Omega_0=1$, $P_m$ is larger than 0.29 for all models, with a
mean value of order 0.36. The dependence of the distances upon $\Omega_0$
is probably responsible for this effect. As $\Omega_0$ goes up, the
distances become shorter, and correspondingly there are fewer lens planes
near $z\sim1$, reducing the ``averaging-out'' of the density contrast 
discussed above. 

Figure 12 shows that for $\Omega\leq0.5$, $P_m$ does not seem to depend upon 
$\lambda_0$. For $\Omega_0=0.7$, however, $P_m$ is significantly larger for 
$\lambda_0=0$ models than $\lambda_0>0$ models. This is very strange,
since the $\lambda_0=0$ models and flat models resemble each other more
closely as $\Omega_0$ increases. This effect is probably spurious, 
a consequence of poor statistics. As $\Omega_0$ increases, more experiments
must be performed in order to obtain good statistics. This was our
motivation for performing a large number of experiments for the
Einstein-de~Sitter model. However, we did not performed a particularly
large number of experiments for the $\Omega_0=0.7$ models, which
we regard as the least interesting models
presented in this paper.

\epsfysize=6.5cm
\hskip3.3cm\epsfbox{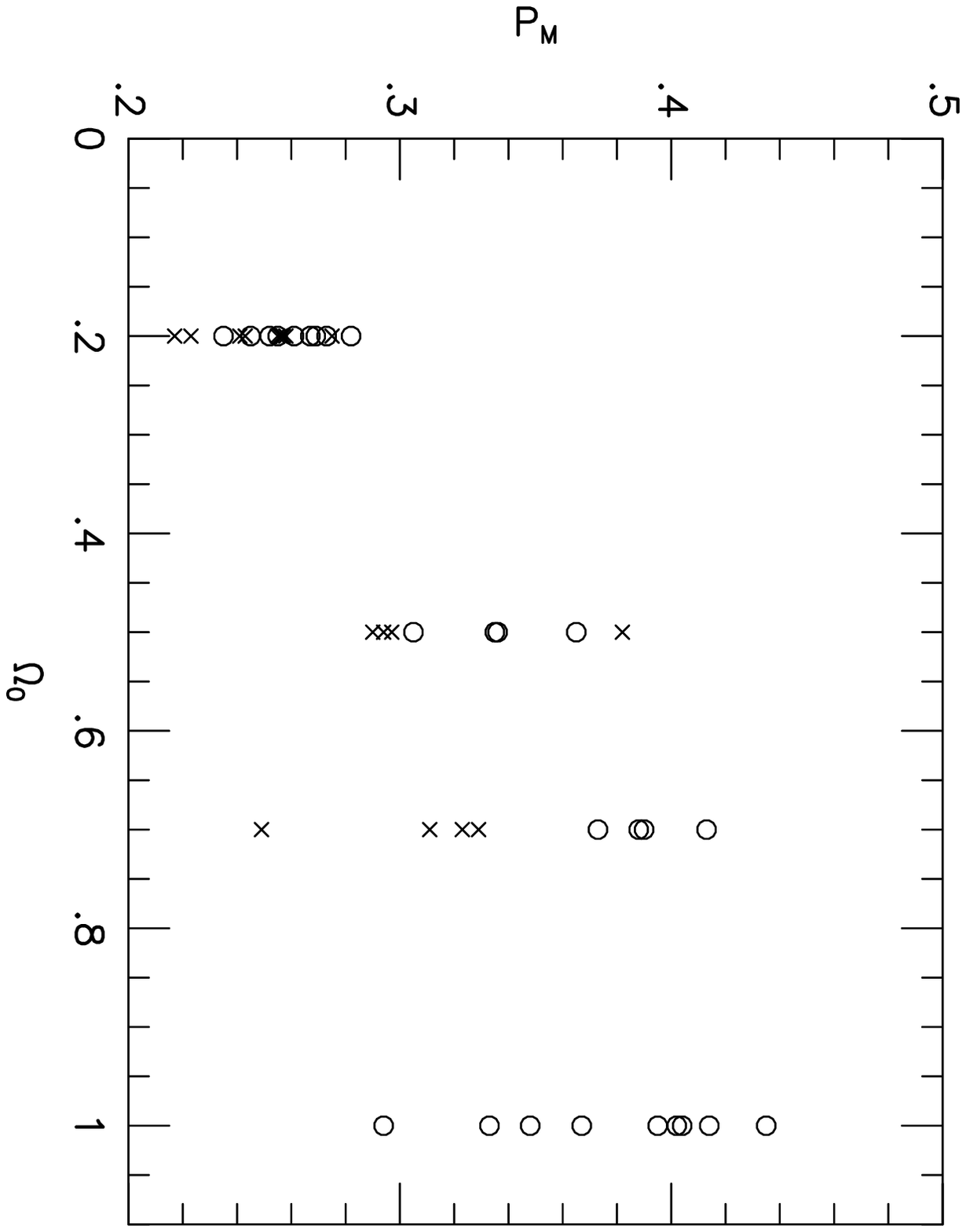}

\medskip

{\narrower\noindent 
Figure 12: Magnification probability $P_m$ vs. $\Omega_0$.
Open circles: flat models (including Einstein-de~Sitter); 
crosses: open models.\par}

\bigskip\smallskip

\ctr{\bf 5.3.\quad Classification of the Images}

\medskip

In Figure 13, we plotted, for illustration purpose, some of the
images we have encountered. Each panel shows the image of a single 
circular source of angular diameter $1''$. 
We have plotted the location of the light rays on the image
plane, and enlarged the dots representing these rays until the
images look continuous. Individual rays can be seen along the edges
of the images. In the absence of lensing, each image would be circular and 
would contain 190 light rays.

Figure~13a shows the most common case: a single image, magnified and
sheared. In such case, there is no galaxy along the line of sight.
The magnification is caused by the background matter 
located near the beam, while the shear is mostly caused by
the background matter and galaxies distant from the beam.
Figures~13b and 13c illustrate
the two most common cases of double image. Figure~13b 
shows a case of strong lensing. In such cases, the magnification is always
large, in the range $\mu=3-7$, the brightness ratio between
the bright image and the faint one is small, of order of a few, 
and the brightest
image forms an arc, while the faintest image is more compact and 
invariably shows a spike that points toward the other image. 
This is a consequence of the particular density profile we assume for
galaxies (eq.~[5]).
Figure~13c is a case of weak lensing. In such cases,
the magnification is small, of order $\mu=1.2-1.3$, and the
brightness ratio is very large. The effect of lensing is to ``carve'' a
second, faint image out of the brighter one. Figure~13d shows an Einstein ring,
caused by a lensing galaxy nearly aligned with the source. For such rings,
the magnification is always very high, sometimes larger than 10. Figure~13e
shows a very different, and unusual, Einstein ring. The ring is pinched at 
two different locations, where the width is only one light ray, and the
magnification is quite small, less than 2. This case is actually similar to the
double image case shown in Figure~13c, and one 
can go from one case to the other
continuously by shifting the location of the source.
Figures~13f--13h show cases of triple image. The images can form either
a circular pattern (as in Fig.~13f), in which case the brightness ratios 
between the images are
small, or a linear pattern (as in Figs.~13g and~13h), in which case one image
is usually much brighter than the other two, that image being 
located either in the
middle, as in Figure~13g, or on the edge, as in Figure~13h. 
The latter one is similar to some triple images generated by
Makino \& Tomita (1995, their Figs. 6b and 6d).
Figures~13i and~13j show cases of quadruple image. These cases 
suggest that several galaxies are involved in the lensing.
Figure~13k shows an interesting case of a double image, one of them being a 
ring. This is actually a combination of the cases shown in Figures~13b and~13d.
There are two galaxies near the line of sight, one responsible for forming
the ring, and the other responsible for 
forming the second image. Figure~13l shows a very rare case of an Einstein
ring with two holes. This requires an alignment between the source and
2 intervening galaxies, each of them producing a hole. 
Figures~13m and~13n show even rarer cases of an Einstein ring with 2 holes and
a secondary image, and an Einstein ring with 3 holes, respectively.
We have encountered only one of each of these cases in all of our experiments.

\bigskip

\epsfysize=16cm
\hskip1.3cm\epsfbox{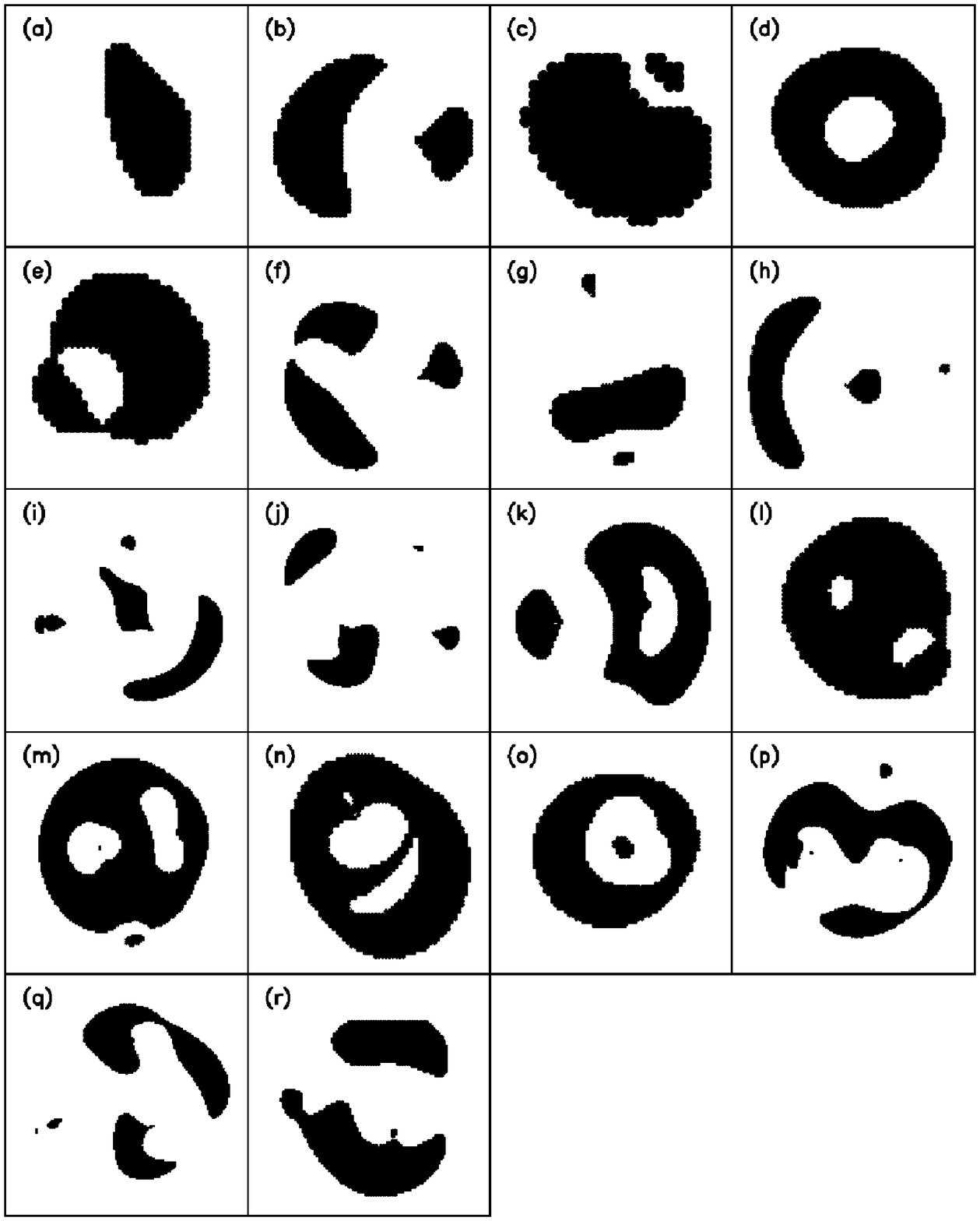}

\medskip

{\narrower\noindent 
Figure 13: Images of lensed circular sources, for a few interesting cases:
(a) single image; (b) and (c) double image; (d) and (e) Einstein ring;
(f), (g), and (h) triple image; (i) and (j) quadruple image;
(k) Einstein ring with secondary image; (l) double Einstein ring;
(m) double Einstein ring with secondary image; (n) triple Einstein ring;
(o) Einstein ring with central image; (p), (q), and (r) complex
images caused by multiple lensing.\par}

\bigskip

Figure~13o shows an Einstein ring with a central spot. 
The presence of this spot is a consequence
of the particular model we use for the galaxies. We do not represent galaxies
as singular isothermal spheres, but rather as nonsingular isothermal spheres
with a central core. Such objects can produce three images if the angular
separation between the object and the source is sufficient small (see
SEF, pp. 244 and 396, and \S5.5.1 below). In this case, the three images are
located on the line going through the source and the lens on the
celestial sphere, one image on each side of the lens, and the third image in
the core of the lens. If the lens and the source are aligned, there
are no preferred directions, and the two outer images will turn into a ring,
while the third image, located in the core of the galaxy, will form a ``spot.''
This explains the existence of images like the one displayed in Figure~13n,
but it does not explain the existence of rings {\it without} spot, such
as the one displayed in Figure~13d. In our experiments, less than 1\% of
the rings have a central spot. This is simply a resolution
effect. As SEF point out, for lenses modeled as
nonsingular isothermal sphere, the third image, located in the core, is 
very faint. If the brightness of that image is less than one light ray, we 
cannot resolve it. Indeed, we often found rings containing a spot composed of
less than five rays, which we regarded as being underresolved.

Finally, in Figures~13p--13r, we plotted some
of the strangest images we encountered. These complex
images result from lensing by several galaxies, combined with shear caused 
by the background matter.

The reader should keep in mind that most of these cases are extremely rare.
99.7\% of the images fall in the category illustrated by Figure~13a,
a single image, magnified and sheared. Of the remaining cases, the vast 
majority of them fall in the categories illustrated by Figures~13b--13d.
The cases shown in Figures~13e--13r are all very rare. In \S5
we will discuss in more detail the probability of occurrences of
these various cases.

\bigskip\smallskip

\ctr{\bf 5.4.\quad The Shear Distributions}

\medskip

There are several ways to compute the shear caused by gravitational
lensing. The most direct way is to compute iteratively
the magnification matrix along each light ray (SEF, \S9.1.2), and
compute its eigenvalues. This approach is not very practical for our
experiments. We consider beams composed of light rays separated by 
$0.064''$. The magnification matrix would tell us the properties
(rotation and deformation) of a source having that size. We are, however,
considering sources of angular diameter $1''$, containing, in the absence
of magnification, 190 light rays. We could always try to somehow combine 
the magnification matricies for all the rays inside an image to infer
its properties, but there is a much simpler approach. Since we are
actually resolving the shapes of individual images, we
can estimate the shear simply by computing the aspect ratio of the images.

To compute the aspect ratio of an image, we first compute the
geometrical center of the image on the image plane,
$${\bf r}_{{\rm gc},i}={1\over N_i}\sum_k {\bf r}_{k,i}\,,\eqno{(28)}$$

\noindent where ${\bf r}_{k,i}$ is the location of light ray $k$ in image $i$,
and $N_i$ is, as usual, the number of rays in image $i$. We then compute the 
2-dimensional quadrupole tensor of the image,
$${\bf Q}=\sum_k({\bf r}_{k,i}-{\bf r}_{{\rm gc},i})
({\bf r}_{k,i}-{\bf r}_{{\rm gc},i})=\left[\matrix{A & B \cr B & C \cr}\right]
\,,\eqno{(29)}$$

\noindent where the last equality defines the components $A$, $B$, $C$.
The aspect ratio of the image is obtained from the
eigenvalues $\lambda_1$, $\lambda_2$ of ${\bf Q}$, as follows,
$${a_1\over a_2}=\left\{[A+C+[(A-C)^2+4B^2]^{1/2}\over[A+C-[(A-C)^2+4B^2]^{1/2}
\right\}^{1/2}\,.\eqno{(30)}$$

\noindent where $a_1$ and $a_2$ are the long and short ``axes'' of the image.
One interesting property of this expression is that in the case of an
elliptical image, $a_1$ and $a_2$ are the semi-major and semi-minor
axes of the ellipse, respectively. The resulting shear distributions are 
plotted in Figures 14--17. Error bars have the same meaning as in Figures 6--9.

\bigskip\smallskip

\ctr{\sl 5.4.1.\quad The $\sigma_8$ Dependence}

\medskip

Figure~14 shows the shear distributions for models with the
same values of $\Omega_0$, $\lambda_0$, and $H_0$, and different values of
$\sigma_8$. In the absence of lensing, the distributions would be 
$\delta$-functions located at $a_1/a_2=1$. As for the magnification 
distributions, the larger the departure from a $\delta$-function is,
the stronger the effect of lensing is. The top panel shows various 
Einstein-de~Sitter models with different values of $\sigma_8$. As $\sigma_8$
increases, the peak of the distribution decreases while the high-tail
of the distribution increases. This was expected, since the large-scale
structure, whose amplitude is measured by $\sigma_8$, 
is the primary origin of the shear. Interestingly, the distributions
are much less sensitive to the value of $\sigma_8$
for the other models. For the $\Omega_0=0.2$, $\lambda_0=0$ model, the
curves in the high-tail regions are very similar and cross each other
several times, while for the $\Omega_0=0.2$, $\lambda_0=0.8$ model
the cases $\sigma_8=0.8$ and $\sigma_8=1.0$ are almost undistinguishable.
Therefore, the dependence of the shear distribution upon $\sigma_8$
is more pronounced for models with large $\Omega_0$. We found a
similar trend in Figure~6 for the magnification distributions. 

\bigskip

\epsfysize=14.3cm
\hskip2cm\epsfbox{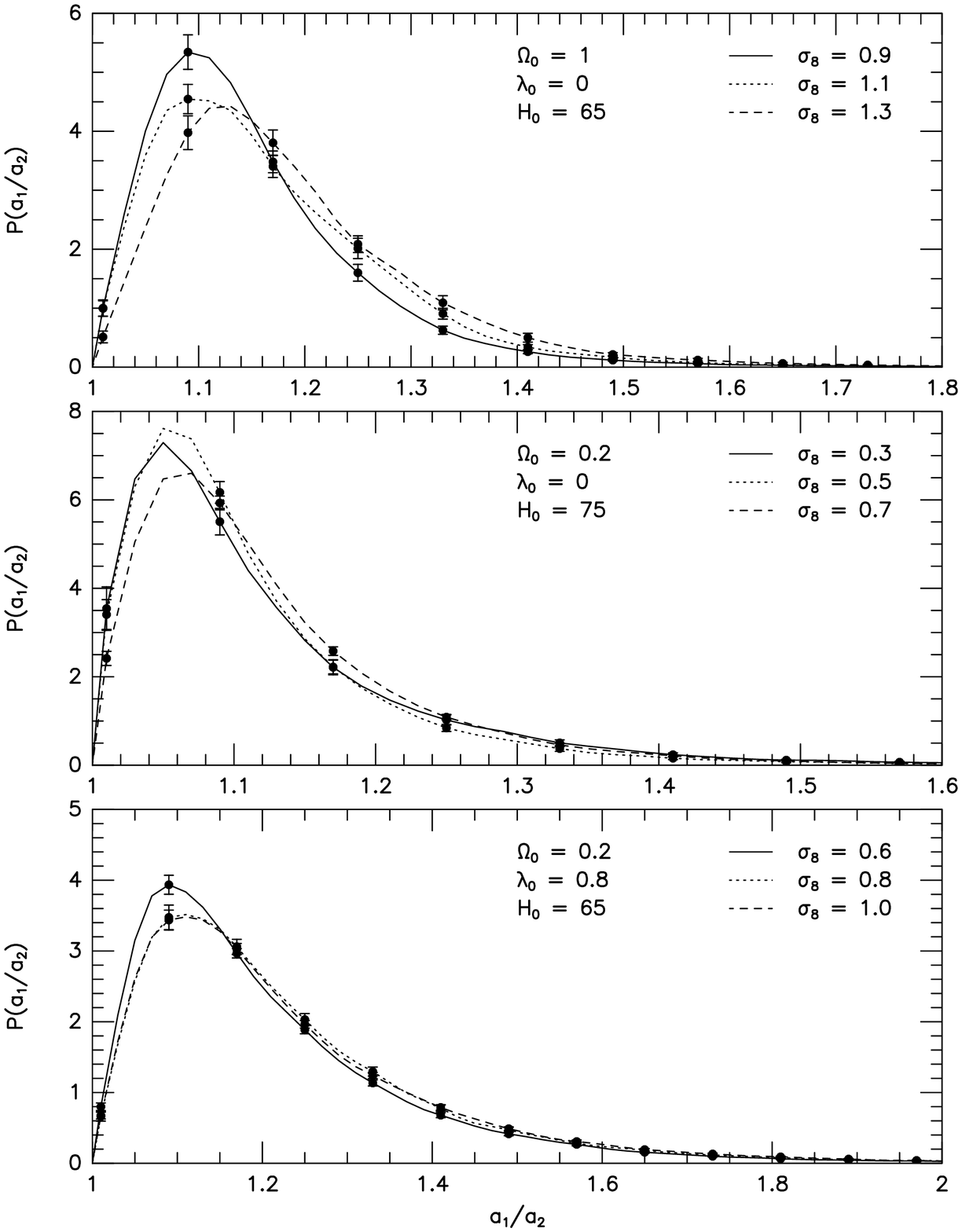}

{\narrower\noindent 
Figure 14: Distributions of aspect ratios for various combinations
of $\Omega_0$, $\lambda_0$, and $H_0$, showing the effect of varying 
$\sigma_8$. Error bars indicate the $1-\sigma$ uncertainty $\Delta P$ on 
the mean, for a few representative bins. See text for details.\par}

\bigskip\smallskip

\ctr{\sl 5.4.2.\quad The $H_0$ Dependence}

\medskip

Figure~15 shows the shear distributions for models with the
same values of $\Omega_0$, $\lambda_0$, and $\sigma_8$, and different values 
of $H_0$. The curves in each panel are very similar, except for the
case $H_0=55{\rm\,km\,s^{-1}Mpc^{-1}}$ in the top panel, for which the 
statistics are poor. We do not find any particular trend, and the ordering
of the curves with $H_0$ is not even monotonic. As for the magnification 
distributions shown in Figure~7, the absence of dependence upon $H_0$ results
from competing effects. With larger $H_0$, the mean background density
is higher, increasing the effects of lensing, but the cosmological
distances are shorter, decreasing the effects of lensing.

\epsfysize=14.3cm
\hskip2cm\epsfbox{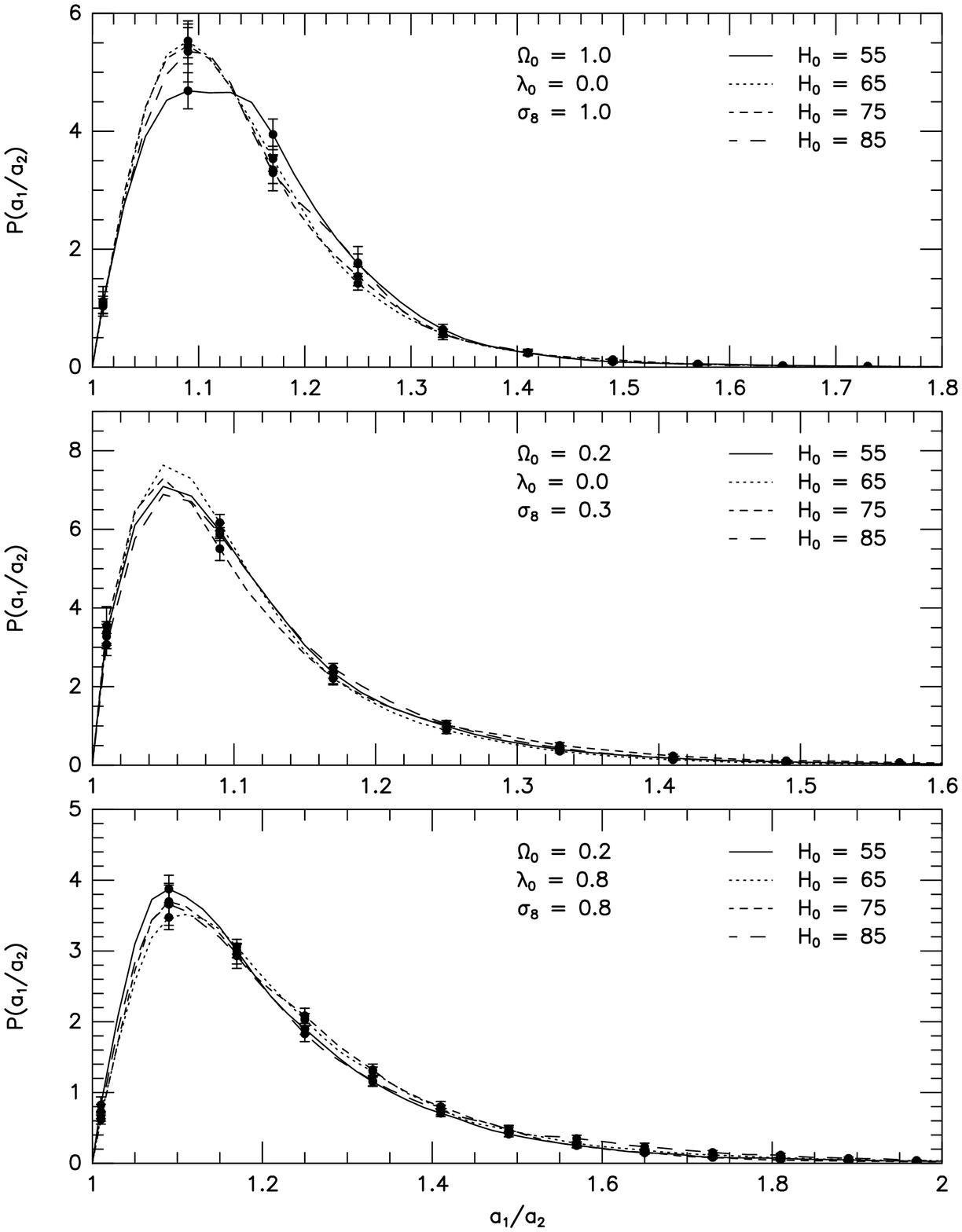}

{\narrower\noindent 
Figure 15: Distributions of aspect ratios for various combinations
of $\Omega_0$, $\lambda_0$, and $\sigma_8$, showing the effect of varying
$H_0$. Error bars indicate the $1-\sigma$ uncertainty $\Delta P$ on 
the mean, for a few representative bins. See text for details.\par}

\bigskip\smallskip

\ctr{\sl 5.4.3.\quad The $\Omega_0$ Dependence}

\medskip 

The top panel of
Figure~16 shows the shear distributions for two 
matter-dominated models ($\lambda_0=0$) with the
same values of $H_0$ and $\sigma_8$, and different values of
$\Omega_0$. This situation is similar to the one encountered in
\S5.1.3 for the magnification distributions. All elements of
gravitational lensing --- cosmological distances, mean background
density, large-scale structure --- depend upon $\Omega_0$, but the
dependence upon the mean background density dominates, and consequently the
distribution is wider (that is, lensing is stronger) for models
with larger $\Omega_0$. The bottom panel of Figure~16
shows comparisons between flat models ($\Omega_0+\lambda_0=1$). 
This is a totally different
situation. As $\Omega_0$ decreases, $\lambda_0$ increases, and the dependence
of the cosmological distances upon $\lambda_0$ becomes the dominant effect,
resulting in stronger lensing for models with smaller $\Omega_0$.

\bigskip\smallskip

\ctr{\sl 5.4.4.\quad The $\lambda_0$ Dependence}

\medskip

Figure~17 shows the shear distributions for two models with the
same values of $\Omega_0$, $H_0$, and $\sigma_8$, and different values of
$\lambda_0$. As $\lambda_0$ increases, the distributions become wider, indicating that the effect of lensing is stronger. A larger value
of $\lambda_0$ results in larger cosmological distances, which is
clearly the dominant effect here. The effect of introducing
a finite value $\lambda_0$ is very large for the $\Omega_0=0.2$ model,
but becomes weaker for larger values of $\Omega_0$, as the flat and 
open models become more similar.

\bigskip

\epsfysize=9.7cm
\hskip2cm\epsfbox{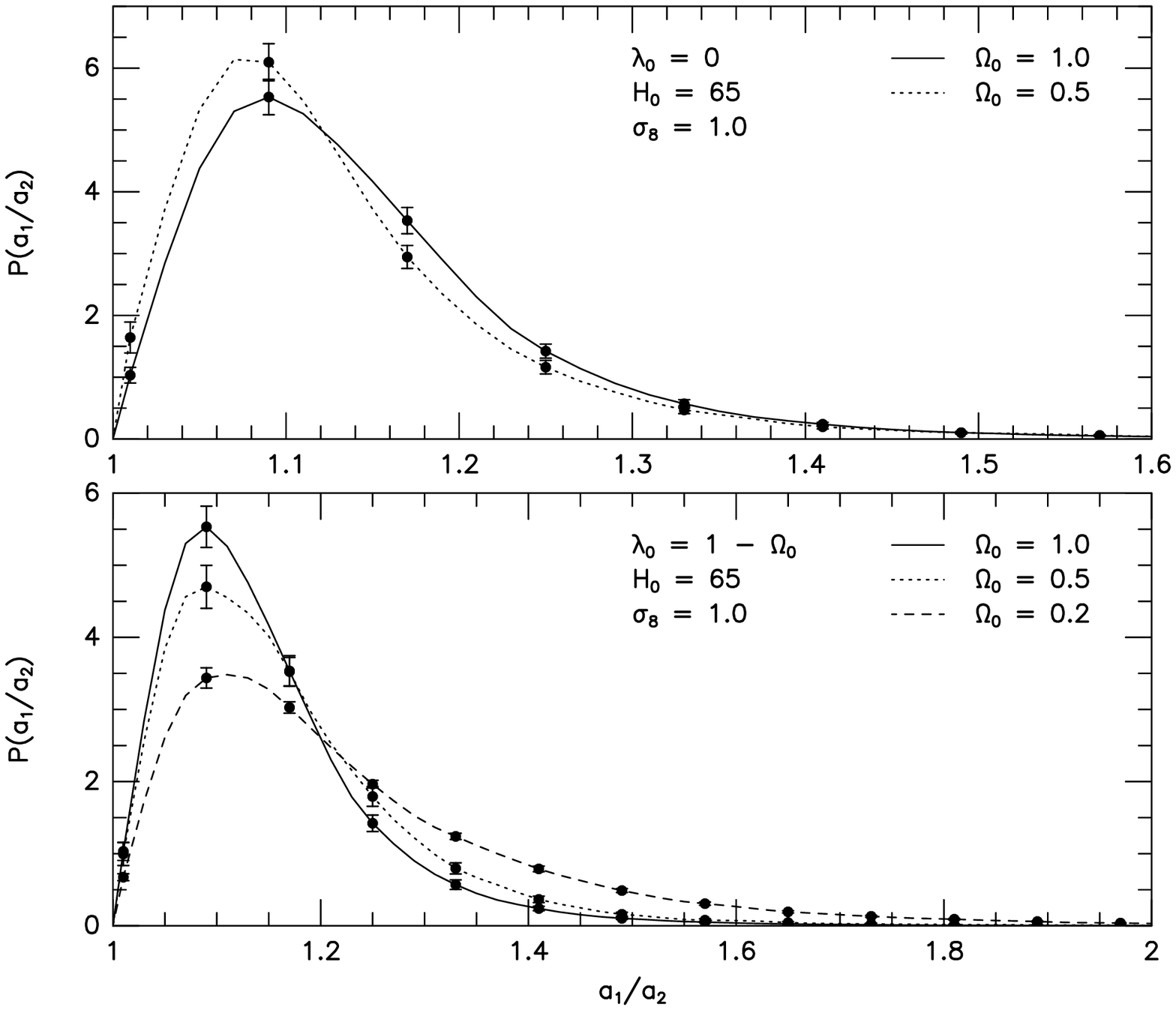}

{\narrower\noindent 
Figure 16: Distributions of aspect ratios for various combinations
of $\lambda_0$, $H_0$, and $\sigma_8$, showing the effect of varying
$\Omega_0$. Error bars indicate the $1-\sigma$ uncertainty $\Delta P$ on 
the mean, for a few representative bins. See text for details.\par}

\bigskip\smallskip

\ctr{\sl 5.4.5.\quad Large-Scale Structure versus Galaxies}

\medskip

All results presented in this section are consistent with the ones
presented in \S5.1. The magnification distributions and
shear distributions have the same, or similar, dependences
upon the cosmological parameters. This strongly suggests that magnification
and shear are two different manifestations of one single physical
phenomenon: weak lensing. In \S5.1.5, we argued that the effect of galaxies
on the magnification distributions is small, because galaxies cover only
a small fraction of the lens planes. For the shear distributions,
the situation is different. While magnification is primarily caused 
by the matter located near the beam, shear is primarily caused by the matter
located away from the beam. In this case, the cross section of galaxies 
becomes irrelevant. The lack of contribution to the shear from galaxies comes
from two different effects. First, even for the lowest mean background
densities we consider, $\Omega_0=0.2$, the galaxies account for only
24\% of the total mass. Second, as explained in great details in Paper~I,
each galaxy in the simulation resides on top of an extended, smooth,
compensating ``hole''
of negative density, which is introduced to account for the matter that has 
been removed from the intergalactic medium to form this galaxy in the 
first place. This is necessary, otherwise the process of ``adding''
galaxies to the simulation would not conserve mass.
\footnote{$^{15}$}{This is
an improvement over the approach used by Jaroszy\'nski (1991), which involved
an overall, uniform reduction of the mean background density.}
As a result, the total mass galaxy + ``hole'' is zero, and galaxies
have no influence beyond the radius of the hole, which we
fixed at $r_{\rm hole}=1\,\rm Mpc$. Hence, only the (usually) small
fraction of galaxies which are located within $1\,\rm Mpc$ of the beam can
contribute to the shear.

\vfill\eject

\epsfysize=9.7cm
\hskip2cm\epsfbox{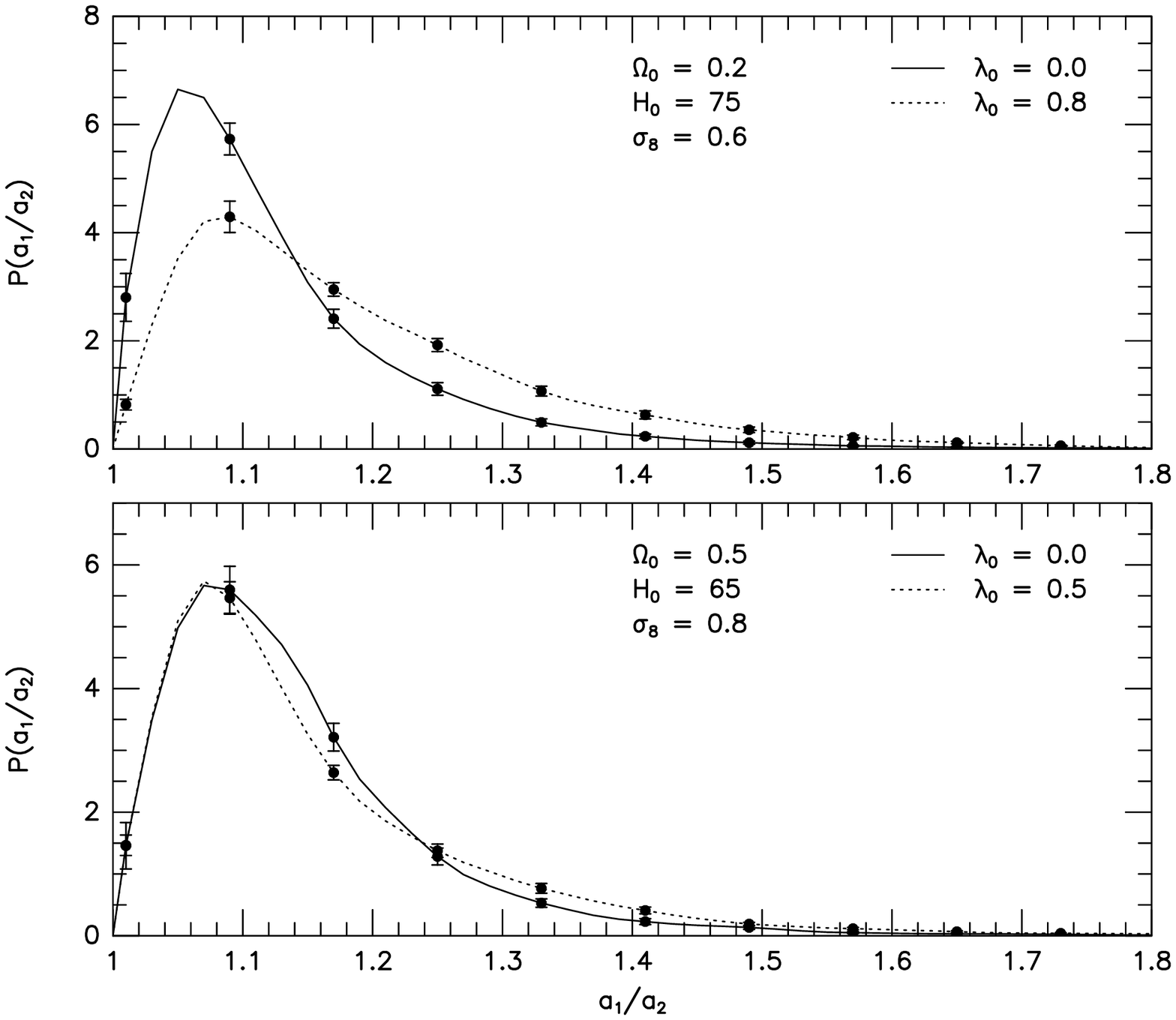}

{\narrower\noindent 
Figure 17: Distributions of aspect ratios for various combinations
of $\Omega_0$, $H_0$, and $\sigma_8$, showing the effect of varying
$\lambda_0$. Error bars indicate the $1-\sigma$ uncertainty $\Delta P$ on 
the mean, for a few representative bins. See text for details.\par}

\bigskip\smallskip

\ctr{\bf 5.5.\quad The Multiplicity of Images}

\medskip

\ctr{\sl 5.5.1.\quad Practical Considerations}

\medskip

For all cases of multiple images found in our experiments, we have computed
the image separations, defined as the angular separation between the
center of the images. In doing so, we had to decide which cases should
actually be identified as multiple images. We have designed an algorithm
with pattern recognition capability to automatically identify multiple
images and compute their separations, as well as the brightness of each image.
It turns out that in a very large number of cases, one of the images contains
only one or two rays. In such cases, that image is demagnified by a factor
of 100, and the brightness ratio between the two images is of order of
100 or more. It is extremely doubtful that an observer could resolve such
a faint image. Indeed, as we will argue in \S5.5.2 below, the vast majority
of observed gravitational lenses probably contain an unresolved image.

To keep the number of cases with multiple images at a manageable level,
we excluded from this study all images containing less than 5 light rays.
Therefore, a double image with one image containing less than 5 rays
is treated as a single image, a triple image is treated as a double image,
and so on. By imposing this restriction, we are postulating that
an image with less than 5 rays could never be resolved. However, we are
not assuming that an image with 5 rays or more is resolvable. The 5-ray
limit was introduced mostly for convenience, to facilitate the
analysis, but as we will show in \S5.5.2 below, the number of double
images we predict is much too large to agree with observations, indicating that
a realistic cutoff would have to be much larger than 5 rays. 

\bigskip\smallskip

\ctr{\sl 5.5.2.\quad Theory, Simulations, and Observations}

\medskip

We define $P_n$ as the probability that a randomly selected source will have
$n$ images. In our experiments, each cell on the source plane is a potential
location for a source, and cells are given equal probability. To
compute $P_n$ we simply divide the number of cases with $n$ images
by the total number of cells included in the analysis. We have performed
a total of 3,798 experiments. For most experiments, 841 cells are included
in the analysis. In some cases, however, cells located near the edge
of the beam must be rejected (see \S3.3). Combining all experiments for
all models, we have a total of 3,137,675 cells, each of them representing
a potential source. Excluding any image that is composed of less than 5 rays,
we found 10,728 double images, 126 triple images, and 6 quadruple images.
No cases with five images or more were found. This contradicts the {\it 
odd-number theorem} (SEF, p. 172, Theorem 1), as well as observations,
which find $N_2\sim N_4\gg N_3$ (Kochanek et al. 1998). 
This is a consequence of limited resolution.
In most realistic lens models, whenever multiple images are produced,
one image is always very faint (see, e.g., SEF, pp. 58 and 175), often
too faint to be observed. Furthermore, for
lens models such as nonsingular isothermal spheres (the model used in 
this paper for galaxies), three images at most can be produced, and the
faintest one is located in the core of the lensing galaxy, making it
very difficult to observe. Our experiments do not have such limitations,
but are limited by the number of light rays used. An unlensed source
contains 190 light rays, and therefore images with a magnification
$\mu<1/190=0.0053$ are unresolved. 

\bigskip\smallskip

\ctr{\sl 5.5.3.\quad Double Images}

\medskip

For each model, we computed the probability $P_2$ of finding a double image.
The results are plotted in Figure~18. 
There are 43 points in each panel, corresponding to the 43
different cosmological models considered.
The values of $P_2$ are quite high for the reasons explained above.
There is a strong trend of $P_2$ to increase with $\lambda_0$, as
the top right panel of Figure~18 shows. 

\bigskip

\epsfysize=12cm
\hskip1.5cm\epsfbox{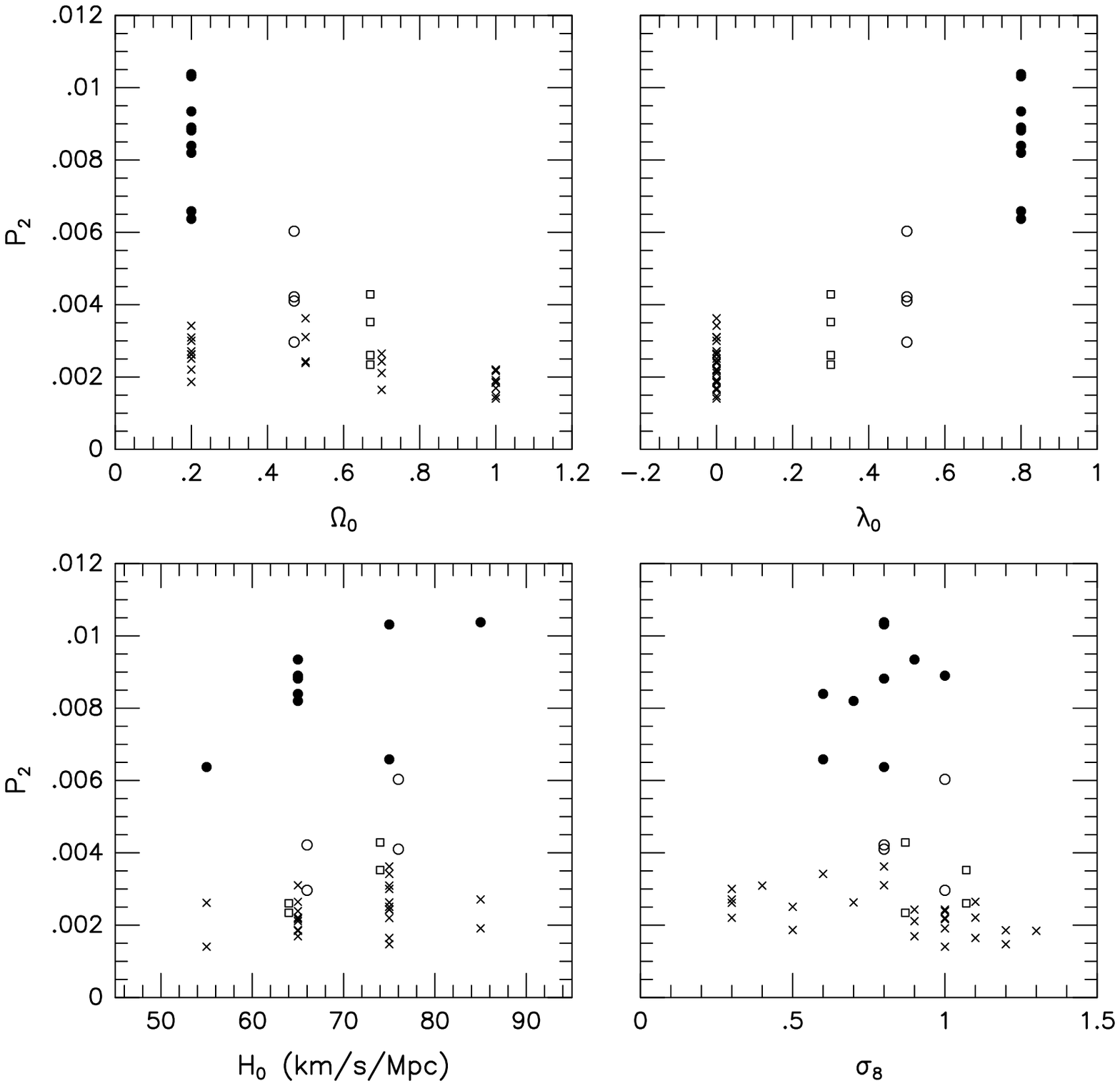}

{\narrower\noindent 
Figure 18: Double-image probability $P_2$, versus $\Omega_0$ 
(top left panel), $\lambda_0$ (top right model), $H_0$
(bottom left panel), and $\sigma_8$ (bottom right panel).
Each point corresponds to one cosmological model. Symbols
indicate the value of the cosmological constant;
crosses: $\lambda_0=0$; open squares: $\lambda_0=0.3$;
open circles: $\lambda_0=0.5$; filled circles: $\lambda_0=0.8$.\par}

\bigskip

In order to study the variations of $P_2$ with the other parameters 
{\it at fixed $\lambda_0$}, we use different symbols
to designate the different values of $\lambda_0$. 
The large scatter in the values of $P_2$
is clearly caused by the dependence of
$P_2$ upon $\lambda_0$, as the various symbols indicate.
On all panels, $\lambda_0=0.8$ models (solid circles) are concentrated at 
the top, while $\lambda_0=0$ models (crosses) are concentrated at the bottom.
Looking at the crosses in the top left panel, and looking
separately at the various symbols in the bottom panels, we find
no obvious trend, indicating that $P_2$ is essentially independent of
$\Omega_0$, $H_0$, and $\sigma_8$ at fixed $\lambda_0$.
These results implies that (i) double images, and multiple images in general,
are caused by galaxies and not by the background large-scale structure,
(otherwise there would be a strong dependence upon $\sigma_8$)
and (ii) the strong dependence of $P_2$ upon $\lambda_0$
indicate that the cosmological distances are the dominant effect
in multiple imaging. 

The most important and interesting property of double images, after
their likelihood of occurrence, is the distribution of
angular separations between the images. This will be the topic of \S5.6
below.

\bigskip\smallskip

\ctr{\sl 5.5.4.\quad Triple Images}

\medskip

We found 126 cases with triple images, out of a total of 3,137,675
potential sources. With such a small number of cases, we cannot make a
quantitative determination of the probability $P_3$ for 
occurrence of triple images, and its relationship with the 
cosmological parameters. Still, we have enough cases to study the
general properties of triple images, and to identify trends. It turns
out that several of the results found for the double images apply to
triple images as well.

We plotted the properties of triple images in Figure~19. The top left
panel shows a histogram of the magnification distribution. 
The magnifications are quite high, usually in the range $\mu=2-7$,
with some cases having $\mu>10$. Triple images are strong
lensing events caused by massive galaxies, though the
tidal field of the background matter and nearby galaxies might play
a key role in determining the shape of the images.

As for the double images, there is a strong trend for triple images to
occur in models with a large cosmological constant. We studied 17 models
with $\lambda_0\neq0$, and found cases of triple images in all of them but
one: the model $\Omega_0=0.7$, $\lambda_0=0.3$, 
$H_0=65{\rm\,km\,s^{-1}Mpc^{-1}}$, $\sigma_8=0.9$, for which we 
have performed
64 experiments. We also studied 26 models with $\lambda_0=0$, and found
cases of triple images in only 13 of them, in spite of the fact that
we have performed more experiments for these models than 
the $\lambda_0\neq0$ models.
The top right panel of Figure~19 shows the frequency of occurrence
of triple images, defined as the number of cases (indicated
by numbers) divided by the number of experiments, versus $\lambda_0$.
About one fifth of the cases (27 out of 126) were found in models with
$\lambda_0=0$, but 71\% of the experiments (2,699 out of 3,798) were performed
with these models. The trend is clear: the frequency of triple images
increases sharply with $\lambda_0$. As for the double images, this indicate
that the cosmological distances are the dominant effect. 

The bottom left panel of Figure~19 is the scatter plot of the brightness
ratios $B_2/B_3$ and $B_1/B_2$, where $B_1$, $B_2$, and $B_3$ are the
brightnesses of the brightest, intermediate, and faintest images,
respectively. The dashed line corresponds to $B_2/B_3=B_1/B_2$.
The brightness ratios can be very large, up to $\sim100$, and for this reason
we decided to use logarithmic scales. Triple images located 
above the dashed line have $B_2/B_3>B_1/B_2$, and 
are composed of ``two bright images and a faint one,'' while triple 
images located below the dashed line have $B_2/B_3<B_1/B_2$, and 
are composed of ``one bright image and two faint ones.'' These latter cases
outnumber the former ones 2 to 1. This shows that
individual galaxies are not solely responsible for the formation of triple
images, since a galaxy modeled as a nonsingular isothermal sphere
could only produce a triple image composed
of two bright images and one faint, possibly unresolved image. Tidal
perturbation by nearby galaxies and by the background matter, and lensing
by several galaxies, must play an important role.

The bottom right panel of Figure~19 is a scatter plot of the minimum angular
separation $s_{\min}$ and maximum angular separation
$s_{\max}$ between the images, in arc seconds. We can identify three 
specific limiting cases: (i) In the {\it circular limit}, defined by 
$s_{\min}=s_{\max}$ (top dashed line), the three images form a circular
pattern, as in Figure~13f. (ii) In the {\it linear limit}, defined
by $s_{\min}=s_{\max}/2$ (bottom dashed line), the three images form a linear
pattern, with the two outside images equidistant from the central one. 
These are the kind of triple images that a nonsingular isothermal
sphere would produce. (iii) In the {\it hierarchical limit}, defined
by $s_{\min}\ll s_{\max}$ (horizontal axis), two images are very close
to each other, with the third image located much farther. These would
invariably correspond to cases of one bright image and two faint ones, since
the two images located near each other would have to be faint in order
not to overlap. As Figure~19 shows, none of these limits seems to be preferred.
The scatter is large, and the only noticeable thing is that the strong
hierarchical limit does not seem to occur very often, as very few
points are located near the horizontal axis. 

\bigskip

\epsfysize=14cm
\epsfbox{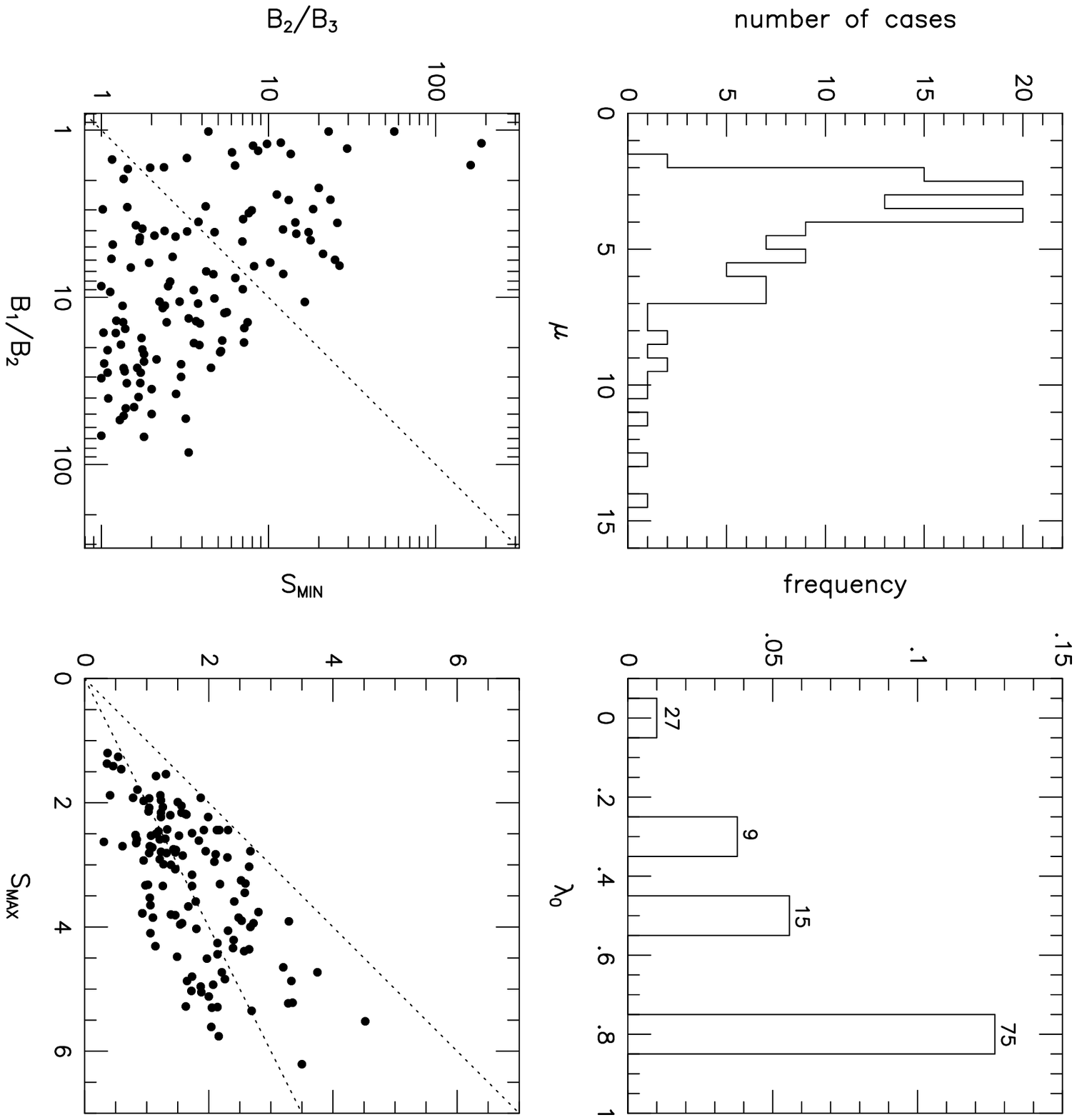}

{\narrower\noindent 
Figure 19: Properties of the triple images. Tof left panel: Histrogram
of the magnification $\mu$. Top right panel: Frequency of occurrence
of triple images versus $\lambda_0$. The numbers indicate the actual
number of cases found. The frequency is that number divided by the
number of experiments. Bottom left panel: brightness ratio $B_2/B_3$ 
(intermediate to faintest) versus brightness ratio $B_1/B_2$ 
(brightest to intermediate). The dashed line corresponds to
$B_2/B_3=B_1/B_2$.
Bottom right panel: minimum angular separation 
$S_{\min}$ versus maximum angular separation 
$S_{\max}$. The dashed lines correspond to the ``circular limit''
$S_{\min}=S_{\max}$ and the ``linear limit''
$S_{\min}=S_{\max}/2$.\par}

\bigskip\smallskip

\ctr{\sl 5.5.5.\quad Quadruple Images}

\medskip

Quadruple images are extremely rare. We only found 6 cases, and two of
them are plotted in Figure~13. All cases were found in models with
$\lambda_0=0.8$, reinforcing the trend of multiple
images being more frequent in models with large cosmological constant. 
The magnifications are $\mu=2.119$, 2.910, 4.846, 8.344, 11.222, and 14.012.
With 3 cases out of 6 having $\mu>8$, quadruple images are
even stronger lensing events than triple images. According to the 
odd-number theorem, each of these cases must be at least a quintuple image
with one faint, unresolved image. This is most likely the case for
the quadruple image shown in Figure~13j. These images forms a circular pattern,
and there is probably a fifth, unresolved image in the center. More complex
cases, such as the one shown in Figure~13i, could very well have more
than 5 images.

\bigskip\smallskip

\ctr{\bf 5.6.\quad The Distribution of Image Separations}

\medskip

\ctr{\sl 5.6.1.\quad Histograms of the Image Separations}

\medskip

For each case with multiple images found in the experiments, we
computed the angular separation between the images.
The center of each image is computed using equation~(28), and
separations are computed between image centers.
Figures~20 and~21 show histograms of the angular separations in arc seconds,
for all models. All histograms are plotted on the same scale to facilitate
comparison. On each panel, we have indicated the total
number $N_2$ of double images found. This number varies among the
different models, first because some models are more likely
to produce multiple images than others, and second because
the number of experiments performed was not the same for all models. 
Models with $N_2<200$ tend to have very noisy distributions,
and many more experiments would be required in order to determine 
the precise shape of these
distributions. Still, several trends are apparent. We are considering sources
with an angular diameter of $1''$. Since one image at least must be brighter
than the unlensed source would be (SEF, p. 172, Theorem 2), the smallest
possible image separation is $0.5''$ corresponding to a double image like
the one shown in Figure~13c. Most histograms in Figures~20 and~21 show
a distributions that rises sharply from $0.5''$ to $1''$, and then drops
slowly at larger separations, with a high-tail that extends to 
separations of order $4''-6''$. 

As in the case of the double-image probability $P_2$ discussed
in \S5.5.3 above, we find no obvious
correlation between the shape of the histograms and the value of 
$\sigma_8$. This again indicates that double images are caused primarily by
direct interaction between the beam and individual galaxies, and not by the
large-scale structure. There is, however, a relationship between the
largest angular separations and the value of $\lambda_0$. For models
with $\lambda_0=0$, the high-tail of the distribution function rarely
extends beyond $4''$, while for $\lambda_0=0.8$ models, the high-tail
often extends to separations of $6''$. As for the probability $P_2$,
the shape of the high-tail depends strongly upon the cosmological distances.
Increasing these distances results in higher image separations for a given
lensing galaxy. This affects the magnification distribution, by extending
the high-tail to higher separation, and also the probability $P_2$, by
``separating'' images that otherwise would have overlapped and been
detected as a single image. Fukugita et al. (1990) found that the image 
separations were independent of the cosmological constant, but only considered
the rms value $\langle s^2\rangle^{1/2}$ of the separation, and not
the actual distributions. Since cases with large separations are rare,
the details of the high-tail can hardly affect the rms value, and therefore
these results are not in contradiction.

In several histograms, especially the ones for $\lambda_0>0$ models,
we see a secondary peak at large separation. Consider for instance the
model $\Omega_0=0.2$, $\lambda_0=0.8$, $H_0=65\rm\,km\,s^{-1}Mpc^{-1}$,
$\sigma_8=0.8$, which is indicated by an asterisk in Figure~20. There
are no double images with separations between $4.00''$ and $4.75''$,
but there are several images with separations larger that $4.75''$. This
might seem like a very small effect that could be dismissed as a statistical
fluctuation, but this feature is found in many histograms, suggesting
that it could actually be real. This could possibly result from a coupling
between galaxies and large-scale structure. Galaxies are predominantly
responsible for multiple imaging. But most galaxies are located inside
clusters, where the density of background matter is high. This background
matter might amplify the lensing effect of the galaxy, resulting in a peak
at high separation angles. This issue requires more investigation.

\bigskip\smallskip

\ctr{\sl 5.6.2\quad A Synthetic Angular Separation Distribution}

\medskip

To gain more insight into the origin and properties of the angular separation
distributions shown in Figures~20 and~21, we will now synthesize an angular
separation distribution using a simple analytical model. We consider
the $\Omega_0=0.2$, $\lambda_0=0.8$, $H_0=65\,\rm km\,s^{-1}Mpc^{-1}$,
$\sigma_8=0.8$ model, and make the following assumptions: (1)  
Lensing is entirely caused by galaxies; we ignore the
lensing by the background matter. (2) Each galaxy acts as if it was alone;
we ignore the tidal effects of nearby galaxies, and the possibility of 
lensing events involving several galaxies. With these assumptions, the 
problem is reduced to studying lensing by isolated, nonsingular isothermal
spheres. A problem of such importance has generated a great deal of
interest in the past, and several analytical results have been derived
(Dyer 1984; Hinshaw \& Krauss 1987; Blandford \& Kochanek 1987; 
Kochanek \& Blandford 1987) which we can now apply to our model. 
We use the notation of SEF, \S12.2.3.

\bigskip

\epsfysize=19cm
\hskip0.5cm\epsfbox{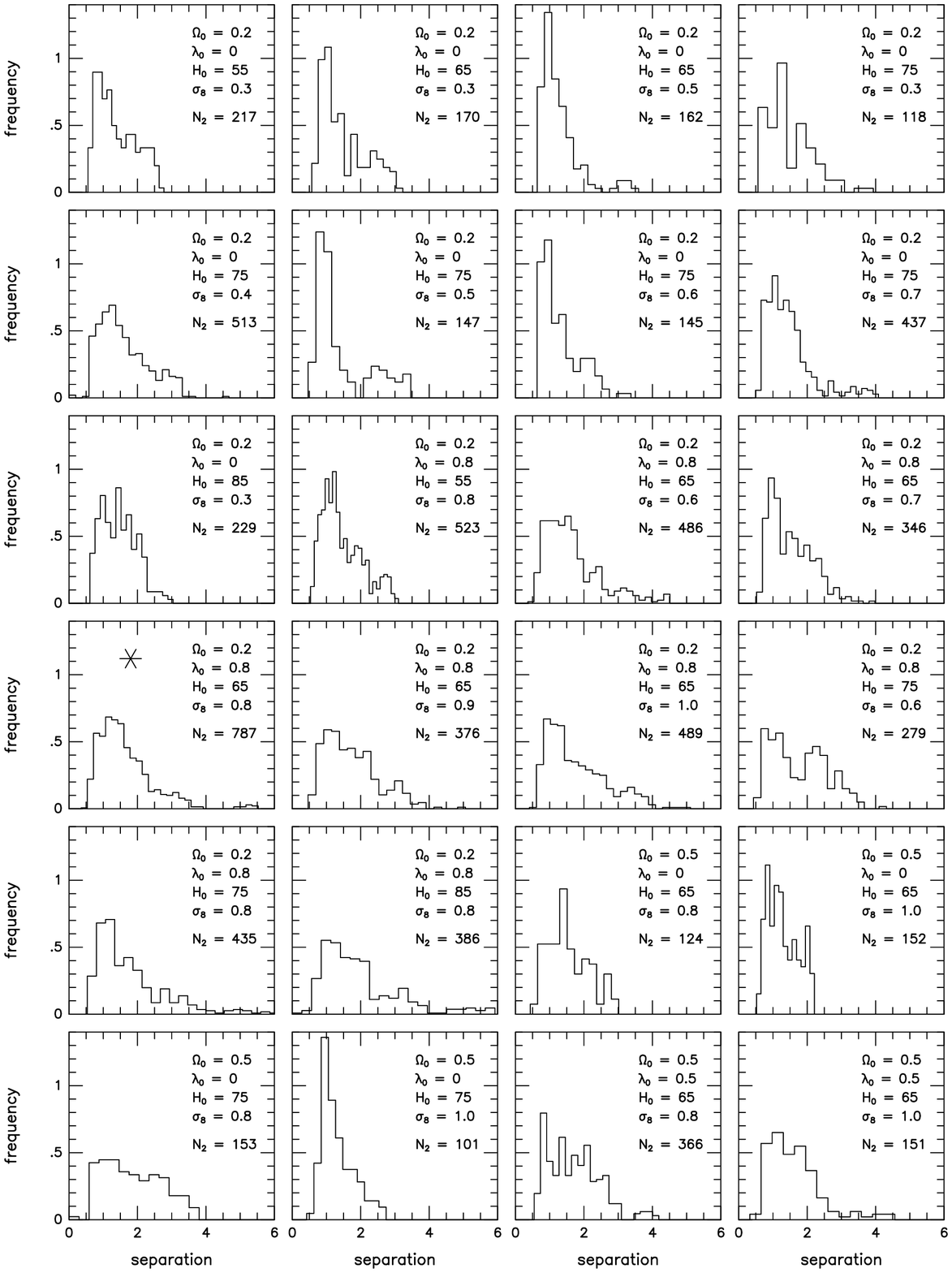}

{\narrower\noindent 
Figure 20: Histograms of the distribution of image separations
in arc seconds. The value of the cosmological parameters
and the number of double images are indicated in each panel
(with $H_0$ in units of $\rm km\,s^{-1}Mpc^{-1}$). The large
asterisk indicates the distribution which is replotted in the 
bottom panel of Fig.~22.\par}

\epsfysize=16cm
\hskip0.5cm\epsfbox{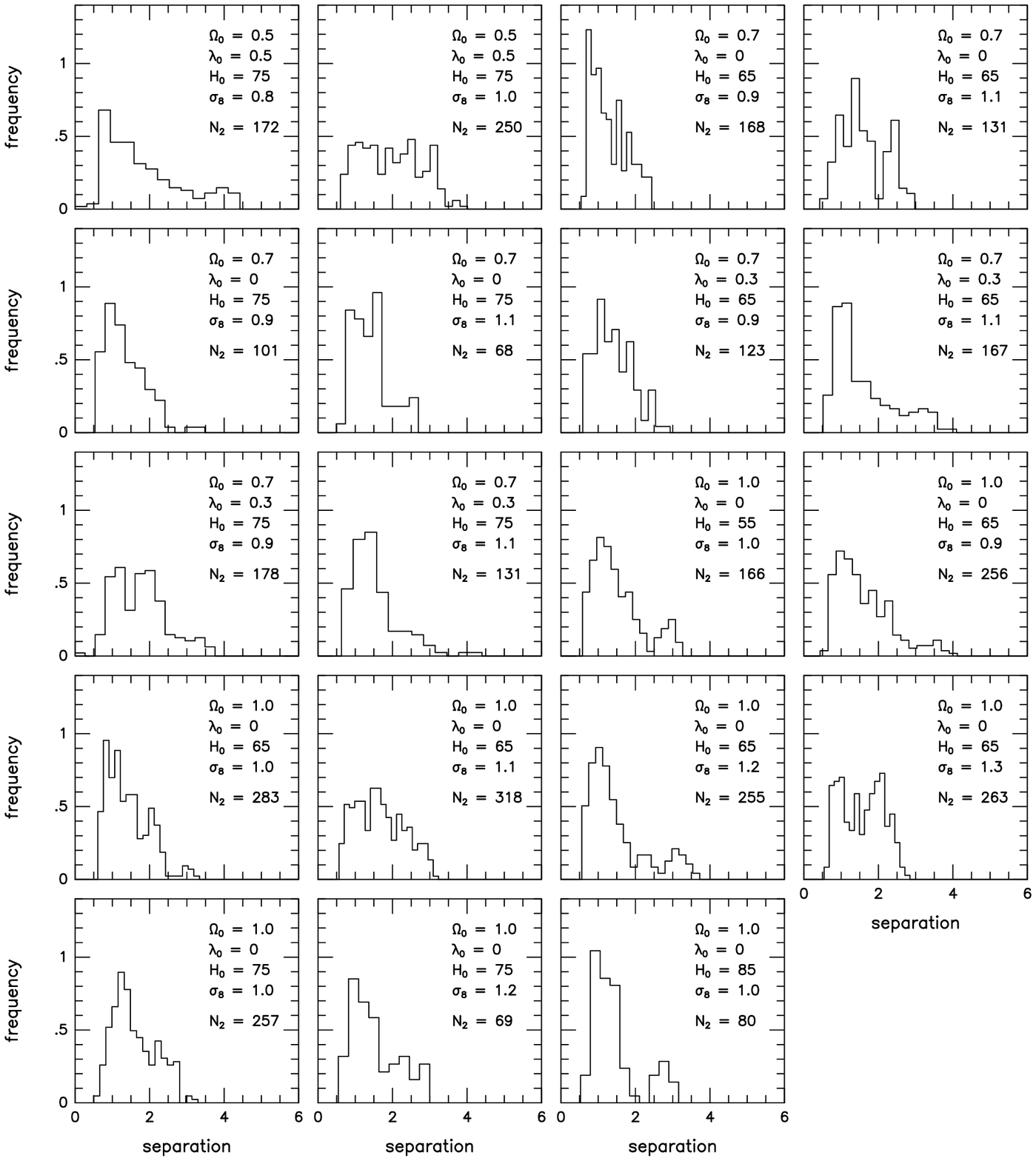}

{\narrower\noindent 
Figure 21: Same as Fig.~20, for different cosmological models.\par}

\bigskip

For each galaxy, we introduce a length scale $\xi_0$, defined by
$$\xi_0=2\pi\left({v\over c}\right)^2{D_LD_{LS}\over D_S}\eqno{(31)}$$

\noindent (SEF, eq.~[8.34a], with $v=\sqrt{2}\sigma_v$), where $v$ is
the circular velocity given by equation~(8). We now define the scaled
quantities $x_c\equiv r_c/\xi_0$ and $y=\eta/\xi_0$, where $r_c$ is
the core radius given by equation~(6), and $\eta$ is the distance between
the source and the optical axis going through the observer and the
center of the lens. We also define, for $x_c<1$, a critical radius 
$y_r\equiv(1-x_c^{2/3})^{3/2}$. The nonsingular isothermal 
sphere has the following properties (SEF, \S12.2.3): (1) If $x_c\geq1$, the
source will have only one image. (2) If $x_c<1$ the source will have
one image if $y\geq y_r$, and 3 images if $y<y_r$. Hence, each lens which
satisfies the condition $x_c<1$ has a cross section for multiple imaging equal
to $\pi y_r^2$.

In our analytical model, we consider as potential lenses all the galaxies
located between the source and the observer. 12\% of the galaxies,
located mostly at redshifts $z<0.1$ or $z>2$, are rejected
as unable to produce multiple images, having scaled core radii
$x_c>1$. The top panel of Figure~22 shows the distribution
$x_c$ for the remaining galaxies. The distribution is bimodal, with most
early type galaxies (ellipticals and S0's) having $x_c<0.05$, and
most spiral galaxies having $0.2<x_c<1$. In our model, spiral galaxies 
have core radii that are typically 10 times larger and circular velocities
that are typically twice smaller than early type galaxies, as Table~1 shows.
Since $x_c\propto r_c/v^2$, we expect the values of $x_c$ to be typically 40
times smaller for early type galaxies than for spiral galaxies, explaining the
bimodal distribution shown in Figure~22. The spread in the distribution 
within each peak is
caused by the dependence of $r_c$ and $v$ upon the luminosity $L$.

\bigskip

\epsfysize=14cm
\hskip3cm\epsfbox{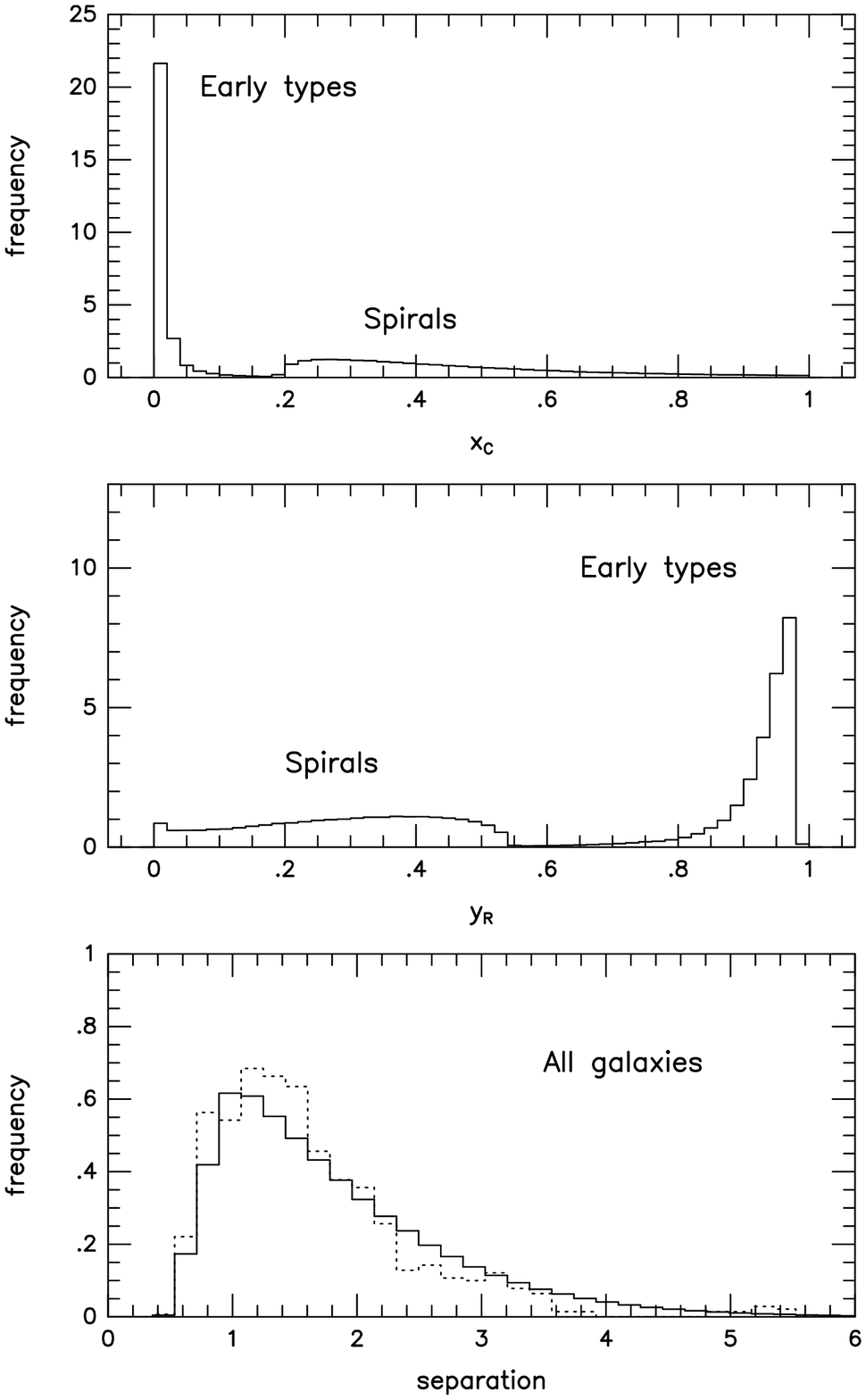}

{\narrower\noindent 
Figure 22: Synthetic angular separation distribution for
$\Omega_0=0.2$, $\lambda_0=0.8$, $H_0=65\,\rm km\,s^{-1}Mpc^{-1}$,
$\sigma_8=0.8$ model. Top
panel: distribution of scaled core radii $x_c$. Middle panel:
distribution of critical radii $y_r$. Bottom panel: synthetic
angular separation distribution (solid line), and actual distribution,
taken from the thirteenth 
panel of Fig.~20 (dotted line). See text for details.\par}

\bigskip

The second panel of Figure~22 shows the distribution of critical radii. Small
values of $x_c$ lead to large values of $y_r$, and vice versa. Early type
galaxies have critical radii concentrated near the maximum possible value
$y_r=1$,\footnote{$^{16}$}{$y_r=1$ would require $r_c=0$, that is, a
{\it singular} isothermal sphere} while spiral galaxies have $y_r<0.55$.
Hence the cross section for multiple imaging is dominated by
early type galaxies. 

If a galaxy modeled as a nonsingular isothermal
sphere produces multiple images (that is, 3 images), the angular separation
$s$ between the two outermost images depends upon the source position $y$,
and has a maximum value given by
$$s_{\max}={2\xi_0\over D_L}(1-x_c^2)^{1/2}\,.\eqno{(32)}$$

\noindent Hinshaw \& Krauss (1987), and
Cheng \& Krauss (1999) showed that the dependence of $s$ on $y$
is weak. Following the suggestion made by SEF (p. 396), we will assume that
whenever multiple images occur, the image separation is of 
order $s\sim s_{\max}$. We can then compute the distribution of image 
separations directly from the distribution of scaled core radii $x_c$. 
In doing so, we found out that many separations are significantly
smaller than the source size, in which case the images strongly overlap
and would be observed as one single image. To build a realistic
distribution, we must impose limits on the smallest possible image
separation that allows individual images to be resolved. We assume that
sources have an angular diameter of $1''$. The smallest possible
image separation is therefore $0.5''$, corresponding to an image 
configuration such as the one shown in Figure~13c. However, with such 
small image separations, the images will often overlap (Fig.~13c is
a particular, ``lucky'' case). We assume that at separations $s<0.5''$
the individual images can never be resolved, that at separations
$s>1''$ they can always be resolved, and that at separations $0.5''<s<1''$
they can sometimes be resolved, with a probability that varies
linearly from 0 to 1 between $s=0.5''$ and $s=1''$. The requirement
that $s$ must exceed $0.5''$ in order to possibly create resolvable
multiple images eliminates as potential lenses
about half of the early time galaxies,
and 97\% of the spiral galaxies. 
Hence, early types galaxies are much more likely to produce resolved
multiple images than spiral galaxies. Not only the early type
galaxies capable of producing resolvable multiple images outnumber the 
spiral galaxies 20 to 1, but their cross sections for multiple imaging
are larger, as the second panel of Figure~22 shows. Notice, however, that
both the numerical simulations and the analytical model treat galaxies as
isothermal spheres. Blain, M\"oller, \& Maller (1999) have shown that the
presence of a galactic disc can significantly increase the cross section
for multiple imaging. Hence, our simulations and analytical model probably 
underestimate the contribution of spiral galaxies to multiple imaging.

We computed the distribution of angular separations, using the above criterion 
for ``resolvability.'' We also gave to each galaxy a weight $w=y_r^2$ to
take the effect of cross section into account. The resulting distribution
is shown in the bottom panel of Figure~22, by the solid line. For comparison,
we plotted the actual distribution (dotted line) for this
particular model, taken directly from the 
thirteenth panel of Figure~20 (indicated by an asterisk). 
The agreement is quite remarkable. The synthetic
distribution reproduces the main features of the actual distribution: 
a sharp rise
a separations $s<1''$, and a slow decline at separations $s>1''$. The
analytical model underpredicts the number of cases with separations
$s\sim1''-1.6''$, and overpredicts the number of cases with separations
$s>2.4''$. This is probably a consequence of ignoring the presence of the
background matter. 

For the particular cosmological model considered in this section, we 
found 787 cases
with double image. Because this number is quite high, we are
confident that the distribution of image separations is fairly well determined
by the experiments. Hence, the bottom panel of Figure~22 serves to
validate the analytical model. Armed with the knowledge that the
analytical model can reproduce, to a reasonable accuracy, the actual
distribution of separations, we intend to performed a much more detailed
study of these distributions and their relationship to the cosmological
parameters, based on the analytical model instead of ray tracing experiments.
This study will be presented in a forthcoming paper (Martel, Premadi,
\& Matzner 2001).

\bigskip\smallskip

\ctr{\bf 5.7.\quad Einstein Rings}

\medskip

Einstein rings are very common in our experiments, but most ot them are
rather unspectacular: The image contains a few hundred rays, and the hole
is made of a small number of ``missing'' rays, often less than 10.
Such holes have an angular diameter $D_{\rm hole}$ of order 
$0.1''$ or less, and it is
doubtful that observations could possibly resolve such small holes.
Only the most massive galaxies can produce spectacular rings with
hole angular diameters of order $1''$, such as the one
shown in Figure 13d, but since only a small fraction of
galaxies are very massive, such spectacular rings are quite rare.

We need to decide where to draw the line between holes that are too small
to be resolved, and holes that are not. This is obviously a very
subjective decision, especially since the possibility of resolving small holes
depends upon the details of observation and
instrumentation. We decided, quite arbitrarily,
to exclude from this study rings with a hole diameter $D_{\rm hole}<0.64''$.
Because this choice is arbitrary, we cannot derive precise statistics of the
occurrence of rings and its dependence upon the cosmological parameters.
We can only describe the intrinsic properties of the rings, and express
their relationship with the cosmological parameters in terms of general 
trends.

The properties of the Einstein rings are summarized in Figure~23. The top
left panel shows a histogram of the number of rings found versus their
magnification $\mu$. This shows that rings are usually high-magnification
events. The histogram peaks at a magnification $\mu\sim5$, and then
drops slowly with magnification. Several rings have a magnification $\mu>10$, 
and the largest magnification we encountered was $\mu=24.097$
(it is the double ring shown in Fig.~13m). This
justifies our use of the term ``spectacular'' to designate such rings.

\bigskip

\epsfysize=15.5cm
\hskip2cm\epsfbox{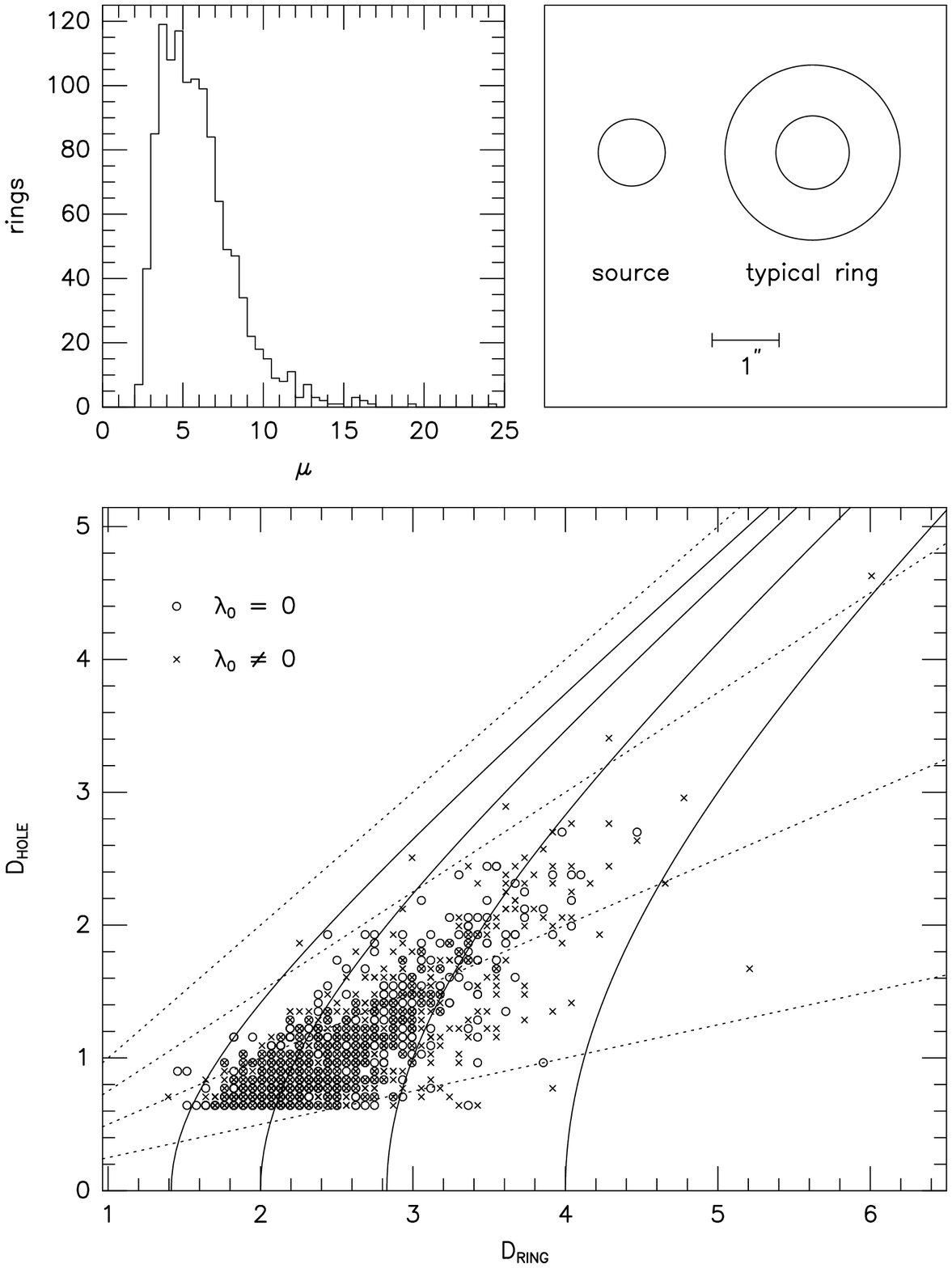}

{\narrower\noindent 
Figure 23: Properties of the rings. Top left panel: histogram of the
number of rings versus magnification $\mu$. Top right panel: 
schematic plot showing the image of an unlensed source and
a ``typical ring,'' obtained by averaging the properties of all the
rings found in the simulations. The horizontal bar indicates 1 arc second.
Bottom panel: scatter diagram of
the hole diameter $D_{\rm hole}$ versus ring diameter $D_{\rm ring}$
measured in arc second. The different
symbols indicate models with and without
a cosmological constant.
The solid curves, form top to bottom, correspond to magnifications $\mu=2$,
4, 8, and 16. The dotted lines, from left to right, correspond
to thickness parameters $T_{\rm ring}=0.00$, 0.25, 0.50, and 0.75.\par}

The top right panel of Figure~23 shows a ``typical'' ring, obtained by
averaging the ring diameters $D_{\rm ring}$ (the outer diameter
of the images) and the hole diameters 
$D_{\rm hole}$ over all rings found in all experiments. The average
hole diameter is $\langle D_{\rm hole}\rangle=1.09''$, slightly larger than
an unlensed source. The average ring diameter is
$\langle D_{\rm ring}\rangle=2.62''$.

In the bottom panel of Figure~23, we plot the distribution of ring
diameters and hole diameters in arc seconds. The solid curves are
contours of constant magnification, from left to right: $\mu=2$,
4, 8, and 16. Most rings have magnifications between $\mu=4$
and $\mu=8$ as we saw also in the histogram. The circles and crosses
indicate models with $\lambda_0=0$ and $\lambda_0\neq0$, respectively.
Rings with $\mu<8$ are found in comparable numbers
in both cases, but rings with $\mu>8$ are
found predominantly in models with $\lambda_0\neq0$, and particularly
in models with $\lambda_0=0.8$. This is the most important trend we
found. The largest magnifications we have
encountered, $\mu=19.446$ and $\mu=24.097$,
were for models with $\lambda_0=0.8$.
We also found a weaker trend: $\mu$ tends to increase with $H_0$.
The dashed lines are, from top to bottom,
lines of constant thickness parameter $T_{\rm ring}=0.00$, 0.25, 0.50, and
0.75, where the thickness parameter is defined by
$$T_{\rm ring}\equiv{D_{\rm ring}-D_{\rm hole}\over D_{\rm ring}}\,.
\eqno{(33)}$$

\noindent An infinitely thin ring has $D_{\rm hole}=D_{\rm ring}$
and $T_{\rm ring}=0$; an infinitely thick ring has $D_{\rm hole}=0$
and $T_{\rm ring}=1$. Figure~23 shows that most rings are quite thick,
having $T_{\rm ring}$ in the range $0.5-0.75$. The typical ring shown
in the top right panel has a thickness parameter $T_{\rm ring}=0.58$. 

\bigskip\smallskip

\ctr{\bf 6.\quad DISCUSSION}

\medskip

The algorithms used for generating the density structures in the universe
and for computing the trajectories of light rays are based on a certain
number of approximations. In this section, we review these various 
approximation, and their possible effect on the results.

The \#1 approximation of any multiple 
lens-plane algorithm is that a continuous distribution of matter can 
be approximated by a finite number of lens planes, and that the 
results converge if the number of planes is sufficiently high. This 
issue was investigated by Lee \& Paczynski (1990), and more
recently by Martel et al. (2000). This latter paper
showed that for sources located at
redshift $z=3$, 14 lens planes are sufficient to
achieve convergence. In this paper, we used between 26 and 69 planes,
depending on the model. Hence, the convergence of our multiple
lens-plane algorithm is not an issue.

Our algorithm for locating galaxies inside dense regions, which
is described in detail in Paper I, ignores the existence
of small-scale correlations, which lead to
galaxy pairs such as the Local Group. 
However, as Figure 10 (bottom panel) shows,
pairs of nearby galaxies are actually quite common in our
simulations. Most of these pairs are not real physical pairs, 
being made of galaxies at different distances seen in projection,
but this does not matter since only the 
projected surface density enters in the lens equation.
The projected distributions of galaxies would hardly look any 
different if pairs were explicitly included in the algorithm.

The relation given by equation (8), relating the circular velocity
to the luminosity, might break down in the limit of small 
luminosities. This would not affect the magnification and shear 
distributions, which are determined primarily by the background
matter, but it could affect the distribution of
image separations. However,
using the equations given in \S5.6.2, we can easily show that
the high-luminosity galaxies have larger cross sections
for multiple imaging, and produce larger separations.
With a Schechter luminosity function, low-luminosity
galaxies are exponentially more abundant, but this does not affect
the results because these galaxies produce separations that are 
much too small to be resolved. As we indicate
in \S5.6.2, imposing a minimum separation of $0.5''$ ``eliminates''
half of the early type galaxies and 97\% of the spiral ones, and 
most of these galaxies are low-luminosity.

Galaxies evolve. Not only their number
density changes as a result of
merging, but their individual density profiles might change
also. We decided to ignore these effects in the current paper.
Our method for locating galaxies in the
computational volume is only approximate, and one might 
question the relevance of adding detailed
modeling of the merging history and galaxy evolution, 
considering the uncertainties already present in the algorithm.
We believe that our neglect of galaxy evolution 
and merging does not affect our results significantly. First, 
the weak lensing (magnification and shear) is unaffected by 
the galaxies. As for the strong lensing, the effect can be 
important. However, one critical point
is that this paper presents a
{\it comparative} study of cosmological models. Some of the lensing 
properties might change if the algorithm is modified. But these 
changes would not necessarily affect the trends seen in the variations 
of these properties with the cosmological parameters. To give a 
specific example: consider the double image probability $P_2$ plotted 
in Figure~18. Would the values of $P_2$ be different if we had used
different models galaxies? Most certainly.
Would the trends revealed by Figure~18 ($P_2$ increasing strongly with
$\lambda_0$, with no dependence on $\Omega_0$, $H_0$ and
$\sigma_8$) be different? Very unlikely.
The focus of this paper was to determine the relationship between the lensing
properties and the cosmological parameters.
We intend to present a study of the effect of galaxy merging, and
of considering various radial density profiles of galaxies, 
in a forthcoming paper. The focus of that paper will be different, and 
we will certainly not consider 43 different cosmological models.

We superpose on top of each galaxy a ``hole'' of
negative density to represent the matter that has been
removed from the intergalactic medium by the galaxy
formation process (Paper I). This
is certainly the most uncertain part of our algorithm. This
uncertainty simply reflects our lack of understanding of the galaxy
formation process. At large scales, the density structures are 
dominated by the distribution of background matter, and the formation
of these structures in CDM universes is well understood and 
accurately simulated by the $\rm P^3M$ algorithm. At small scale, the
density structures are dominated by galactic halos, and we represent 
these halos using analytical models that are in good agreement
with observations, at least at low redshifts. It is the 
intermediate scale, immediately above the galactic scale, which is 
poorly understood. Galaxies must clearly remove matter from the
intergalactic medium when they form, but the effect of this removal
on the density structure of the intergalactic medium
around galaxies is not known.
Yet, it cannot be ignored. Adding galaxies to the system using a
Monte-Carlo method increases the total mass of the system, and 
therefore the mass of the background matter must be reduced somehow
in order to conserve mass. By superposing a hole on top of each galaxy,
we assume that when a galaxy forms, the mass that accumulates
in that galaxy comes from its vicinity. This is highly uncertain. 
Fortunately, the effect of the holes on the lensing properties of
galaxies is small. When the beam hits a galaxy,
it encounters 3 different density structures:
the galaxy itself, the hole superposed on top of the galaxy,
and the background matter.
Of all these three components, the hole is by far the least important one. 
The (negative) mass of the hole is equal to the mass of the galaxy,
which is much smaller than the mass of the nearby background matter,
even for models with $\Omega_0=0.2$. Furthermore, the holes are 
``spread'' over radii much larger than the galaxies themselves. These 
holes should be seen as a small correction to the density of the 
background, to account for the removal of matter by the galaxy 
formation process. Notice also that according to the recent
paper by Cheng \& Krauss (1999), the effect of the 
background matter on the image separation caused by galaxies is 
at most 10\%, and the presence of the hole introduces only a 
small correction to this value.

We assume throughout this paper that sources have an angular diameter of $1''$.
To test the effect of the source size on the results, we selected
a particular model, $\Omega_0=0.2$, $\lambda_0=0.8$, 
$H_0=65\rm\,km\,s^{-1}Mpc^{-1}$, $\sigma_8=0.8$, and
performed an additional 101 experiments using a smaller beam,
$10.95''\times10.95''$ instead of $21.9''\times21.9''$. 
With the same number of cells in the beam, this amounts to considering 
sources of angular
diameter $0.5''$. We found that the magnification distribution
$P(\mu)$ is independent of the source size, the distributions for $1''$
sources and $0.5''$ sources
being nearly identical. Hovewer, the image multiplicities
are reduced. From $(P_2,P_3,P_4)=(0.00882,0.00009,0.00001)$ for
$1''$ sources, they dropped to 
$(P_2,P_3,P_4)=(0.00616,0.00006,0.00000)$ for
$0.5''$ sources. This was expected, since a smaller source does not sample 
as well
the inhomogeneities in the density structure which are responsible
for lensing.

\bigskip\smallskip

\ctr{\bf 7.\quad SUMMARY AND CONCLUSION}

\medskip

We have studied the propagation of light in inhomogeneous
universes, for 43 different {\sl COBE\/}-normalized CDM
models with various combinations of $\Omega_0$, $\lambda_0$,
$H_0$, and $\sigma_8$. 
We have performed a total of 3,798 numerical experiments, using a multiple 
lens-plane algorithm that enabled us to study the properties of weak
and strong lensing simultaneously.
Each experiments consisted of propagating a square beam of
angular size $21.9''\times21.9''$, composed of $341\times341$ light rays, 
from the observer to a source plane located at redshift $z=3$. 
Our main results are the following:

(1) At fixed $\Omega_0$, $\lambda_0$, and $H_0$, the magnification
distribution depends upon $\sigma_8$. As $\sigma_8$ increases, the low-tail
of the magnification distribution shifts toward lower magnifications, 
because light rays are more likely to propagate through mostly underdense
regions. The high-tail of the magnification distribution is
hardly affected. This result indicates that it is
the background matter, and not galaxies, that are primarily responsible
for the magnification of sources.
 
(2) At fixed $\Omega_0$, $\lambda_0$, and $\sigma_8$, the magnification
distribution becomes narrower as $H_0$ increases, reflecting the 
fact that cosmological distances become shorter. This trend is
rather weak, because the distance effect is partly compensated
by the fact that the mean background density increases with $H_0$. 

(3) At fixed $H_0$ and $\sigma_8$, the low-tail of the magnification
distribution shifts to lower values (that is, larger demagnification)
as $\Omega_0$ increases for $\lambda_0=0$ models, and shifts to higher
values
as $\Omega_0$ increases for $\Omega_0+\lambda_0=1$ models. In the case
$\lambda_0=0$, the dependence of the mean background density upon $\Omega_0$
dominates over the respective dependences of the cosmological distances
and large-scale structure upon $\Omega_0$. 
In the case $\Omega_0+\lambda_0=1$, the
dominant effect is the dependence of the cosmological distances upon
$\lambda_0$.

(4) At fixed $\Omega_0$, $H_0$, and $\sigma_8$, the magnification
distribution becomes wider (that is, stronger lensing) as 
$\lambda_0$ increases, because of the increase in the cosmological distances.
The effect is particularly large for models with $\lambda_0=0.8$.

(5) The magnification probability $P_m$ is almost independent of 
$\sigma_8$, for
any combination of $\Omega_0$, $\lambda_0$, and $H_0$, indicating that
$P_m$ does not depend strongly on the amount
of large-scale structure. Our interpretation is that the beam
travels through several underdense and overdense regions whose
effects mostly cancel out. Increasing $\sigma_8$ makes the underdense
regions more underdense and overdense regions more overdense,
but their effects still mostly cancel out. The only trend we found
is that $P_m$ increases with $\Omega_0$ at fixed $\lambda_0$, $H_0$,
and $\sigma_8$. This suggest that $P_m$ depends primarily on the mean
background density.

(6) The shear distribution has essentially the same dependences upon the
cosmological parameters as the magnification distribution.
At fixed $\Omega_0$, $\lambda_0$, and $H_0$, the shear distribution
becomes wider with increasing $\sigma_8$. At fixed
$\Omega_0$, $\lambda_0$, and $\sigma_8$, the shear distributions is very
unsensitive to the value of $H_0$. At fixed $H_0$ and
$\sigma_8$, the shear distribution becomes wider with increasing $\Omega_0$ 
for $\lambda_0=0$ models, and and narrower with increasing $\Omega_0$
(and correspondingly decreasing $\lambda_0$) for $\Omega_0+\lambda_0=1$ models.
At fixed $\Omega_0$, $H_0$, and $\sigma_8$, the shear
distribution becomes wider as $\lambda_0$ increases.

(7) The similarities found between the properties of the magnification
distribution and shear distribution suggests that both phenomena
have the same origin: weak lensing. The dependence of the magnification
distribution upon $\sigma_8$ is a clear indication that the large-scale
structure in the background matter, and not individual galaxies, are
responsible for determining this distribution. This occurs
because galaxies cover only
a small fraction of each lens plane, and are not too likely to be hit by
the beam. This argument does not apply to the shear distribution,
however, since shear is not caused by the matter being directly hit by the
beam, but rather by the distant matter. In this case, the absence of 
significant contribution from galaxies originate from their small mass
(at most 24\% of the total mass is in galaxies), and by the presence,
in the algorithm, of a compensating underdensity on top of each galaxy
to account for the mass removed from the background matter during
the galaxy formation process. The strong dependence of the magnification
and shear distribution upon $\lambda_0$ indicate that while the 
large-scale structure is responsible for these effects, the magnitude
of these effects depends strongly upon the
cosmological distances.

(8) The double-image probability $P_2$ increases strongly with $\lambda_0$.
We found no clear dependence upon $\Omega_0$, $H_0$,
and $\sigma_8$. The absence of dependence upon $\sigma_8$ indicates
that individual galaxies, and not the background matter
are responsible for forming double images, which constitute examples of
strong lensing. The strong dependence upon $\lambda_0$ indicates that,
again, the dominant effect is the cosmological distances.

(9) The distribution of image separations has properties similar to
the probability $P_2$: a strong dependence upon $\lambda_0$ and no
dependence upon $\sigma_8$. The distribution
rises sharply from $0.5''$, half the angular diameter of the source,
to $1''$, and then drops slowly, down to separations of order
$4''$ for $\lambda_0=0$ models and $6''$ for $\lambda_0=0.8$ models. In many
cases, we also found a small, secondary peak is the distribution at
separations of order $5''$. 

(10) Using an analytical model, which assumes that multiple images are
entirely caused by galaxies, we have generated a synthetic distribution
of image separations for one of the model. This synthetic distribution
reproduces the actual distribution obtained from the simulations remarkably
well, indicating that the assumption of galaxies being responsible
for generating multiple images is correct. The only discrepancies are at
separations of $1''-1.6''$, where the synthetic distribution is too low,
and at separation larger than $2.4''$, where the synthetic
distribution does not drop fast enough. These discrepancies probably 
result from
the presence of the background matter, which is ignored in the analytical 
model.

(11) We only found 126 cases of triple images and 6 cases of quadruple
images. With such small numbers, we cannot do any precise determination
of the probabilities $P_3$ and $P_4$, but we can identify several trends.
Triple and quadruple images are strong lensing, 
high-magnification events caused by galaxies. 
Their magnification distribution peaks
around $\mu=4$, and the high-tail extends to values larger than $\mu=10$. 
These cases are predominantly found in $\lambda_0>0$ models, indicating
again that the effect of the cosmological distances dominates. Triple
images come out in a variety of pattern, including circular and linear
patterns. About 2/3 of the triple images are made of one bright image and
two faint ones.

(12) It is difficult to determine precise statistics for the occurrence of 
Einstein rings. Rings are very common, but we rejected most of them because
the ``hole'' was very small. We only included in the analysis rings with
hole diameters larger than $0.64''$. Rings are the most extreme cases of
strong lensing found in our experiments. Most rings have magnifications
in the range $\mu=4-8$, and the magnification
distribution extends to $\mu=25$. The 
brightest image found in any given experiment is almost always a ring.
Rings are usually quite thick, with a thickness comparable to the
angular size of the source. Rings are found both in $\lambda_0=0$
models and $\lambda_0>0$ models, but bright rings, with $\mu>8$,
are found predominantly in $\lambda_0>0$ models.

We can summarize these results as follows: (1) The cosmological distances play 
a critical role in nearly every aspect of gravitational lensing, both weak
and strong. Consequently, the properties of gravitational lenses depend
much more strongly upon the cosmological constant $\lambda_0$ than any
other cosmological parameter. (2) Magnification and shear are examples
of weak lensing caused primarily by the distribution of background
matter, with negligible contribution form galaxies. Consequently, these
effect are sensitive to the value of the rms density fluctuation
$\sigma_8$. (3) Multiple images and rings are examples of strong lensing,
caused by direct interaction with galaxies, with at most a small
contribution from the background matter. Consequently, the properties
of multiple images are independent of $\sigma_8$. They are determined by
the cosmological distances, which depend primarily upon $\lambda_0$, and
by the details of the galactic models, which are usually independent of the
cosmological parameters. Therefore, observations
of weak lensing can be used to determine
the cosmological constant and the {\it unbiased} density structure of the
universe (that is, without having to assume some biasing factor
between luminous and dark matter), while observations of strong lensing
can be used to determine the cosmological constant and the internal
structure of galaxies and clusters. (4) The dependences upon $H_0$ and
$\Omega_0$ are not as simple, because varying these parameters affects
gravitational lensing in several ways that often partly cancel
each other. For instance, the small dependence of the magnification
and shear distribution upon $H_0$ results from the competing effects
of increasing cosmological distance while reducing the mean background
density. Determining $\lambda_0$ and $\sigma_8$ from observations seems much
more promising than determining $\Omega_0$ and $H_0$.

Our experiments consider only sources located at redshift $z=3$, 
in which case
most of the matter responsible for lensing is located at redshifts $z\sim1$.
This is not an important limitation of our study. As Figure~5 shows,
the angular diameter distances vary only weakly with redshift in the
limit $z>1$. Hence, our conclusions remain valid for sources located
at larger redshifts.

In conclusion, our study
shows that it is difficult to single out the effect of each
particular
cosmological parameter. This supports the idea of conducting this
simultaneous
cosmological parameter survey.

\bigskip
 
This work benefited from stimulating discussions with
Andrew Barber, Christopher Fluke, Takashi Hamana, Peter Schneider,
and an anonymous referee.
We are pleased to acknowledge the support of NASA Grant NAG5-2785,
NSF Grants PHY93~10083 and ASC~9504046,
Grant 3658-0624-1999 from the Texas Advanced Research Program,
the University of Texas High Performance Computing Facility
through the office of the vice president for research.
PP gratefully acknowledges the Postdoctoral Fellowship
from the Japan Society for the Promotion of Science and
the hospitalities of the Astronomical Institute of Tohoku
University, and the Centre for Relativity of the
University of Texas at Austin.

%

\parindent=0pt

\def\hh {\hangindent=20pt\hangafter=1}

\bigskip

\ctr{\bf REFERENCES}

\medskip

\hh
Albrecht, A., \& Steinhardt, P. 1982, Phys.Rev.Lett., 48, 1220

\hh
Asada, H. 1997, ApJ, 485, 460

\hh
Babul, A., \& Lee, M. H. 1991, MNRAS, 250, 407

\hh
Bacon, D. J., Refregier, A., \& Ellis, R. S. 2000, MNRAS, submitted (astro-ph/0003008)

\hh
Bardeen, J. M., Bond, J. R., Kaiser, N., \& Szalay, A. S. 1986,
ApJ, 304, 15

\hh
Bartelmann, M., \& Schneider, P. 1991, A\&A, 248, 349

\hh
Bartelmann, M., Huss, H., Colberg, J. M., Jenkins, A., \& Pearce, F. R.
1998, A\&A, 330, 1

\hh
Blain, A. W., M\"oller, O., \& Maller, A. H. 1999, MNRAS, 303, 432

\hh
Blandford, R. D., \& Kochanek, C. S. 1987, ApJ, 321, 658

\hh
Blandford, R. D., \& Nayaran, R. 1986, ApJ, 310, 568



\hh
Bloomfield Torres, L. F., \& Waga, I. 1996, MNRAS, 279, 712 

\hh
Bucher, M., Goldhaber, A. S., \& Turok, N. 1995, Phys.Rev.D, 52, 3314

\hh
Bunn, E. F., \& White, M. 1997, ApJ, 480, 6

\hh
Burles, S., \& Tytler, D. 1998, ApJ, 499, 699

\hh
Charlton, J. C., \& Turner, M. S. 1987, ApJ, 313, 495

\hh
Cheng, Y.-C. N., \& Krauss, L. M. 1999, ApJ, 514, 25

\hh
Chiba, M., \& Futamase, T. 1999, Prog.Th.Phys, 133, 115

\hh
Chiba, M., \& Yoshii, Y. 1997, ApJ, 489, 485

\hh
Chiba, M., \& Yoshii, Y. 1999, ApJ, 510, 42

\hh
Cooray, A. R., Quashnock, J, M., \& Miller, M. C. 1999, ApJ, 511, 562

\hh
Couchman, H. M. P., Barber, A. J., \& Thomas, P. A. 1999, MNRAS, 308, 180

\hh
Davis, M., Efstathiou, G, Frenk, C. S., \& White, S. D. M. 1985, ApJ,
292, 371

\hh
Dressler, A. 1980, ApJ, 236, 351

\hh
Dyer, C. C. 1984, ApJ, 287, 26

\hh
Efstathiou, G., Ellis, R. S., \& Peterson, B. A. 1988, MNRAS, 232, 431
(EEP)

\hh
Falco, E. E., Kochanek, C. S., \& Mu\~noz, J. A. 1998, ApJ, 494, 47

\hh
Falco, E. E., Shapiro, I. I., Moustakas, L. A., \& Davis, M. 1997,
ApJ, 484, 70

\hh
Filippenko, A. V., \& Riess, A. G. 1998,
Physics Reports, 307, 31

\hh
Fluke, C. J., Webster, R. L., \& Mortlock, D. J. 1999, MNRAS, 306, 567

\hh
Fry, J. N. 1985, Phys.Lett.B, 158, 211

\hh
Fukugita, M., Futamase, T., \& Kasai, M. 1990,
MNRAS, 246, 24P

\hh
Fukugita, M., Futamase, T., Kasai, M., \& Turner, E. L. 1992,
ApJ, 393, 3

\hh
Gilliland, R. L., Nugent, P. E., \& Phillips, M. M. 1999, ApJ, 521, 30

\hh
Guth, A. 1981, Phys.Rev.D, 23, 347

\hh
Hamana, T., Martel, H., \& Futamase, T. 2000, ApJ, 529, 56

\hh
Hinshaw, G., \& Krauss, L. M. 1987, ApJ, 320, 468

\hh
Hockney, R. W., \& Eastwood, J. W. 1981, Computer Simulation
Using Particles (New York: McGraw-Hill)

\hh
Hu, W., \& Sugiyama, N. 1996, ApJ, 471, 542

\hh
Im, M., Griffith, R. E., \& Ratnatunga, K. U. 1997, ApJ, 475, 457

\hh
Jaroszy\'nski, M. 1991, MNRAS, 249, 430

\hh
Jaroszy\'nski, M. 1992, MNRAS, 255, 655

\hh
Jaroszy\'nski, M., Park, C., Paczy\'nski, B., \& Gott, J. R. 1990,
ApJ, 365, 22

\hh
Jain, B., Seljak, U., \& White, S. 2000, ApJ, 530, 547


\hh
Keeton, C. R., \& Kochanek, C. S. 1997, ApJ, 487, 42

\hh
Kochanek, C. S. 1992, ApJ, 384, 1

\hh
Kochanek, C. S. 1996a, ApJ, 466, 638

\hh
Kochanek, C. S. 1996b, ApJ, 473, 595

\hh
Kochanek, C. S., \& Apostolakis, J. 1988, MNRAS, 235, 1073

\hh
Kochanek, C. S., \& Blandford, R. D. 1987, ApJ, 321, 676

\hh
Kochanek, C. S., Falco, E. E., Impey, C., Lehar, J., McLeod, B.,
\& Rix, H.-W. 1998, CASTLE Survey, http://cfa-www.harvard.edu/castles

\hh
Krauss, L. M., \& White, M. 1992, ApJ, 394, 385

\hh
Kundi\'c et al. 1997, ApJ, 482, 75

\hh
Linde, A. D. 1982, Phys.Lett B., 108, 289

\hh
Linde, A. D. 1995, Phys.Lett B., 351, 99

\hh
Linde, A. D., \& Mezhlumian, A. 1995, Phys.Rev.D, 52, 6789

\hh
Maoz, D., \& Rix, H.-W. 1993, ApJ, 416, 425

\hh
Makino, N., \& Tomita, K. 1995, PASJ, 47, 117

\hh
Marri, S., \& Ferrara, A. 1998, ApJ, 509, 43

\hh
Martel, H. 1995, ApJ, 445, 537

\hh
Martel, H., \& Matzner, R. 2000, ApJ, 530, 525

\hh
Martel, H., Premadi, P., \& Matzner, R. 1998, ApJ, 497, 512

\hh
Martel, H., Premadi, P., \& Matzner, R. 2000, ApJ, 537, 28

\hh
Martel, H., Premadi, P., \& Matzner, R. 2001, in preparation

\hh
Mart\'\i nez-Gonz\'alez, E., Sanz, J. S., \& Cay\'on, L. 1997, ApJ, 484, 1

\hh
Munshi, D., \& Coles, P. 2000, preprint (astro-ph/0003354)


\hh
Paczy\'nski, B., \& Wambsganss, J. 1989, ApJ, 337, 581

\hh
Park, M.-G., \& Gott, J. R. 1997, ApJ, 489, 476


\hh
Postman, M., \& Geller, M. J. 1984, ApJ, 281, 95

\hh
Premadi, P., Martel, H., \& Matzner, R. 1998, ApJ, 493, 10 (Paper I)

\hh
Ratra, B., \& Peebles, P. J. E. 1994, ApJ, 432, L5

\hh
Roos, M. \& Harun-or-Rashid, S. M. 2000, preprint (astro-ph/0003040)

\hh
Schneider, P., Ehlers, L., \& Falco, E. E. 1992, Gravitational
Lenses (Berlin: Springer-Verlag) (SEF)

\hh
Schneider, P., \& Weiss, A. 1988a, ApJ, 327, 526

\hh
Schneider, P., \& Weiss, A. 1988b, ApJ, 330, 1

\hh
Steigman, G., Hata, N., \& Felten, J. E. 1999, ApJ, 510, 564

\hh
Turner, E. L. 1990, ApJ, 365, L43

\hh
Turner, M. S.. 1999, 
In The Third Stromlo Symposium: The Galactic Halo, eds. 
Gibson, B. K., Axelrod, T. S. \& Putman, M. E., 
ASP Conference Series Vol. 165, 431

\hh
van Waerbeke, L., Bernardeau, F., \& Mellier, Y. 1999, 
A\&A, 342, 15

\hh
van Waerbeke, L., et al. 2000, A\&A, submitted (astro-ph/0002500)

\hh
Wambsganss, J., Cen, R., \& Ostriker, J. P. 1998, ApJ, 494, 29

\hh
Wambsganss, J., Cen, R., Xu, G., \& Ostriker, J. P. 1997, ApJ, 475, L81

\hh
Wang, L., Caldwell, R. R., Ostriker, J. P., \& Steinhardt, P. J. 2000,
ApJ, 530, 17

\hh
Watanabe, K., Sasaki, M., \& Tomita, K. 1992, ApJ, 394, 38

\hh
Weinberg, S. 1972, Gravitation and Cosmology (New York: Wiley)

\hh
Weinberg, S. 1989, Rev.Mod.Phys., 61, 1

\hh
White, S. D. M., Frenk, C. S., Davis, M., \& Efstathiou, G. 1987,
ApJ, 313, 505

\hh
Williams, R. et al. 1996, AJ, 112, 1335

\hh
Williams, R. et al. 1998, in preparation

\hh
Yamamoto, K., Sasaki, M., \& Tanaka, T. 1995, ApJ, 455, 412

\hh
Yoshida, H., \& Omote, M. 1992, ApJ, 388, L1

\vfill\eject\end